\documentclass{style/nseJournal}

\usepackage{amssymb}
\usepackage{amsmath}
\usepackage{subcaption}
\usepackage{hyperref}
\usepackage{cleveref}
\usepackage{lineno}
\usepackage{upgreek}

\usepackage{siunitx}
 \DeclareSIUnit\bar{bar}

\usepackage[acronym,xindy,toc,nonumberlist]{glossaries}

\newacronym{tlk}{TLK}{Tritium Laboratory Karlsruhe}
\newacronym{tapir2}{T$_2$ApIR}{Tritium Absorption InfraRed Spectroscopy 2}

\newacronym{lara}{LARA}{Laser Raman}
\newacronym{muRa}{$\upmu$Ra}{micro Raman}

\newacronym{ftir}{FTIR}{Fourier-transform infrared}

\newcommand{\requiredTemperatureLowerLimitKelvin}{\SI{10}{\kelvin}}

\newcommand{\tritiumBoilingPoint}{\SI{24.99}{\kelvin}} % Souers, vorher war hier 24.92 ohne Quellangabe
\newcommand{\orthopara}{ortho/para}

\newcommand{\HTwo}{\texorpdfstring{H$_2$}{H2}} 
\newcommand{\DTwo}{\texorpdfstring{D$_2$}{D2}} 
\newcommand{\TTwo}{\texorpdfstring{T$_2$}{T2}} 
\newcommand{\QTwo}{\texorpdfstring{Q$_2$}{Q2}}

\newcommand{\SZero}[1]{$S_0(#1)$}
\newcommand{\SZeroBranch}{$S_0$-branch}
\newcommand{\QOne}[1]{$Q_1(#1)$}
\newcommand{\QOneBranch}{$Q_1$-branch}

\newcommand{\QPhonon}[1]{$Q_{P:#1}$}

\newcommand{\SOne}[1]{$S_1(#1)$}
\newcommand{\SOneBranch}{$S_1$-branch}

\newcommand{\HTwoSZeroZero}{\SI{354}{\per\centi\meter}}
\newcommand{\HTwoSZeroOne}{\SI{587}{\per\centi\meter}}
\newcommand{\HTwoSZeroFour}{\SI{1246}{\per\centi\meter}}

\newcommand{\TTwoSZeroZero}{\SI{120}{\per\centi\meter}}

\newcommand{\HTwoSOneOne}{\SI{4498}{\per\centi\meter}}

\newcommand{\HTwoQOneZero}{\SI{4161}{\per\centi\meter}}
\newcommand{\TTwoQOneZero}{\SI{2464}{\per\centi\meter}}

\newcommand{\SapphireRamanOne}{\SI{417}{\per\centi\meter}}

\newcommand{\SapphireRamanThree}{\SI{751}{\per\centi\meter}}

\begin{document}

\title{First results of the Tritium Absorption InfraRed Spectroscopy (T$_2$ApIR) experiment} %title of paper

% Use the \addAuthor macro to add authors in the order they should appear. The second argument corresponds to
% the affiliation declared below.
% The corresponding author should be wrapped in \correspondingAuthor
\addAuthor{\correspondingAuthor{{Alexander} Marsteller}}{a}
% The corresponding author's email can be specified using \correspondingEmail
\correspondingEmail{alexander.marsteller@kit.edu}

\addAuthor{Dominic Batzler}{a}
\addAuthor{Beate Bornschein}{a}
\addAuthor{Lutz Bornschein}{a}
\addAuthor{Elisabeth Eckard}{a}
\addAuthor{Florian Hanß}{a}
\addAuthor{Joshua Kohpeiß}{a}
\addAuthor{Daniel Kurz}{a}
\addAuthor{Ralph Lietzow}{a}
\addAuthor{Michael Sturm}{a}
\addAuthor{Tin Vrkic}{a}
\addAuthor{Stefan Welte}{a}
\addAuthor{Robin Größle}{a}

% Affiliations can be added in the order they should appear. For breaks in addresses, use either \\ or \tabularnewline

\addAffiliation{a}{Institute for Astroparticle Physics, Tritium Laboratory Karlsruhe(IAP-TLK),\\ Karlsruhe Institute of Technology (KIT),\\
            {Hermann-von-Helmholtz-Platz 1, 76344 Eggenstein-Leopoldshafen ,Germany}}

% Add keywords to appear in Abstract in the order they should appear
\addKeyword{Tritium Laboratory Karlsruhe}
\addKeyword{cryogenic tritium}
\addKeyword{thermodynamical properties of hydrogen isotopologues}
\addKeyword{ortho-para}
\addKeyword{Raman spectroscopy}
\addKeyword{infrared spectroscopy}

\titlePage

\begin{abstract}

The literature on experimentally verified material properties of tritium is sparse but information about this is crucial in fusion for pellet production (Magnetic Confined Fusion), target fueling (Inertial Confined Fusion), cryogenic distillation, as well as in astroparticle physics for neutrino experiments, and search for rare physics.
To improve on this, the \gls{tapir2} experiment has been designed and built at the \gls{tlk}, and is in its scientific commissioning phase. 
The main focus of this experiment is to enable the investigation of the properties of all six hydrogen isotopologues and their mixtures in the gaseous, liquid, and solid phase, as well as the dynamics of their phase changes. 
In addition, mixtures with noble gases such as xenon and neon can be investigated.
This is achieved using a cryogenic setup capable of reaching less than \requiredTemperatureLowerLimitKelvin{} in a measurement cell that allows optical access for infrared absorption spectroscopy, Raman spectroscopy and a polariscope setup, as well as temperature and pressure measurement.

\end{abstract}

% \noindent Claim Setzungspaper, Fokus auf Beobachtung von zeitlichen Abänderungen 

% \noindent Konzentration auf die Messmethoden

% \noindent Key Messages:
% \begin{itemize}
%     \item  2D Monitoring optisch/Polariskopie
%     \item Intermolekulare Wechselwirkungen im IR (Ortho/Para Bild)
%     \item Raman monitoring im festen+flüssigen, gasförmig ausstehend
% \end{itemize}

% \noindent Aufhänger:
% Analytic für Flüssigkeiten und Festkörper, Pellets und Targets

% \noindent Key Messages:
% \begin{itemize}
%     \item PoP Qualifizierung von Methoden für Real time monitoring
%     \item IR Kalibrierung erweitern auf o/p und tritium
%     \item Methoden müssen noch parallelisiert werden
%     \item Homogenität, Phasenübergange, etc.
% \end{itemize}
% ----------------------------------------------

% ToDo
% \begin{itemize}
%     \item Annealing physik mehr durchdiskutieren --> Daniel
%     \item Ganzer Paragraph zu IR --> Robin / Daniel
% \end{itemize}

\section{Introduction}
\label{sec:introduction}

Tritium (T, \TTwo{}), the radioactive isotope of hydrogen, is of particular interest as an electron source for neutrino mass measurements \cite{Bornschein2008,Sturm2021,Aker2025,Esfahani2017,Amad2025}, as a natural background and as a calibration source for dark matter search \cite{Aprile2020,Meng2021,Adrover2024,Arnquist2024,Aalbers2025}. 
Tritium is also the most promising fuel in fusion for power generation \cite{VanOost2023}. 
In the former case, detailed knowledge of the properties of tritium is necessary to accurately determine the uncertainties on the neutrino mass measurement and the design of the tritium electron source. 
In the latter case, material properties \cite{Wydra2023} inform the design and feasibility of fuel cycle concepts including processes such  as cryogenic distillation, production of pellets for magnetic confined fusion or targets for laser fusion, and finally for the development of analytic systems and concepts for monitoring and accountancy.
And in general cold hydrogen isotopologues are of interest in wide fields of research from \orthopara{}-conversion \cite{Krasch2023} for hydrogen liquefaction to the production of ultra cold neutrons and tests of fundamental interactions like ultra cold neutrons scattering on cold deuterium \cite{Bison2025}.

Current research at \gls{tlk} focuses on the thermodynamic properties of high purity H-D-T-mixtures \cite{Niemes2023,Wydra2023,Priester2023,Wydra2025} and the development of analytical tools \cite{Priester2022,Niemes2021,Aker2020,Sturm2009,Priester2017} for target production, pellet production, cryogenic pumping and cryogenic distillation, and water detritiation \cite{Cristescu2017} for inertial and magnetic confined fusion. 

In cryogenic distillation for isotope separation, the six hydrogen isotopologues (\QTwo{} = \TTwo{}, DT, \DTwo{}, HT, HD, and \HTwo{}) are separated in a cryogenic refraction column by differences in their vapor pressures at a given temperature. 
As the highest boiling isotopologue, with a boiling point of \tritiumBoilingPoint{} at \SI{1}{bar} \cite{Souers1986}, tritium (\TTwo{}) accumulates at the bottom of the column in the liquid phase.
To monitor the concentration of \TTwo{}, as well as residual amounts of HT and DT, a measurement system is required. 
At the \gls{tlk}, infrared (IR) absorption spectroscopy is under investigation as an online and inline monitoring tool and has been successfully calibrated for the
inactive isotopologues with an accuracy of better than \SI{5}{\percent} absolute \cite{Groessle2017, Mirz2020}.  

The production of fuel targets for inertial confined fusion for a commercial fusion power plant requires a fast rate (\SI{10}{\hertz} \cite{Hurricane2025}) of target production in order to achieve the desired output power while keeping overall tritium inventory low \cite{Goodin2006}.
A key step in this process is loading targets with a deuterium-tritium mixture and fixing it, for example via freezing out the tritium. 

In order to address these open questions an experimental setup to calibrate spectroscopic methods, study thermodynamic properties, and investigate the dynamic phase space behavior of all six hydrogen isotopologues (\QTwo{}) at cryogenic temperatures at high density is necessary. 
Such a setup should allow investigating all six hydrogen isotopologues (\QTwo{}) in different phases, mixtures, pressures, \orthopara{} states and at the triple point using infrared and Raman spectroscopy as well as optical analysis. 
Based on previous experience with the inactive hydrogen isotopologues (\HTwo{}, HD, \DTwo{}) \cite{Groessle2017,Groessle2015} the \gls{tapir2} experiment \cite{Krasch2020} has been set up and commissioned at the \gls{tlk}. 

In this work, first results regarding the real-time performance of analytic systems in the \gls{tapir2} setup are presented. 
\section{Experimental Setup}
\label{sec:setup}

The setup of the \gls{tapir2} experiment is described in detail in \cite{Krasch2020} and \cite{Kohpeiss2025}.
In order to ensure safe operation with tritium, the entire setup is enclosed by a glovebox and connected to the \gls{tlk} closed tritium loop infrastructure \cite{Welte2015}.

\begin{figure}[t]
\centering
\includegraphics[width=\textwidth]{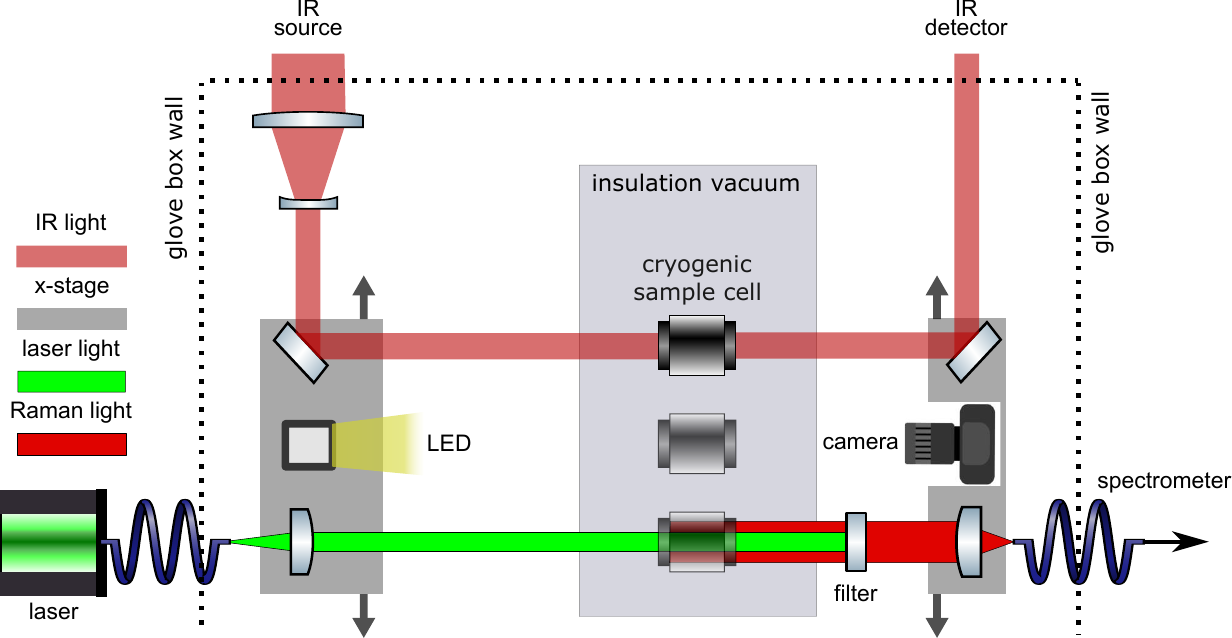}
\caption{
Experimental idea including IR-system, throughpass laser Raman-system and photography in a measurement cell.
The sample cell is included in the figure multiple times to illustrate the photography and Raman beam path.
Only one cell at a fixed position is present in the setup.
}
\label{fig:optical_sketch}
\end{figure}

\gls{tapir2} consists of a cryogenic sample cell with two sapphire windows for optical access in a transmission configuration.
This access can be used for transmission absorption infrared spectroscopy, Raman spectroscopy, and backlight photography (with white light, monochrome light, polarized and non polarized).
A motorized stage enables an automated change between measurement methods, allowing for quasi simultaneous measurement of IR spectra, Raman spectra, and photographs of the cell contents.
A schematic depiction of these optical beam paths in the \gls{tapir2} setup is shown in \cref{fig:optical_sketch}.

The infrared spectroscopy is performed using a \textit{Bruker} \textit{Vertex} 70 \gls{ftir} spectrometer.
Due to the sapphire windows of the cryogenic sample cell and the spectrometer, the accessible range of the infrared spectrum is limited to around \SIrange{2000}{15000}{\per\centi\meter}.
To remove background from atmospheric water, the IR beam path is continuously flushed with nitrogen and further dried with silica gel. 
The infrared interferogram is directed through the cell via mirrors, allowing both the spectrometer and its detector to reside outside of the glovebox, preventing contamination.

For the throughpass Raman spectroscopy system, a \gls{tlk}-developed \gls{muRa} system \cite{Priester2022} is used in a forward Raman configuration.
The excitation laser light is produced by a \textit{rgb lasersystems} \textit{Lambda Beam} \SI{532}{\nano\meter} diode pumped solid state laser with a power of \SI{200}{\milli\watt}.
The produced Raman light is analyzed by one of several exchangable \textit{Ocean Insight} \textit{QePro} spectrometers, which allows the investigation of either the S$_0$-Branch (rotational) or Q$_1$-Branch (ro-vibrational) of the hydrogen isotopologues.
Both the excitation and the Raman light are guided into/out of the glovebox using optical fibers, allowing both the laser and spectrometer to reside outside of the glovebox, preventing contamination.
Inside of the glovebox, a laser clean-up filter before the cell removes excitations from the connecting fiber while a longpass filter with a cut-on at \SI{534.4}{\nano\meter} after the cell removes the \SI{532}{\nano\meter} excitation laser light. 

For optical investigations, a \textit{Canon EOS R} system camera with a \textit{100 mm F/2.8 L IS USM macro} lens was used to take images and videos of the cryogenic measurement cell.

Additionally, linear polarizers mounted in motorized rotation stages can be inserted on both sides of the measurement cell. 
This allows for polarization dependent spectroscopic measurements, as well as for photographic polariscopy.
Polariscopy makes it possible to investigate stress inside solid cell contents via color changes when using a white light back-illumination \cite{Zholonko2011}, and enhances the contrast of density fluctuations in the liquid phase.

\section{Results}

In the following, first results obtained with the \gls{tapir2} experiment are presented.
First, optical investigations into phase change behavior with a focus on crystallization will be shown, followed by spectral data with a focus on dynamic processes.

\subsection{Photography based optical observations of phase change}

\begin{figure}[t]
\centering
\includegraphics[width=0.5\textwidth]{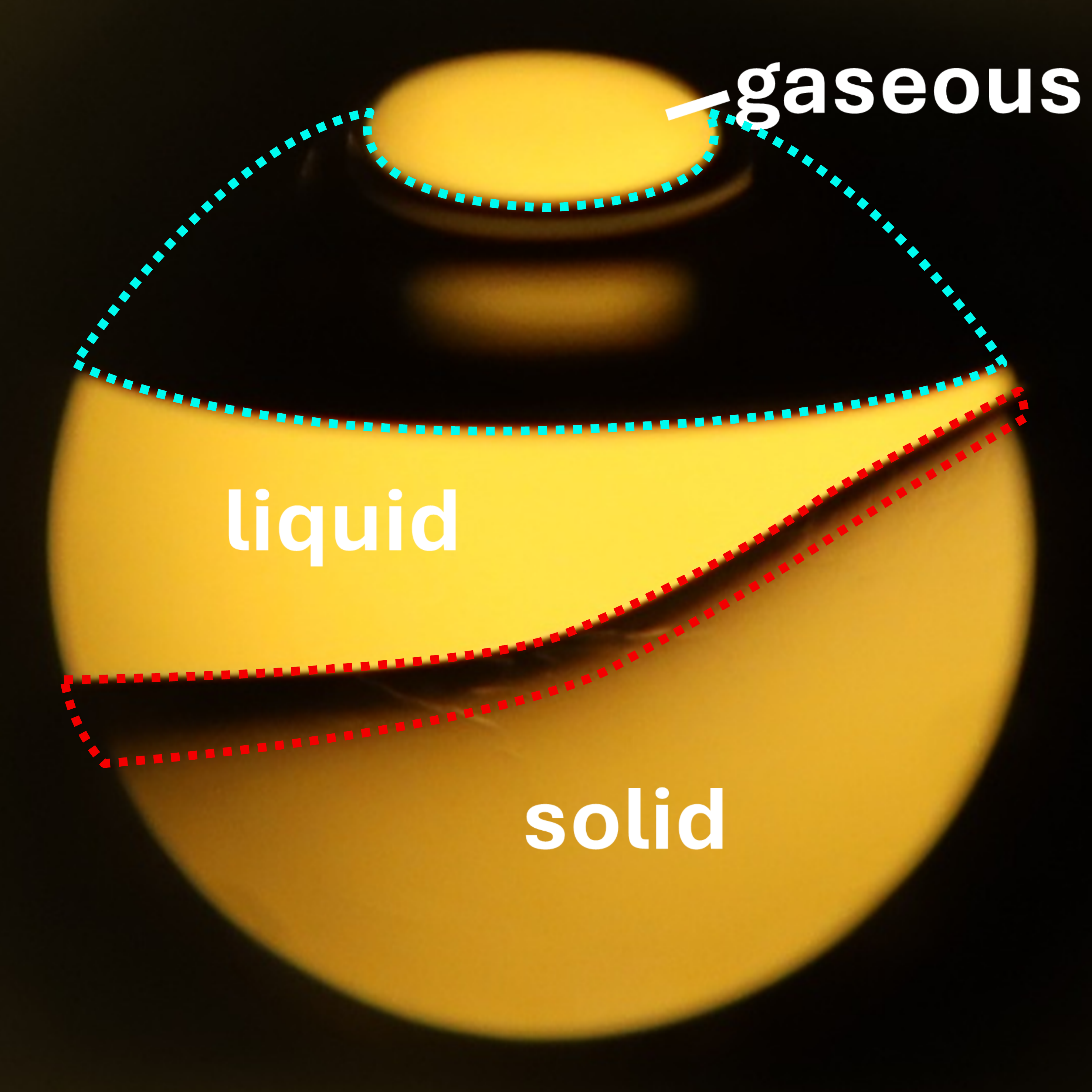}
\caption{
Photograph of the \gls{tapir2} measurement cell content. 
The cell is filled with a system of all three phases of protium (\HTwo{}).
The cyan dotted line shows the meniscus of the liquid gas phase boundary.
The red dotted line shows the saddle shaped liquid solid boundary region.
}
\label{fig:hydrogen_three_phases}
\end{figure}

The cryogenic measurement cell of the \gls{tapir2} experiment can reach the temperatures required to cause hydrogen to liquefy and crystallize.
Via the optical windows of the cell, this phase change can be observed with imaging optics such as a camera using back-illumination from the opposite side of the cell.
An example photograph of the cell filled with a system of solid, liquid, and gaseous protium (\HTwo{}) is shown in \cref{fig:hydrogen_three_phases}.
The interface between gaseous phase at the top of the cell and the liquid phase in the middle (region in cyan dotted line) is visible as a pronounced dark meniscus, which is the result of light of the back-illumination being refracted away from the optical axis of the camera.
The phase boundary between liquid and solid (region in red dotted line) is visible as a dark ridge as well due to the boundary not being perpendicular to the cell windows, again refracting light away.
In the solid phase, several hairline surface features are visible, with the best contrast close to the phase boundary.
The solid is frozen asymmetrically, most likely due to the differences in thermal resistance between the measurement cell and the cryocooler across the interface between them, which causes a slight temperature gradient across the cell.

\subsubsection{Slow crystal growth}

\begin{figure}[t]
\centering
\begin{subfigure}{.17\textwidth}
    \centering
    \includegraphics[width=\textwidth]{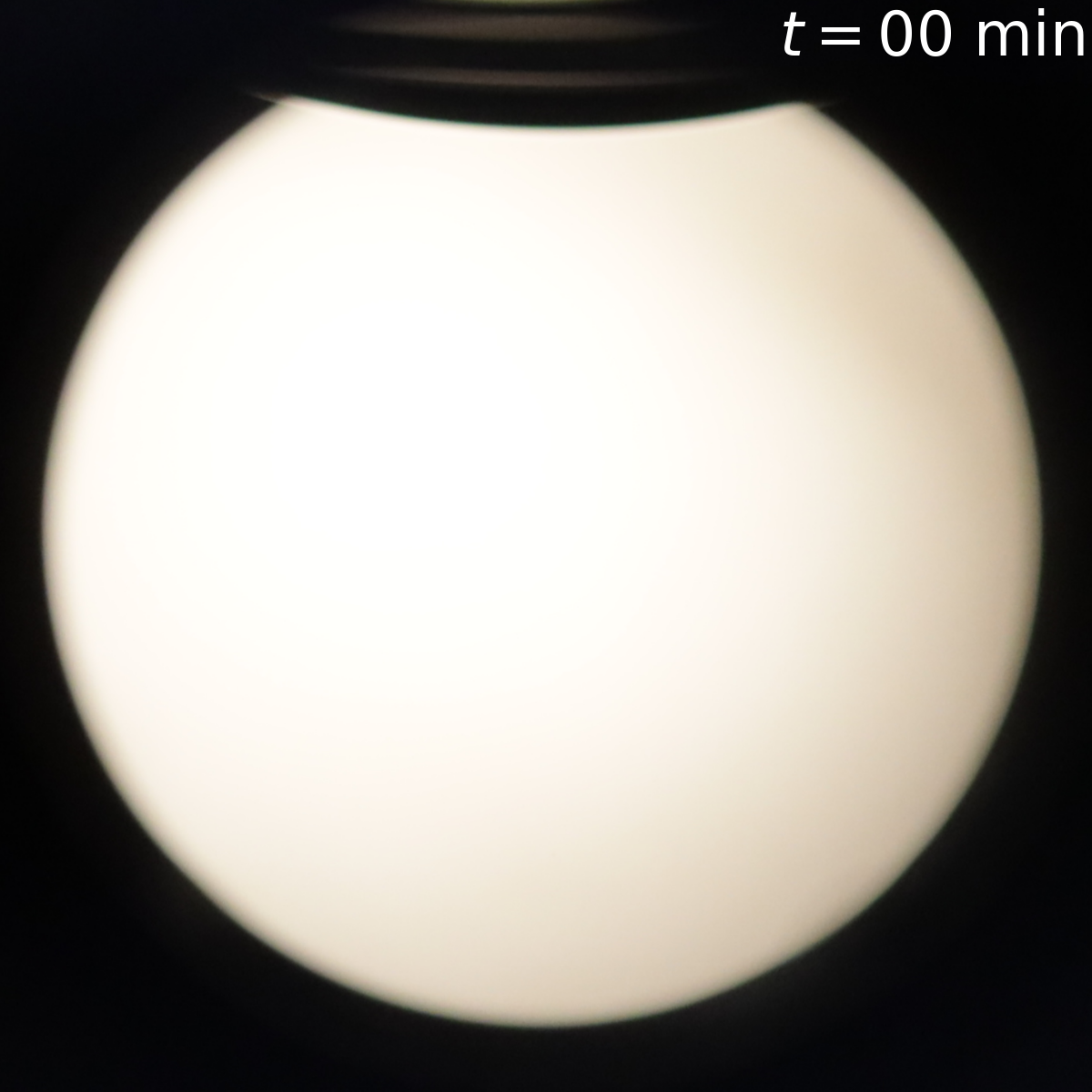}
\end{subfigure}%
\begin{subfigure}{.17\textwidth}
    \centering
    \includegraphics[width=\textwidth]{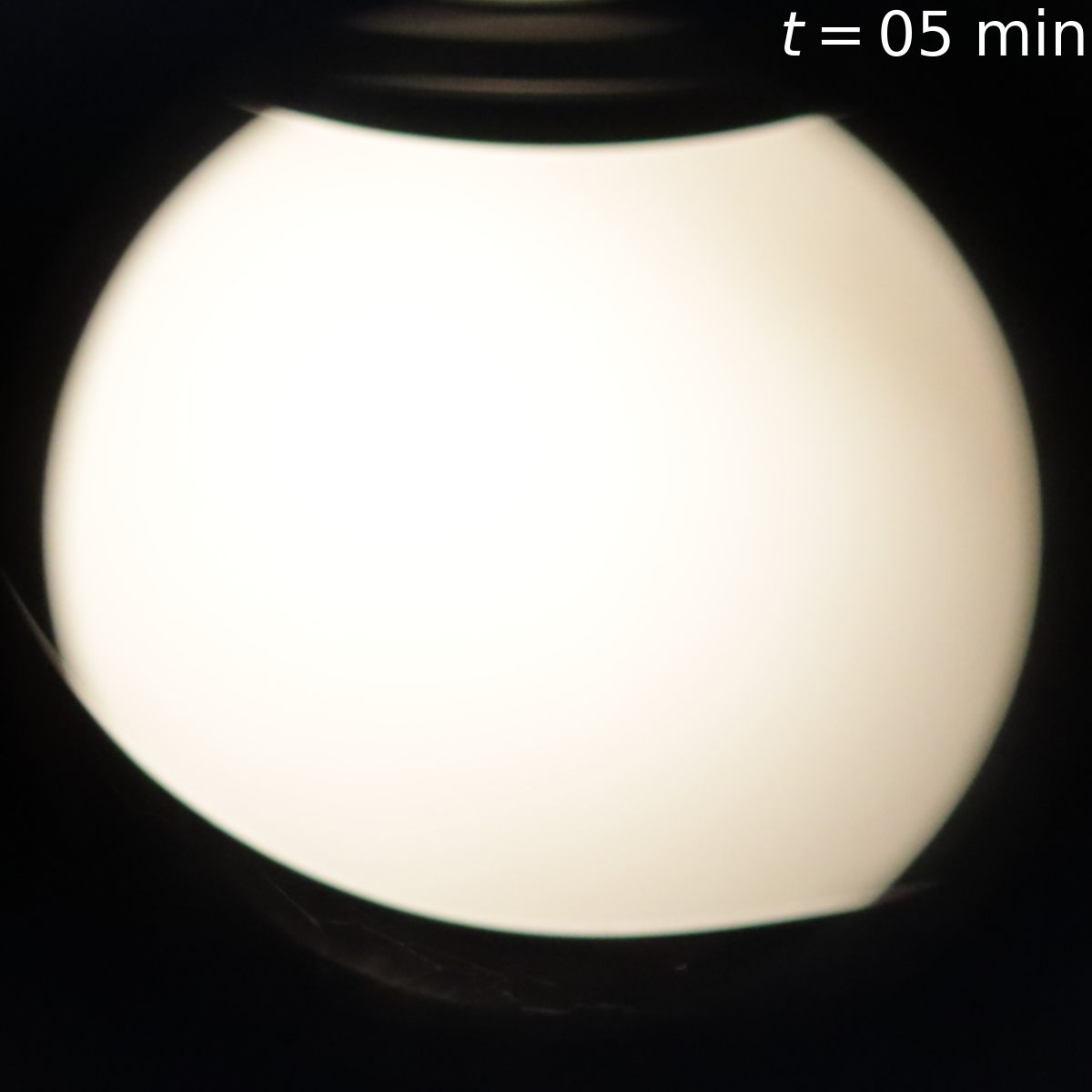}
\end{subfigure}%
\begin{subfigure}{.17\textwidth}
    \centering
    \includegraphics[width=\textwidth]{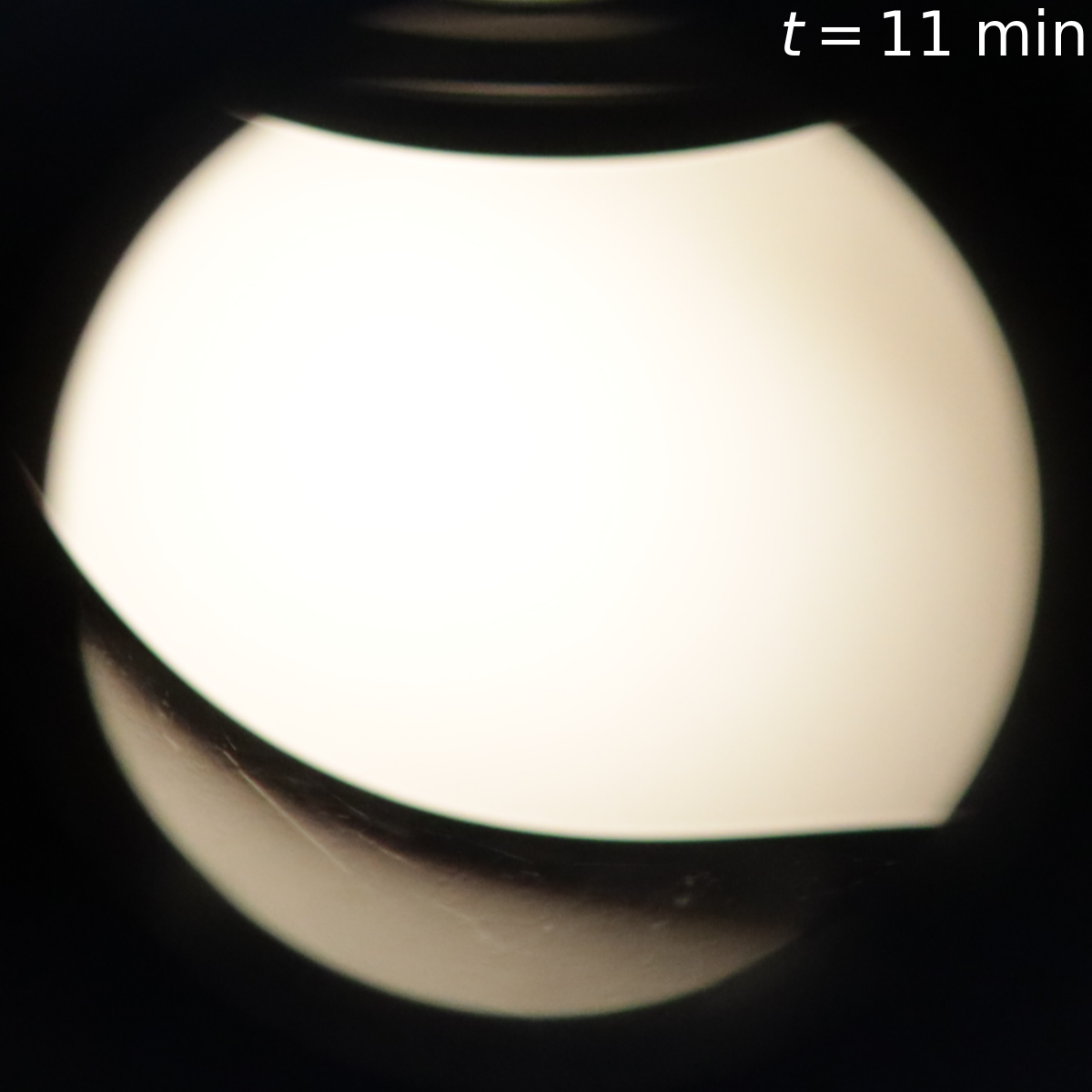}
\end{subfigure}%
\begin{subfigure}{.17\textwidth}
    \centering
    \includegraphics[width=\textwidth]{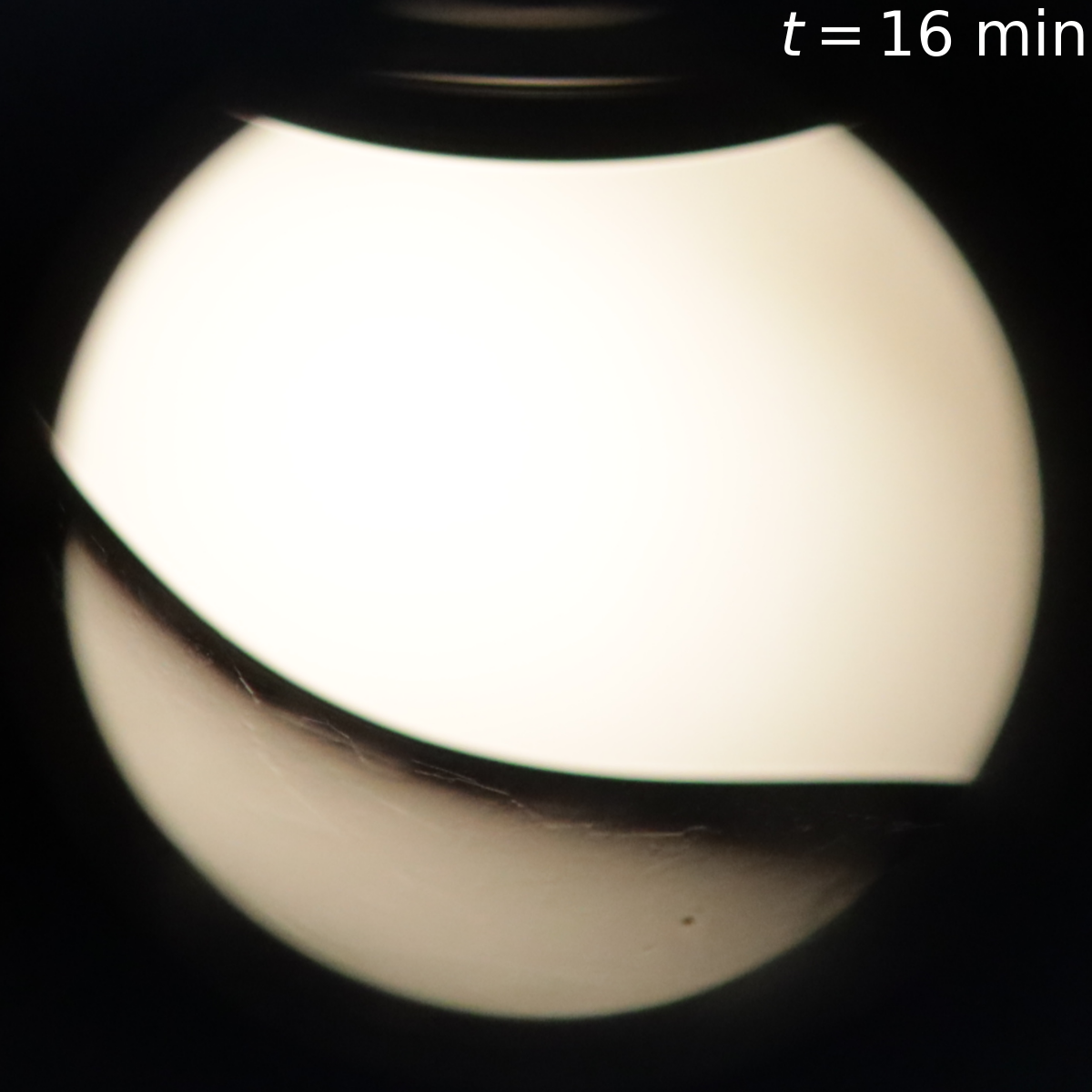}
\end{subfigure}%
\begin{subfigure}{.17\textwidth}
    \centering
    \includegraphics[width=\textwidth]{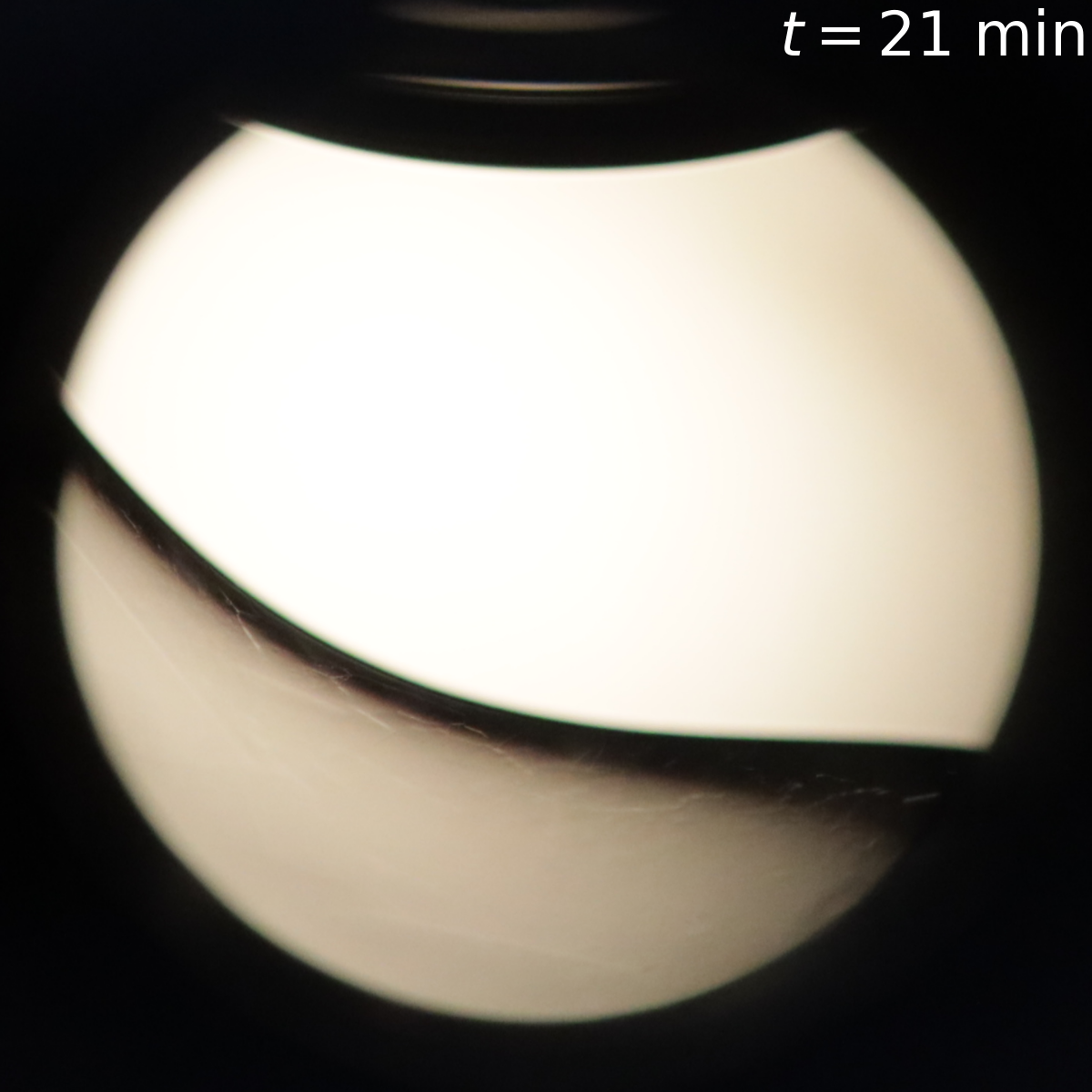}
\end{subfigure}%

\begin{subfigure}{.17\textwidth}
    \centering
    \includegraphics[width=\textwidth]{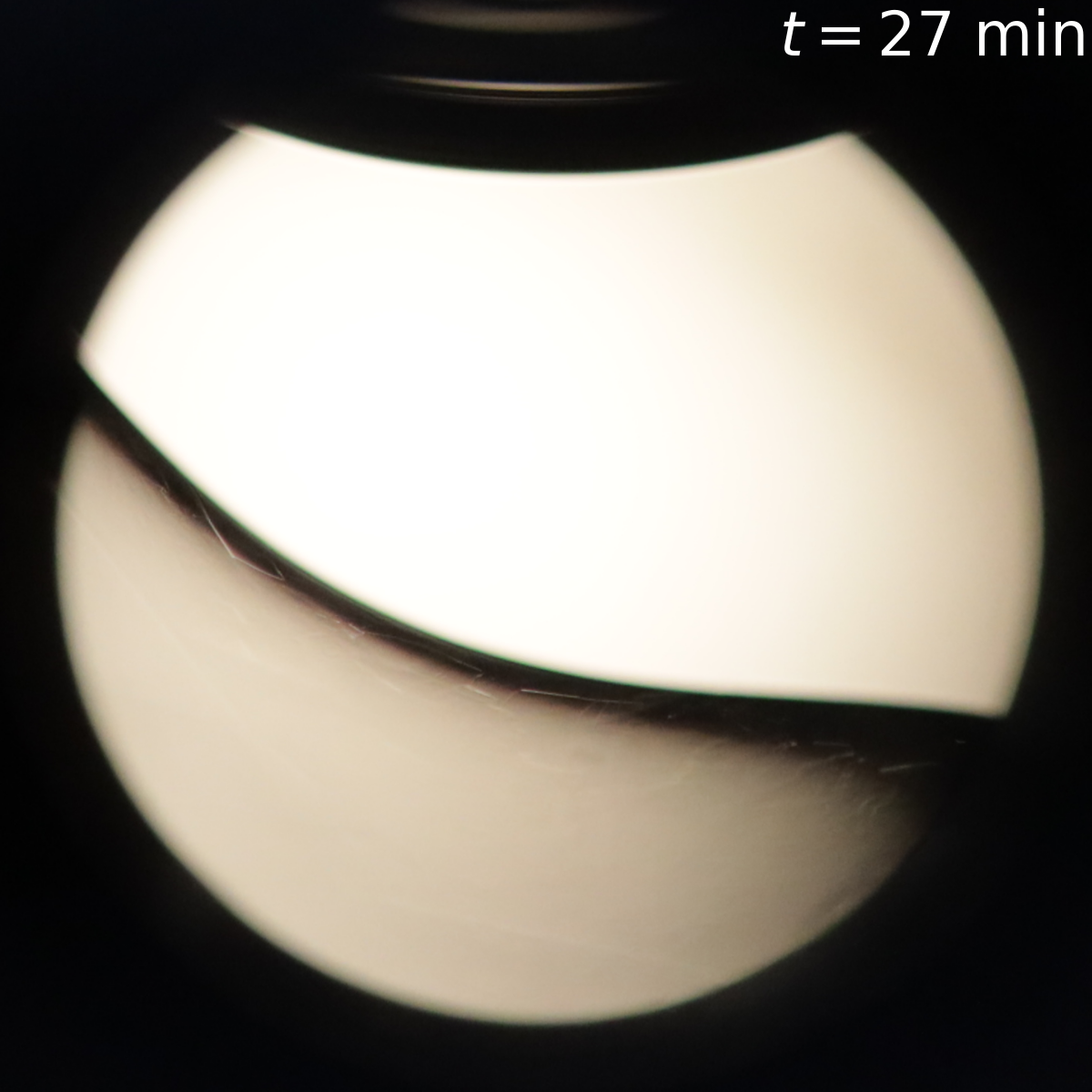}
\end{subfigure}%
\begin{subfigure}{.17\textwidth}
    \centering
    \includegraphics[width=\textwidth]{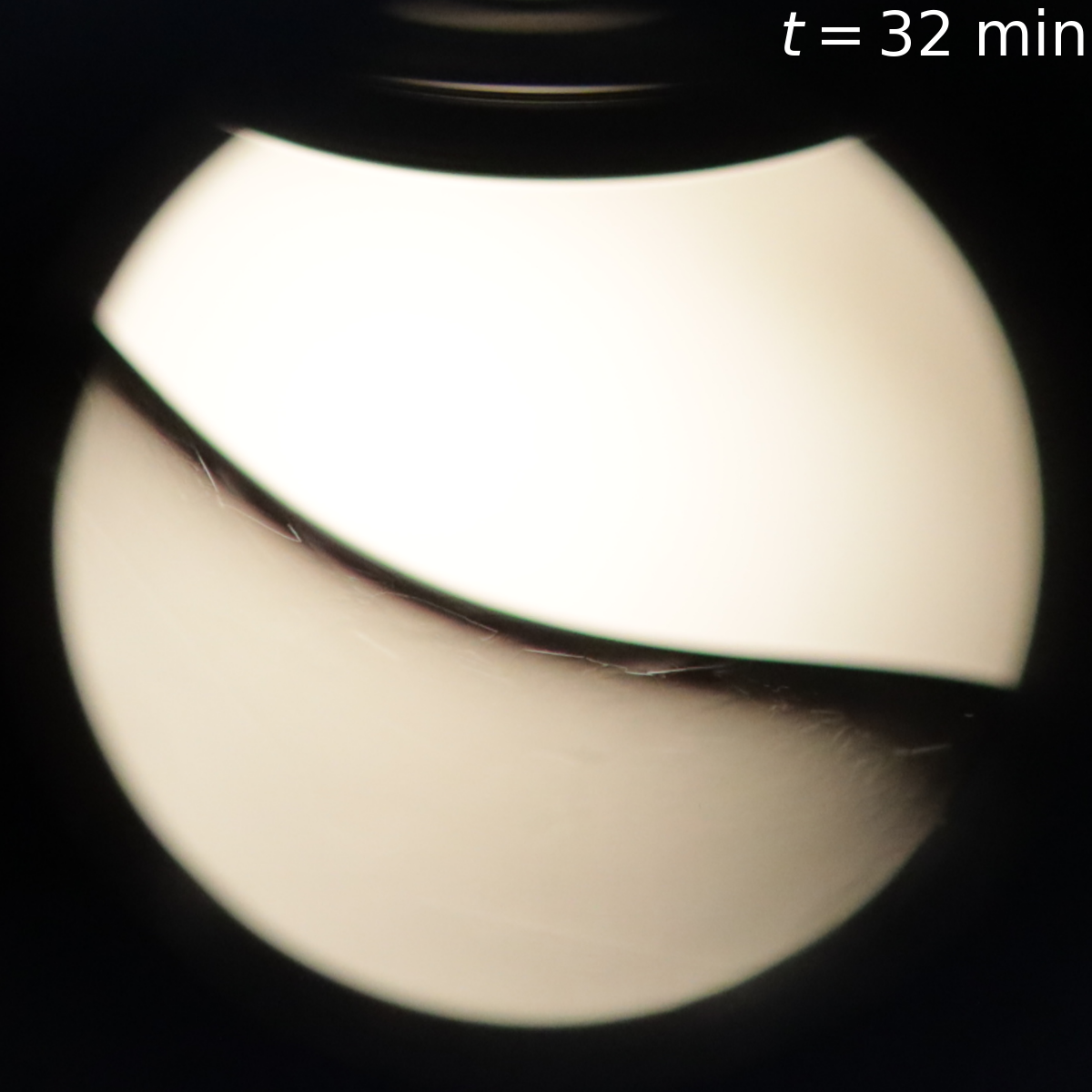}
\end{subfigure}%
\begin{subfigure}{.17\textwidth}
    \centering
    \includegraphics[width=\textwidth]{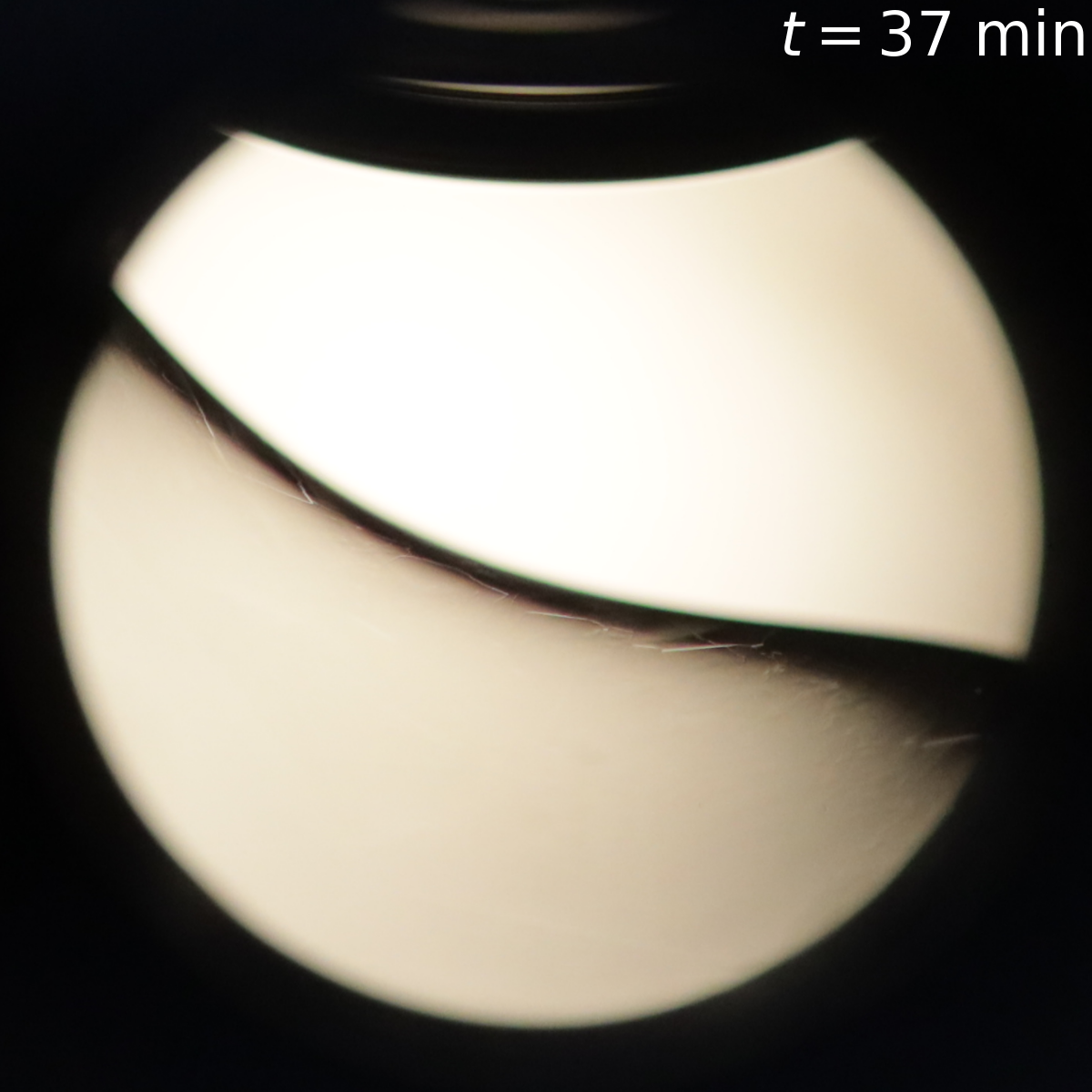}
\end{subfigure}%
\begin{subfigure}{.17\textwidth}
    \centering
    \includegraphics[width=\textwidth]{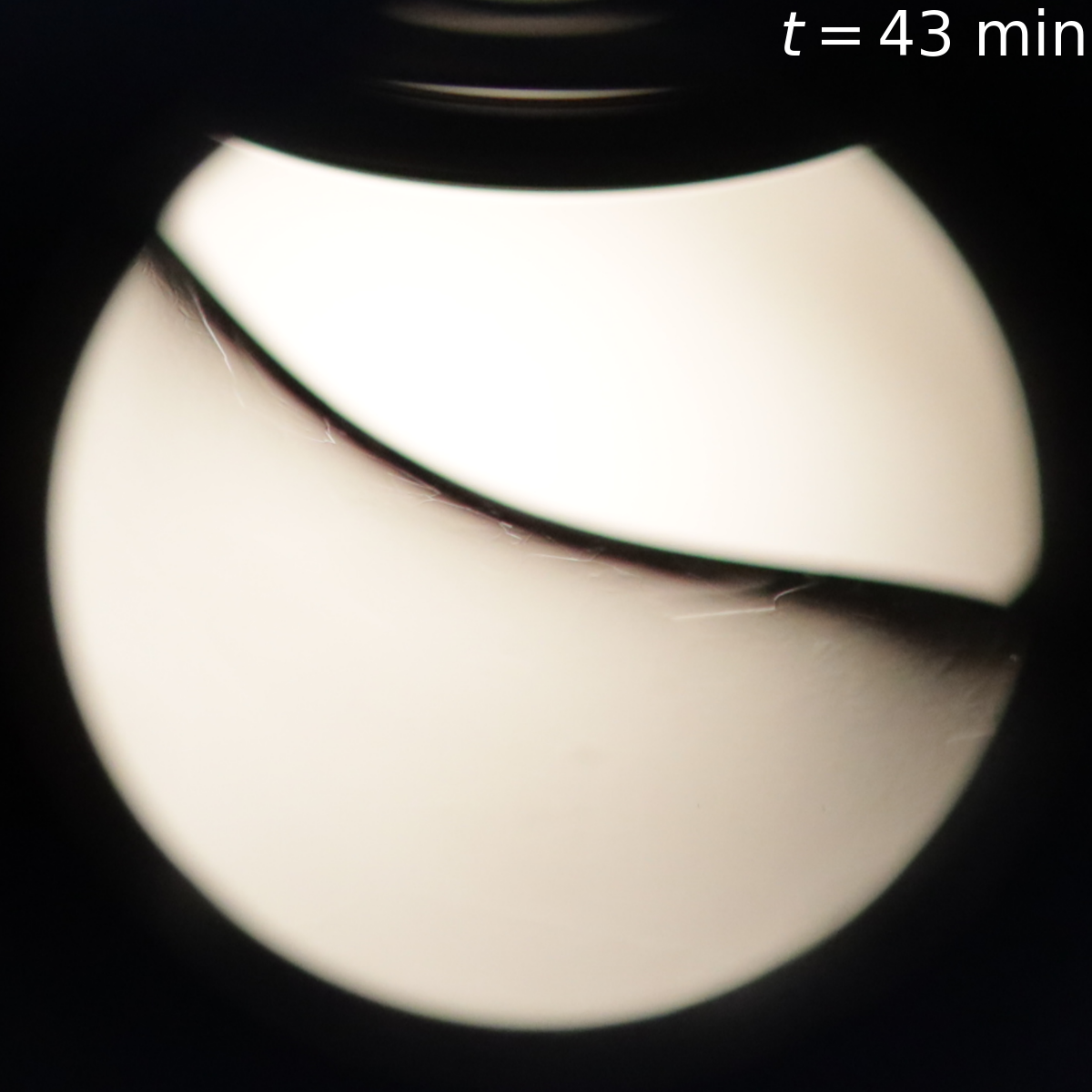}
\end{subfigure}%
\begin{subfigure}{.17\textwidth}
    \centering
    \includegraphics[width=\textwidth]{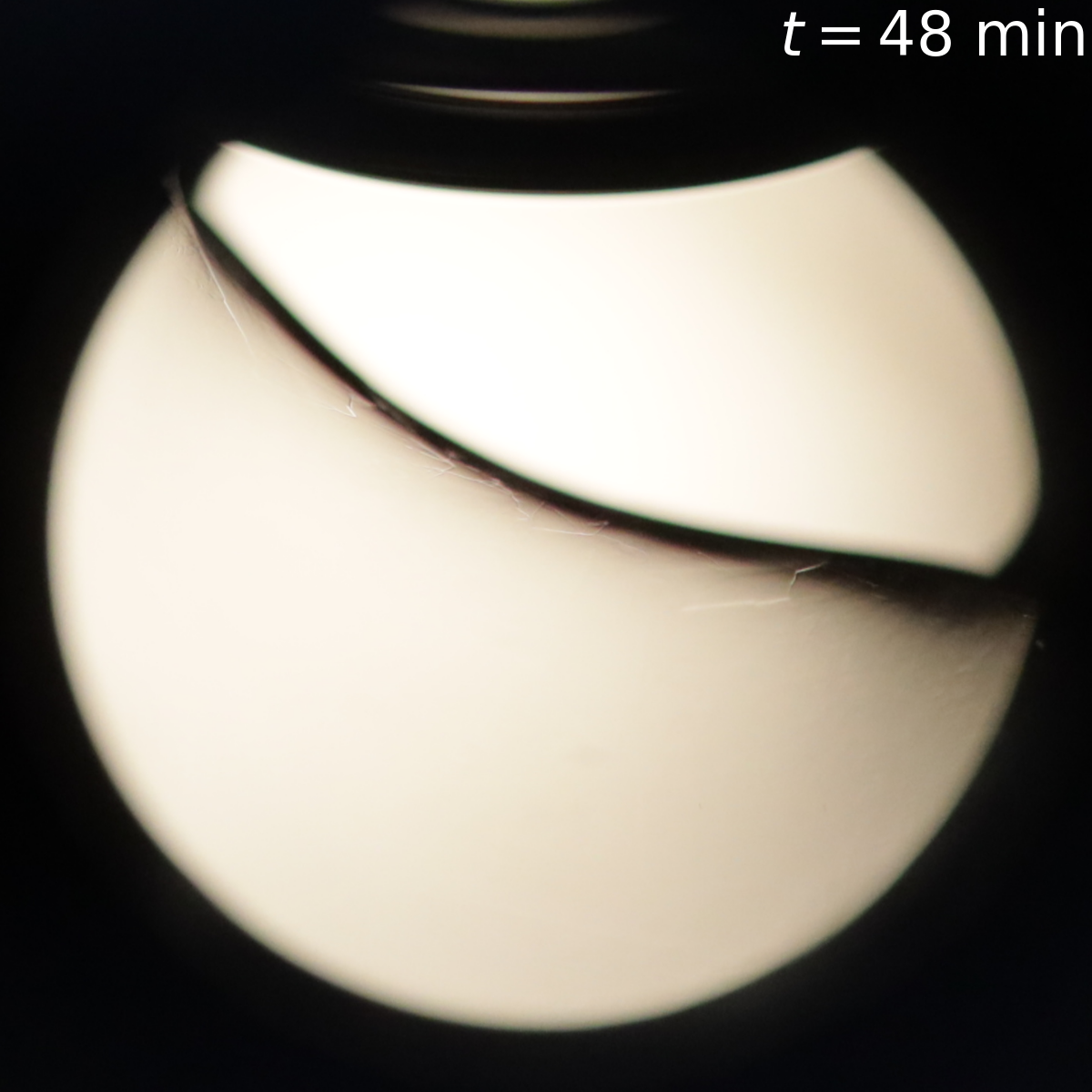}
\end{subfigure}%

\begin{subfigure}{.17\textwidth}
    \centering
    \includegraphics[width=\textwidth]{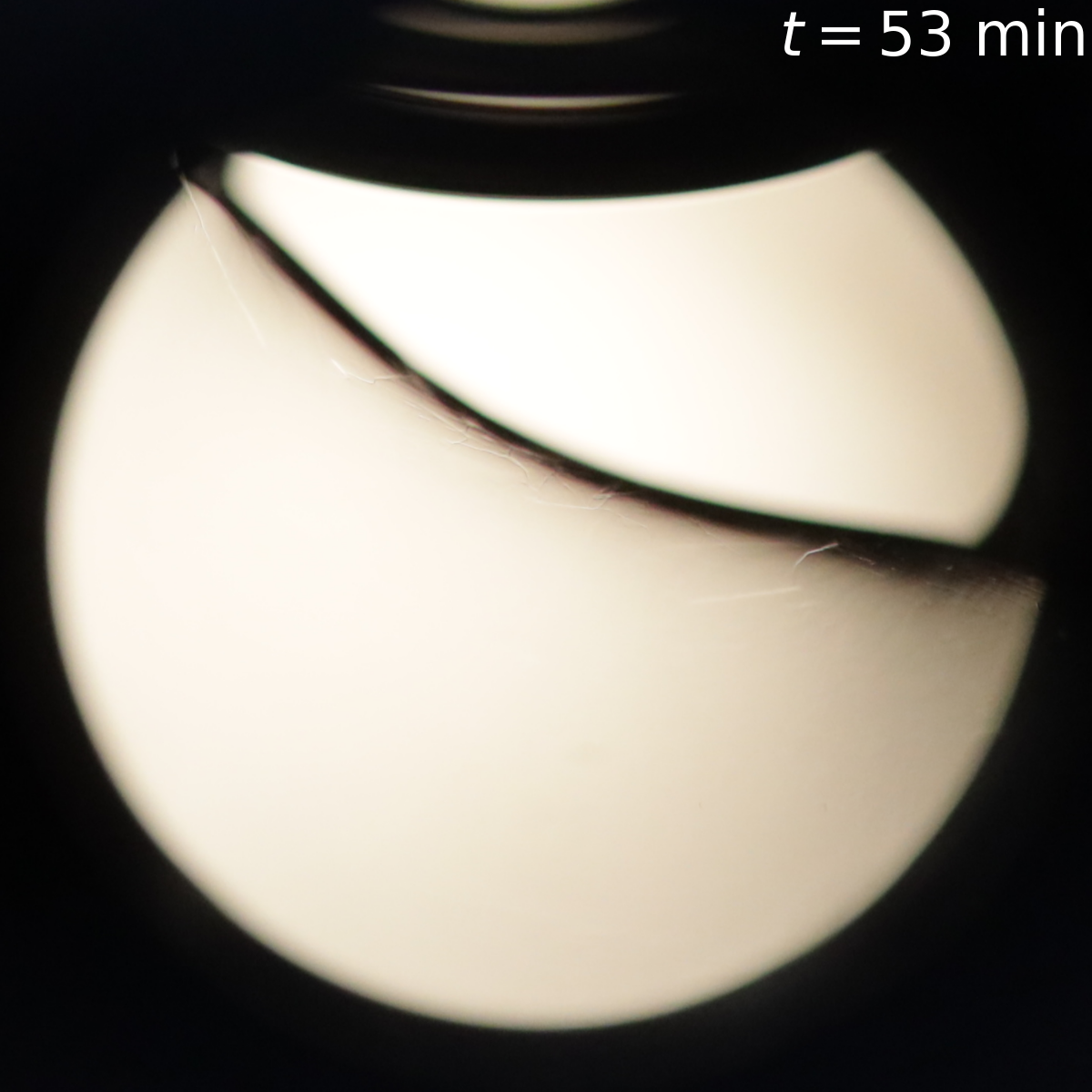}
\end{subfigure}%
\begin{subfigure}{.17\textwidth}
    \centering
    \includegraphics[width=\textwidth]{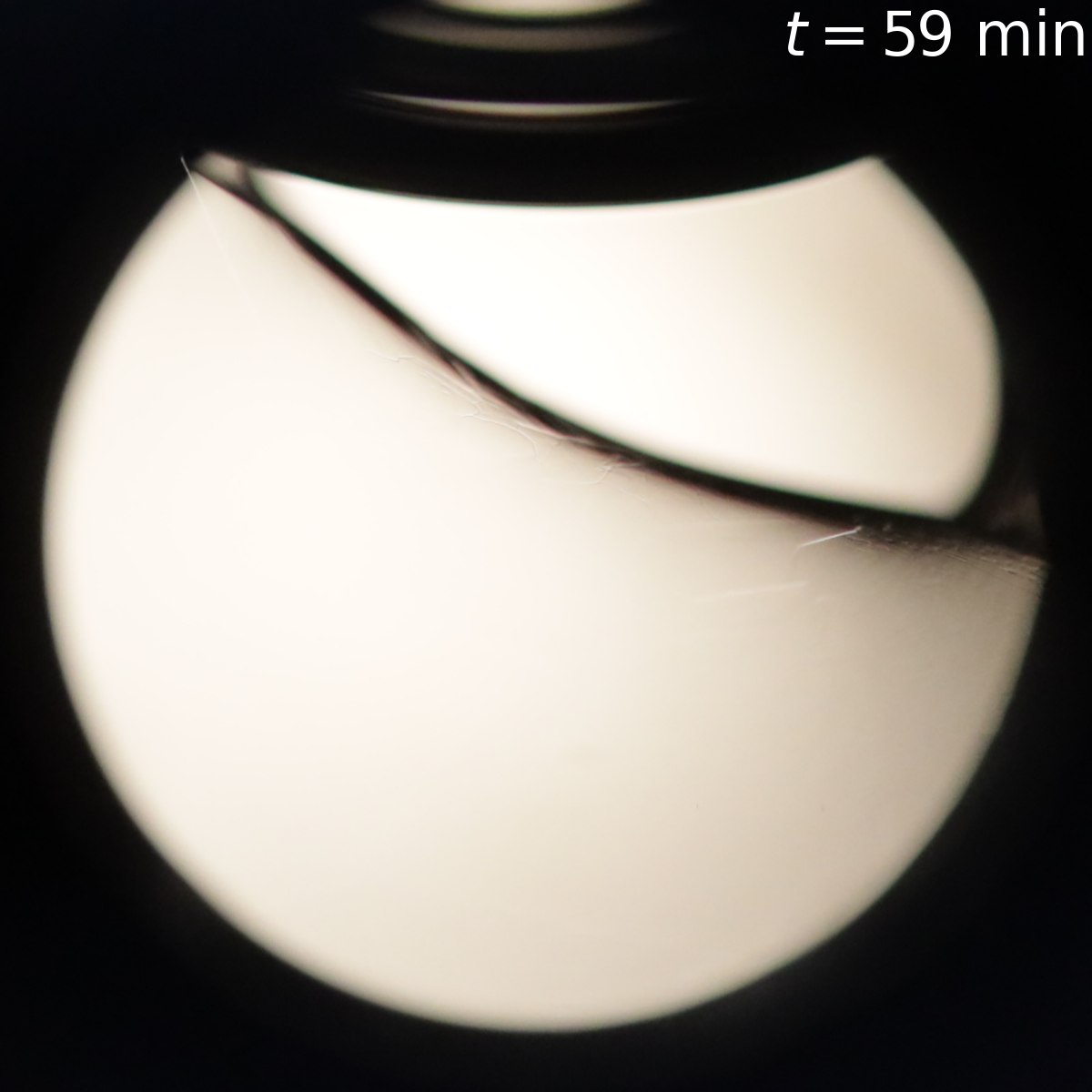}
\end{subfigure}%
\begin{subfigure}{.17\textwidth}
    \centering
    \includegraphics[width=\textwidth]{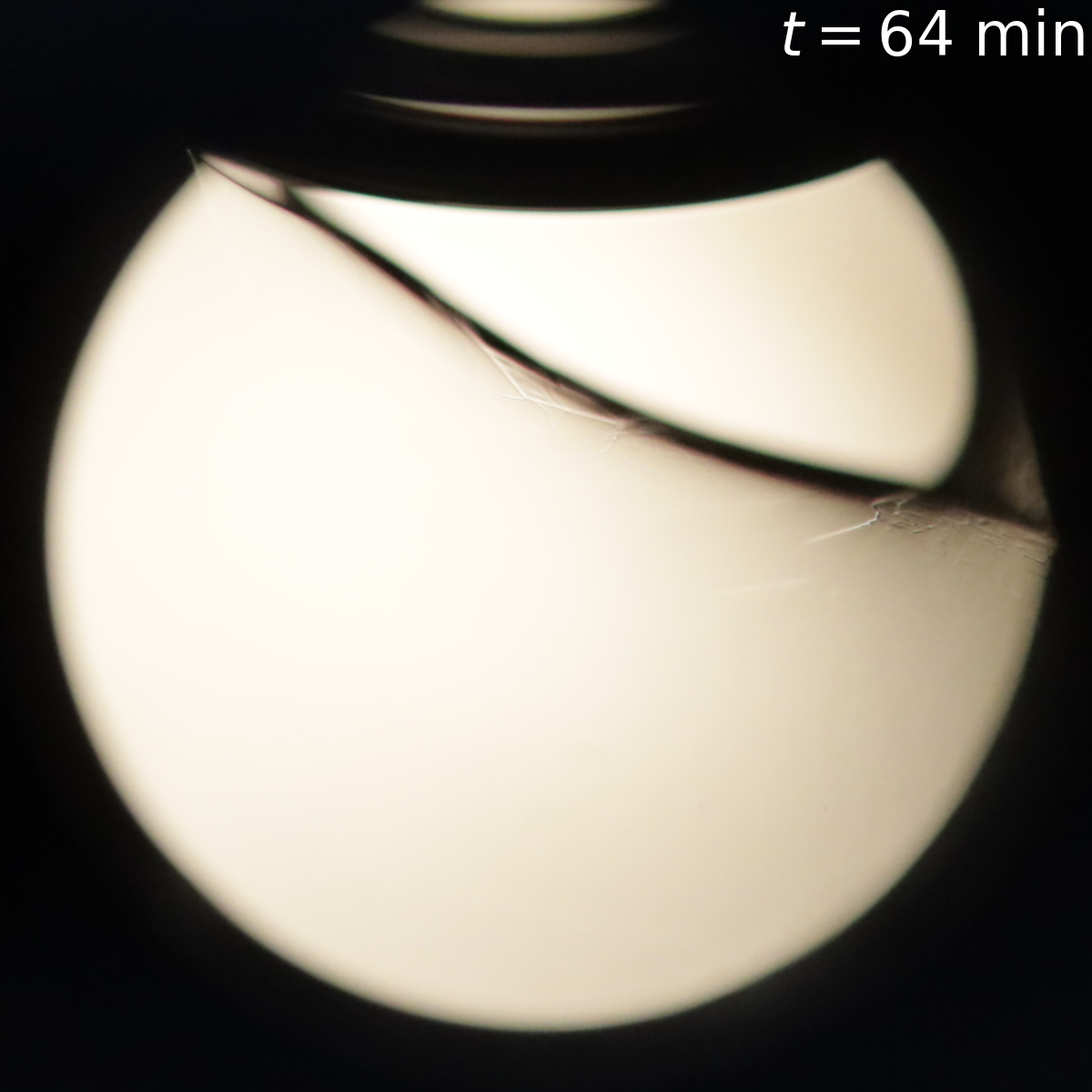}
\end{subfigure}%
\begin{subfigure}{.17\textwidth}
    \centering
    \includegraphics[width=\textwidth]{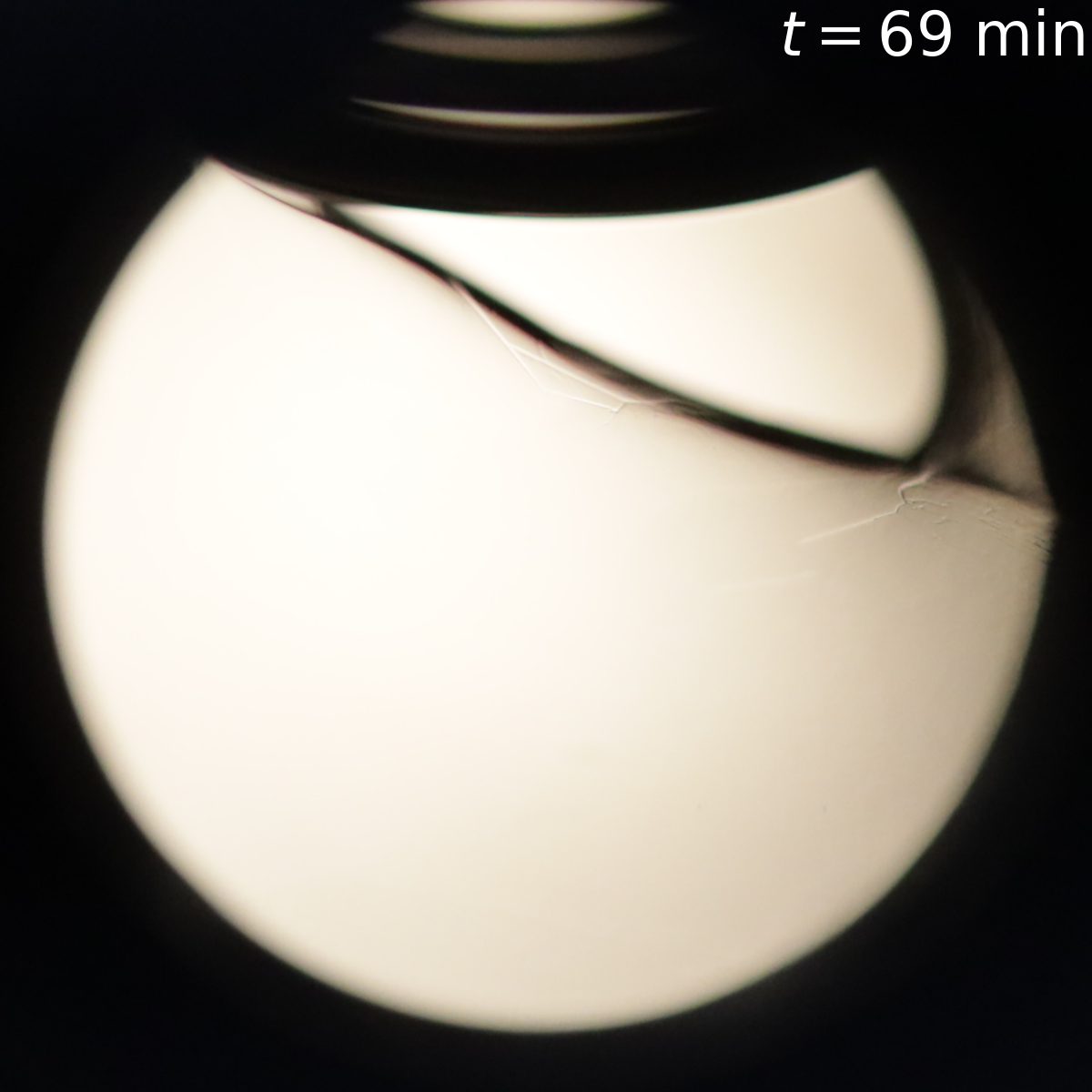}
\end{subfigure}%
\begin{subfigure}{.17\textwidth}
    \centering
    \includegraphics[width=\textwidth]{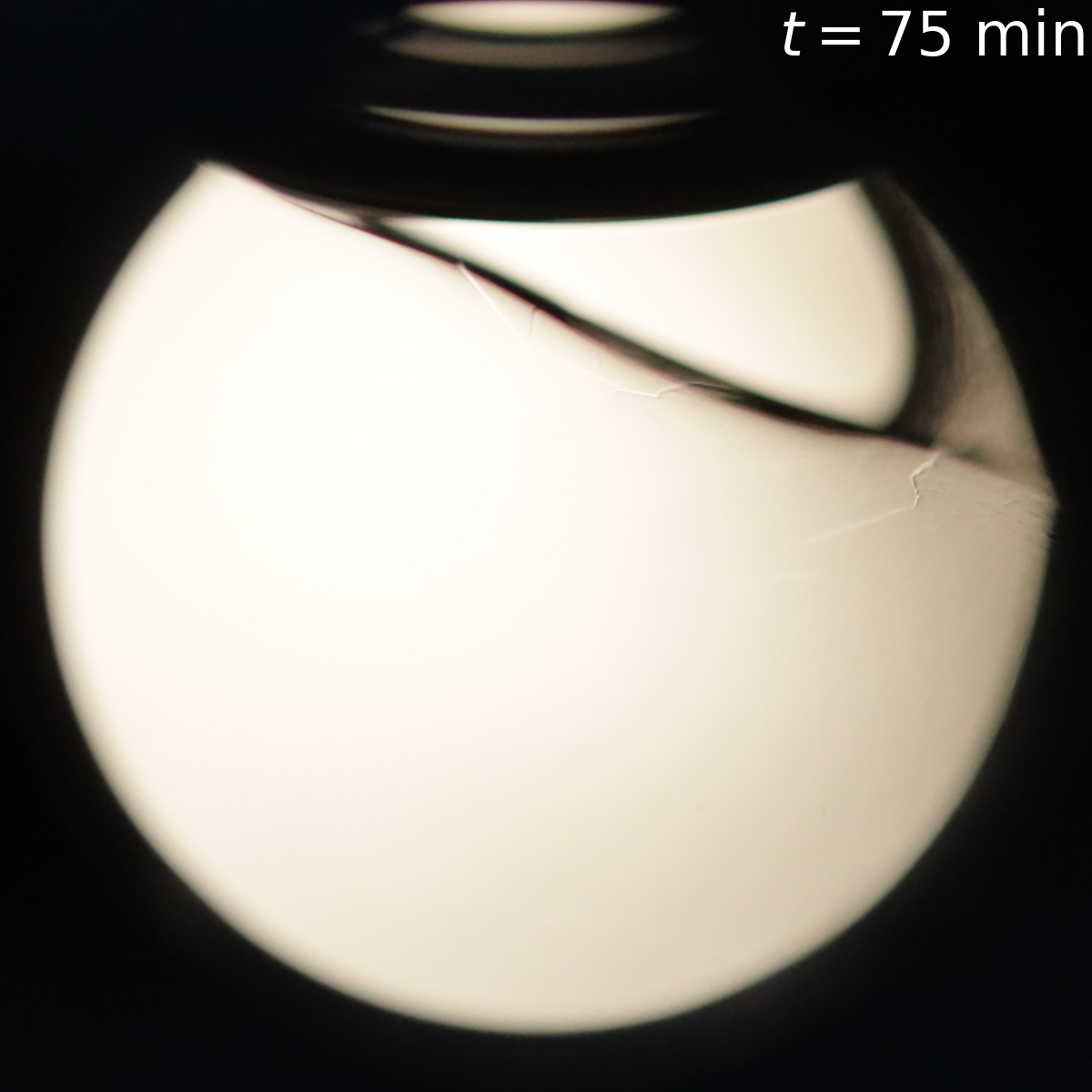}
\end{subfigure}%

\begin{subfigure}{.17\textwidth}
    \centering
    \includegraphics[width=\textwidth]{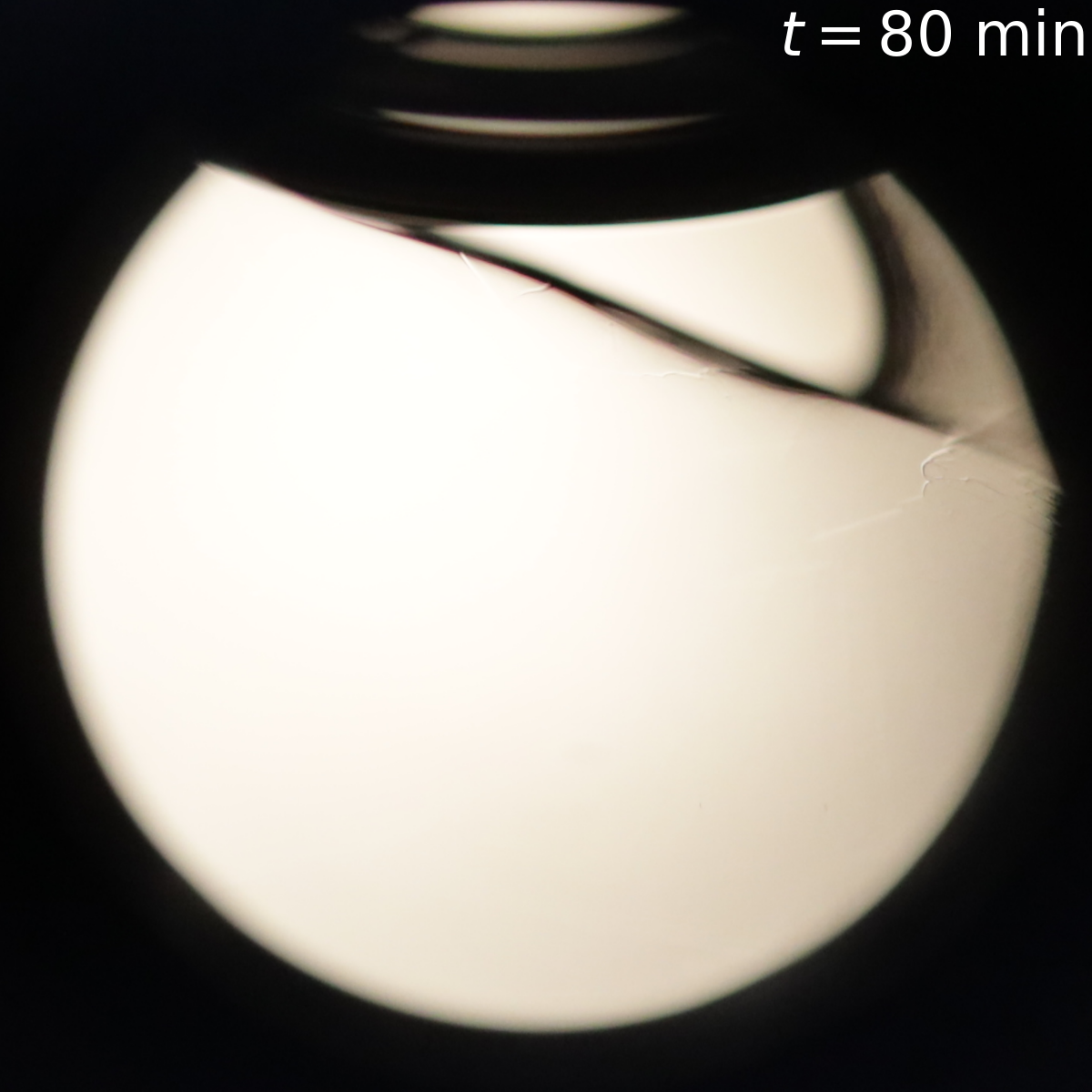}
\end{subfigure}%
\begin{subfigure}{.17\textwidth}
    \centering
    \includegraphics[width=\textwidth]{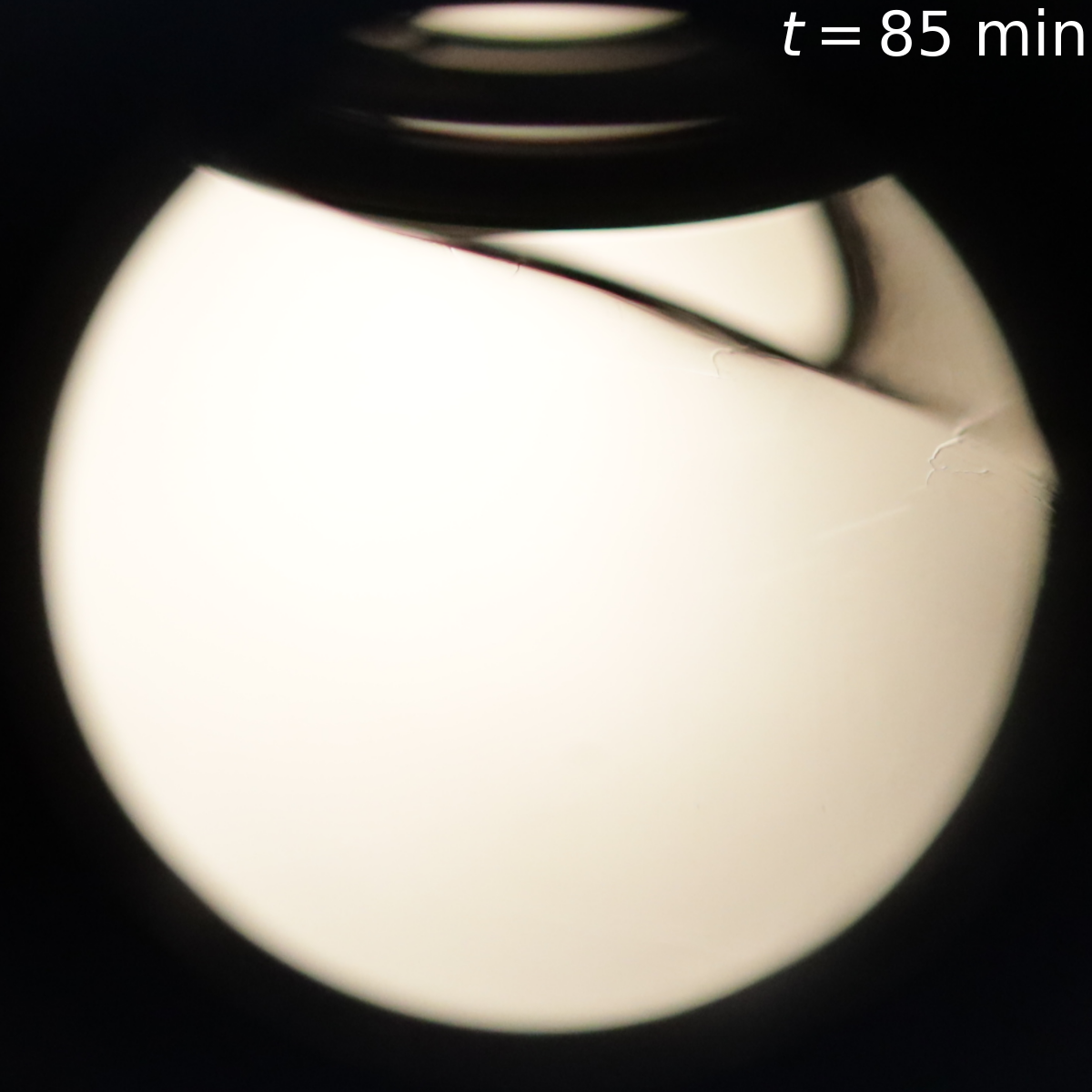}
\end{subfigure}%
\begin{subfigure}{.17\textwidth}
    \centering
    \includegraphics[width=\textwidth]{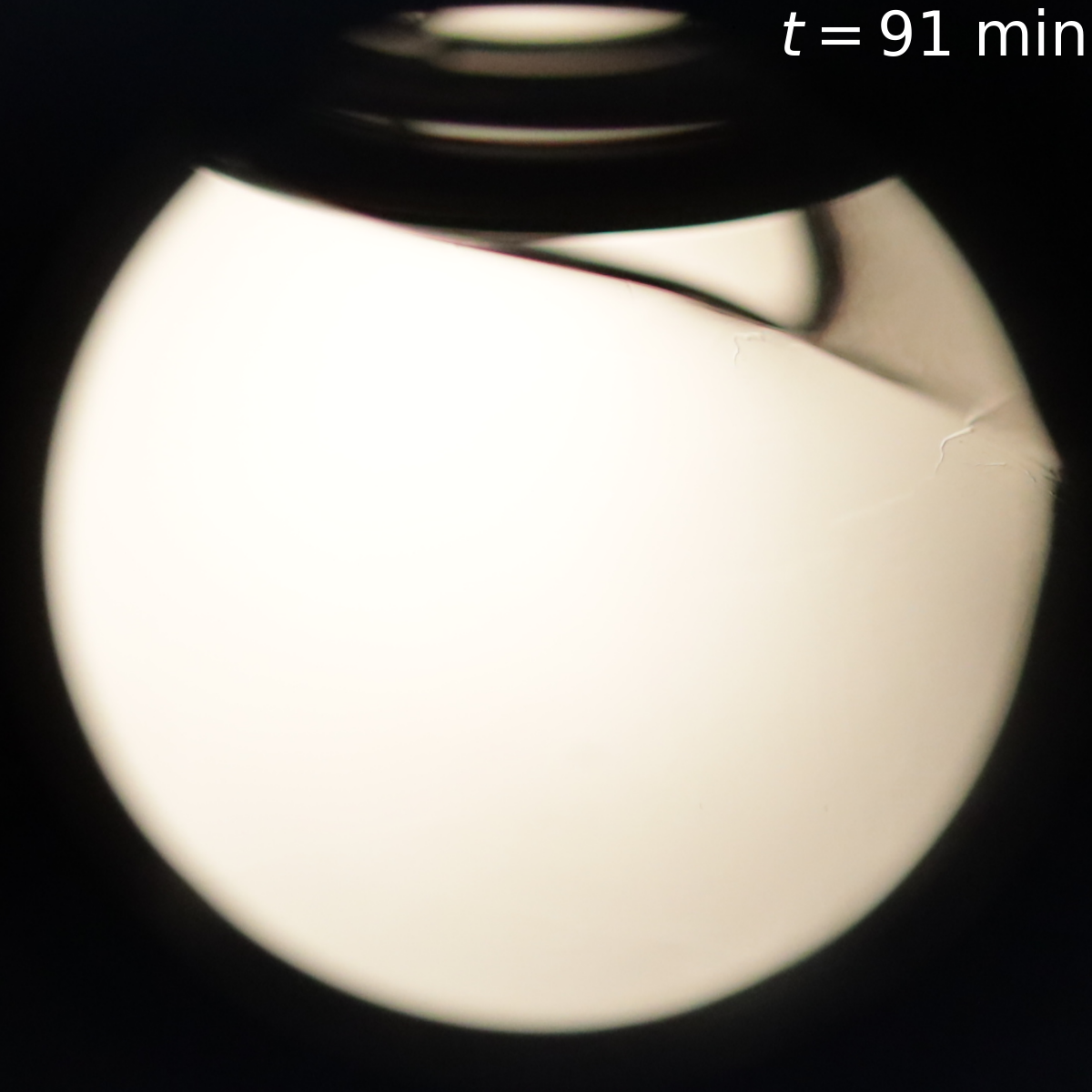}
\end{subfigure}%
\begin{subfigure}{.17\textwidth}
    \centering
    \includegraphics[width=\textwidth]{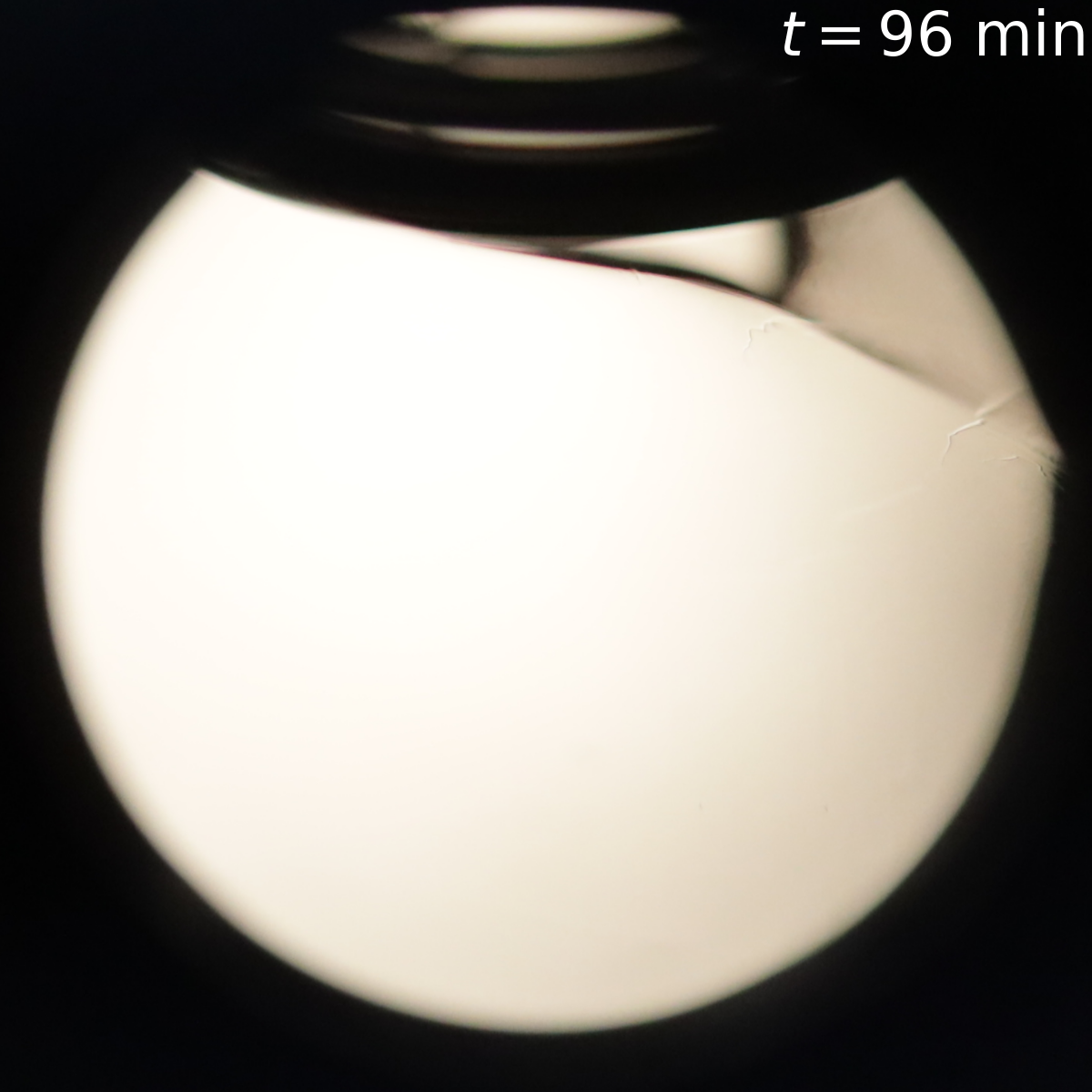}
\end{subfigure}%
\begin{subfigure}{.17\textwidth}
    \centering
    \includegraphics[width=\textwidth]{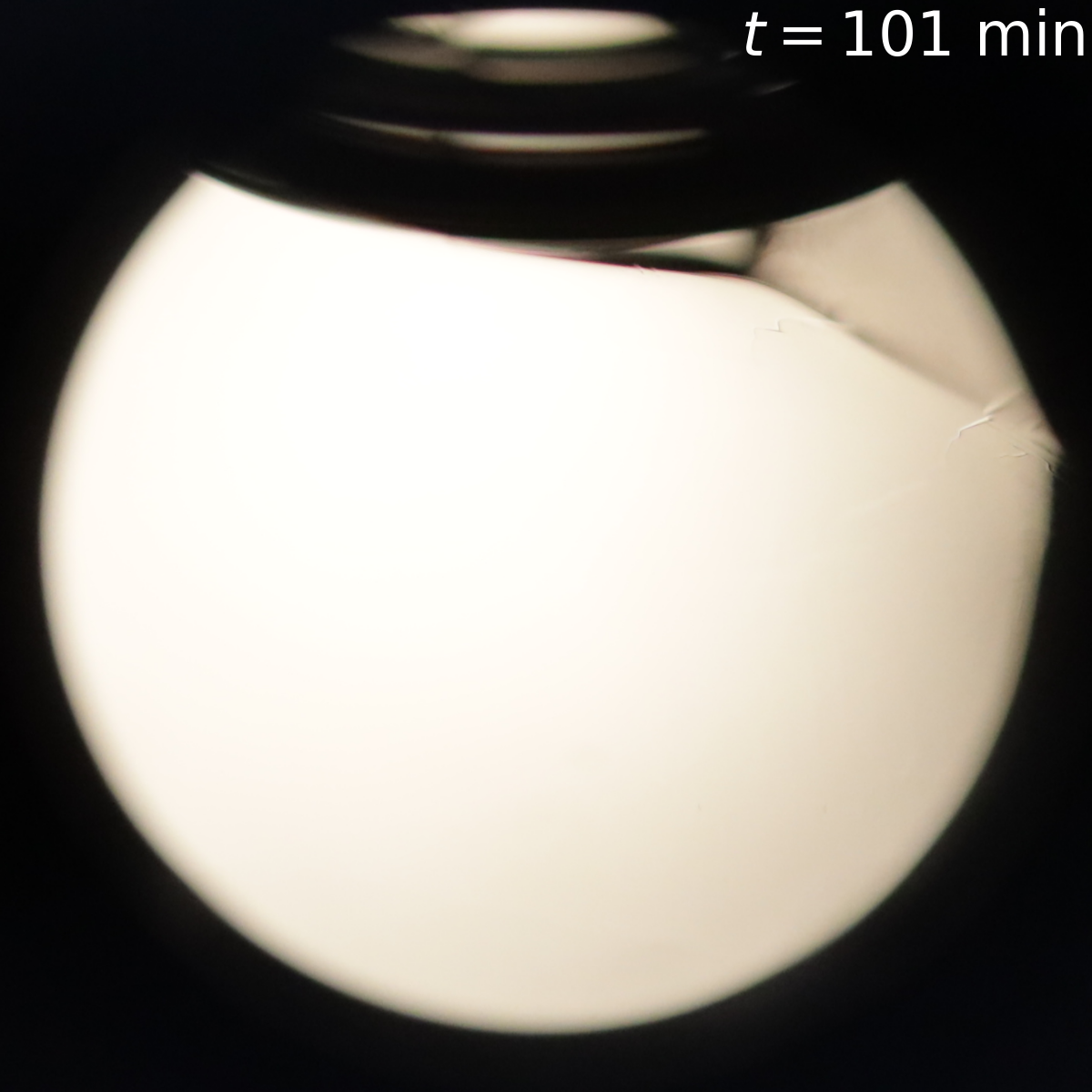}
\end{subfigure}%
\caption{
Slow freezing of liquid \HTwo{}.
Images are sequentially from left to right, then top to bottom.
The entire sequence of images depicts a time span of \SI{\approx100}{\minute}.
}
\label{fig:slow_crystallization}
\end{figure}
For a slow crystal growth two methods are tested and established for \HTwo{} and \DTwo{}. One method is the growth via the liquid phase and the other is desublimation. 
The latter uses the direct transition from to gaseous into the solid phase. 
To grow a hydrogen crystal from the liquid, the cell temperature is initially set to a temperature slightly above the triple point of the used hydrogen isotopologue. 
By condensation the cell fills up with liquid until the desired fill level is reached. 
Subsequently, the temperature is reduced slightly below the corresponding triple point. 
The crystallization then starts at the bottom of the cell and the crystal growths slowly to the top until the whole liquid is transformed into the solid phase as can be seen in \cref{fig:slow_crystallization}. 
This method results in a directly transparent hydrogen crystal which can then be analyzed e.g. by IR spectroscopy. 
However, if the temperature is reduced too far below the triple point, the initially slow crystal growth goes over into a different behavior, described in the following \cref{subsec:rapid_crystal_growth}. 
Therefore, the temperature is a very crucial parameter for this method of crystal growth.

To grow a crystal via desublimation, the sample cell is coupled with a buffer vessel filled with the desired hydrogen isotopologue. 
The initial pressure in this configuration is set to a value below \SI{100}{\milli\bar}. 
The exact value is chosen according to the isotopologue's phase diagram. 
During the process, a slowly growing crystal layer at the lower part of the cell is observed which then grows to a crystal filling out the whole sample cell. 
During this process the pressure inside cell and buffer vessel decreases due to the material deposited in the crystal. The amount of hydrogen in the buffer vessel, however, is large enough, to produce a cell filling crystal. 
In principle, keeping the pressure constant during the growth process is possible by buffering the the buffer vessel directly from the gas supply line.
A subsequent annealing phase is therefore required before IR spectroscopy is possible.

For both methods, the influence of parameters pressure and temperature as well as corresponding gradients needs further investigation. 
So far, both procedures have been performed multiple times with reproducible results.

\subsubsection{Rapid crystal growth}
\label{subsec:rapid_crystal_growth}

\begin{figure}[t]
\centering
\begin{subfigure}{.17\textwidth}
    \centering
    \includegraphics[width=\textwidth]{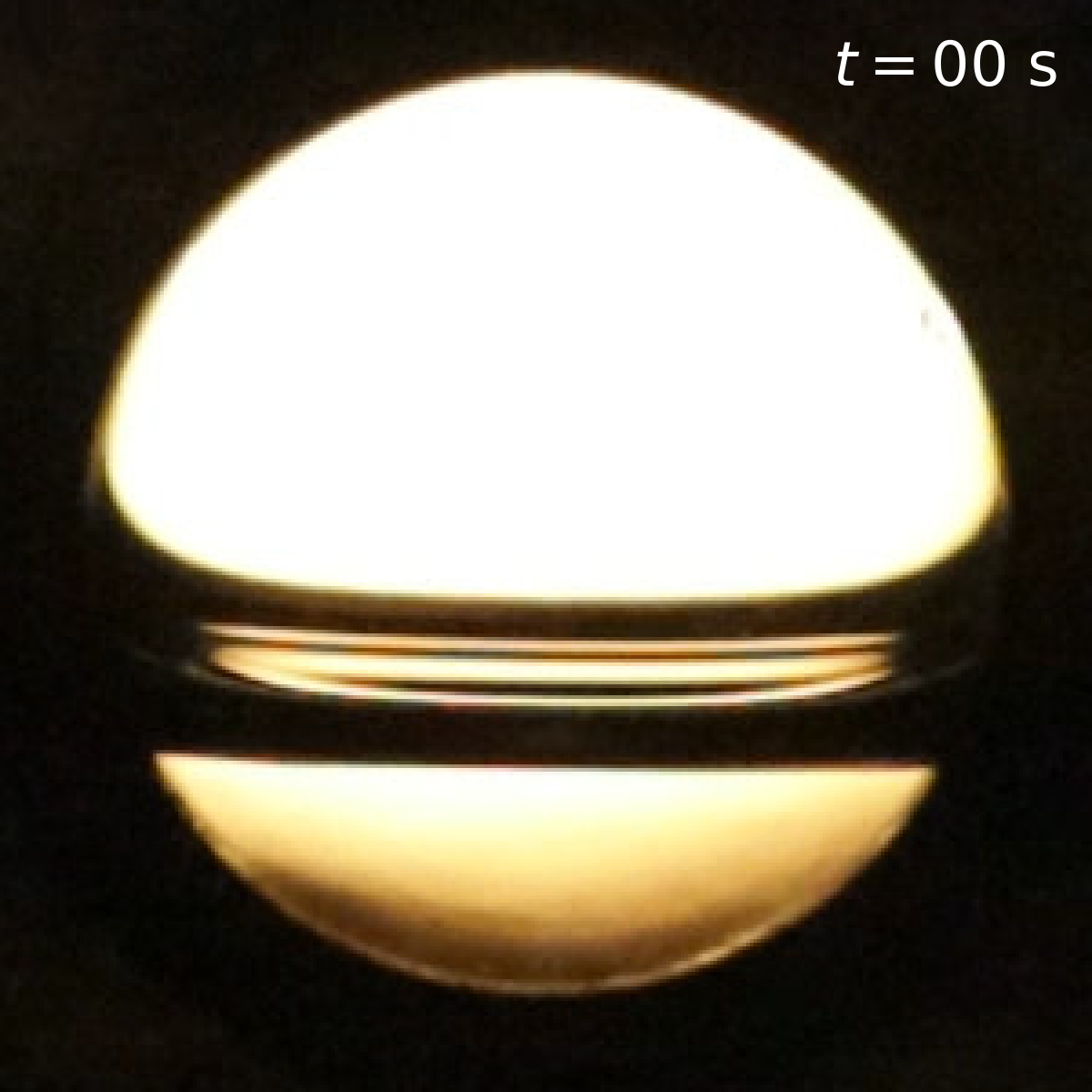}
\end{subfigure}%
\begin{subfigure}{.17\textwidth}
    \centering
    \includegraphics[width=\textwidth]{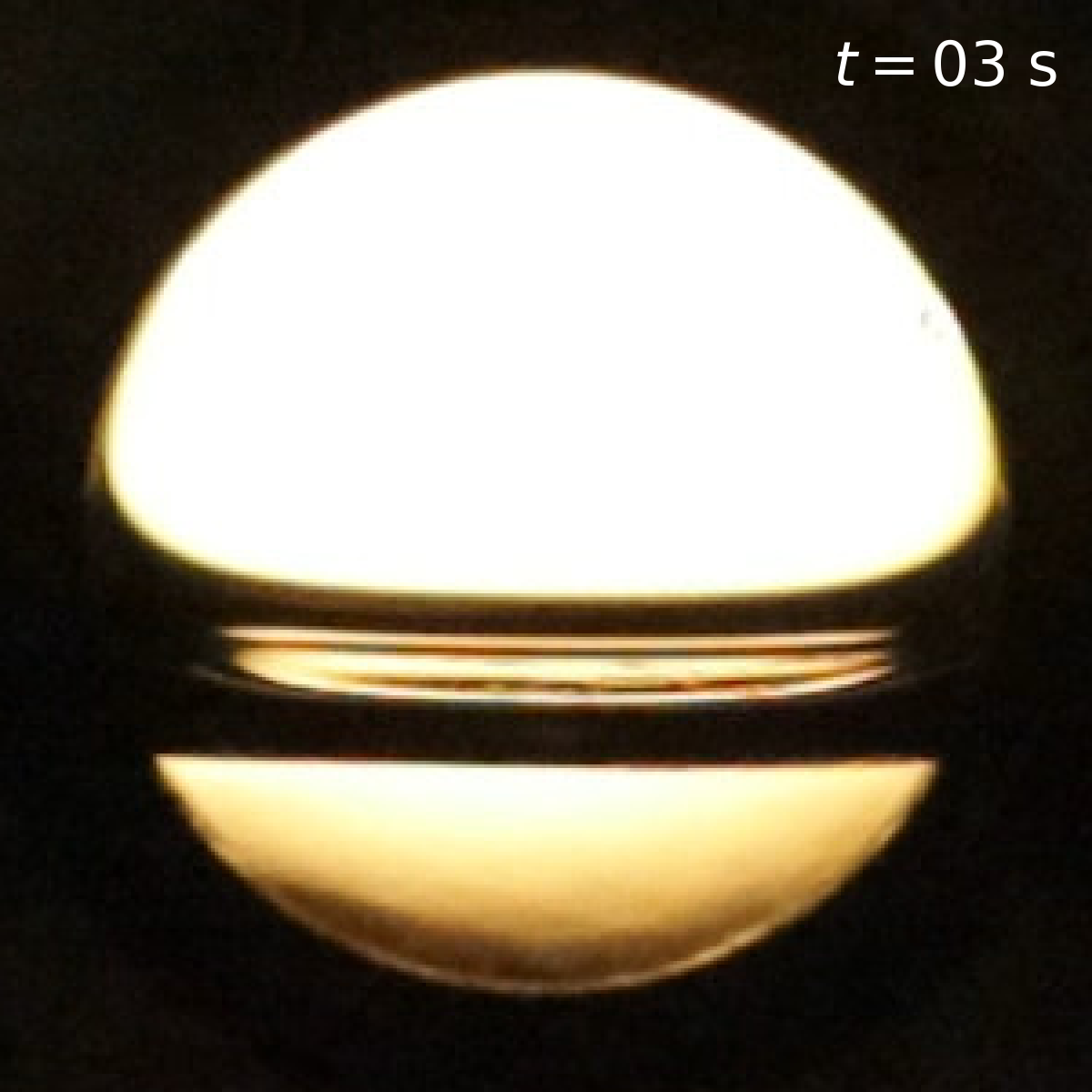}
\end{subfigure}%
\begin{subfigure}{.17\textwidth}
    \centering
    \includegraphics[width=\textwidth]{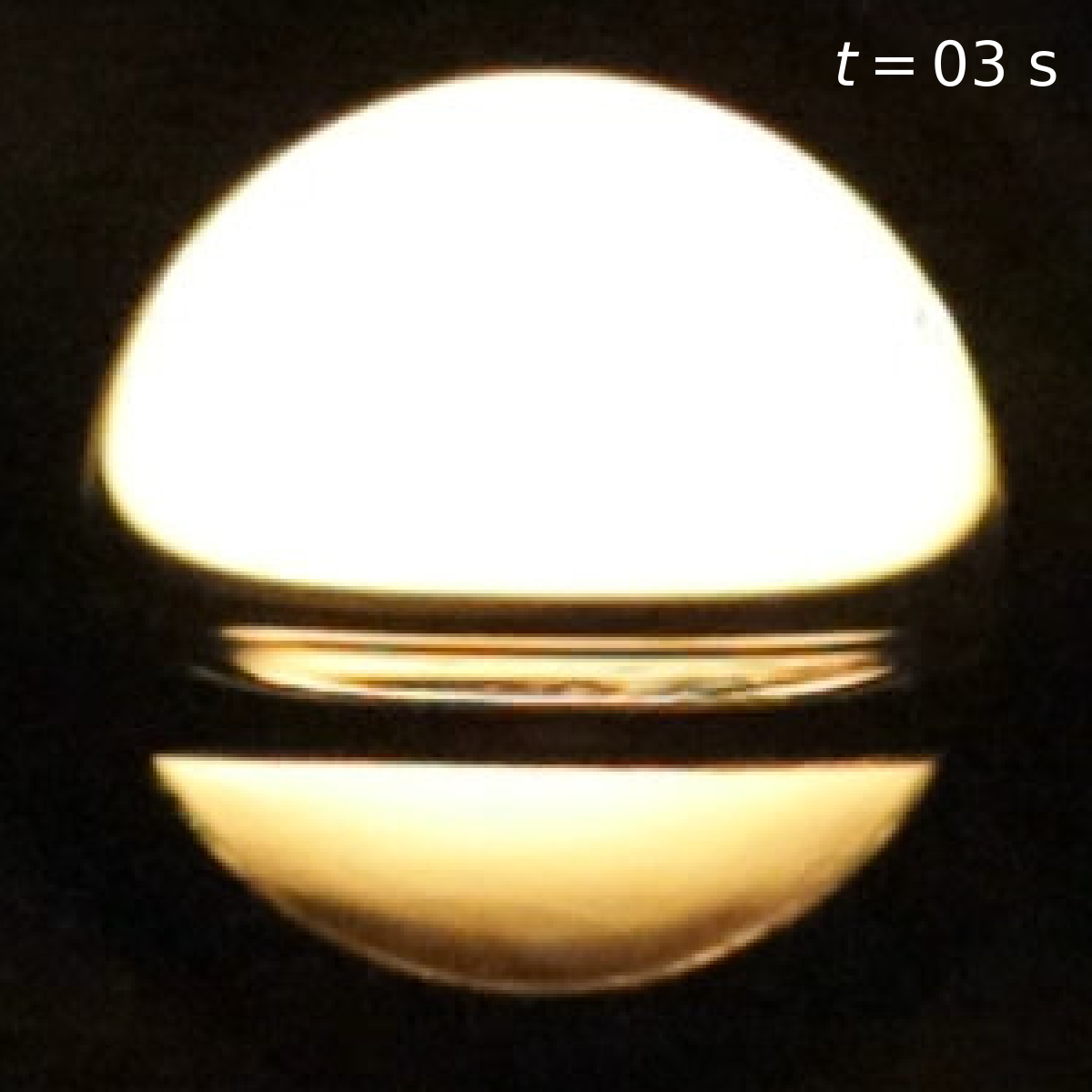}
\end{subfigure}%
\begin{subfigure}{.17\textwidth}
    \centering
    \includegraphics[width=\textwidth]{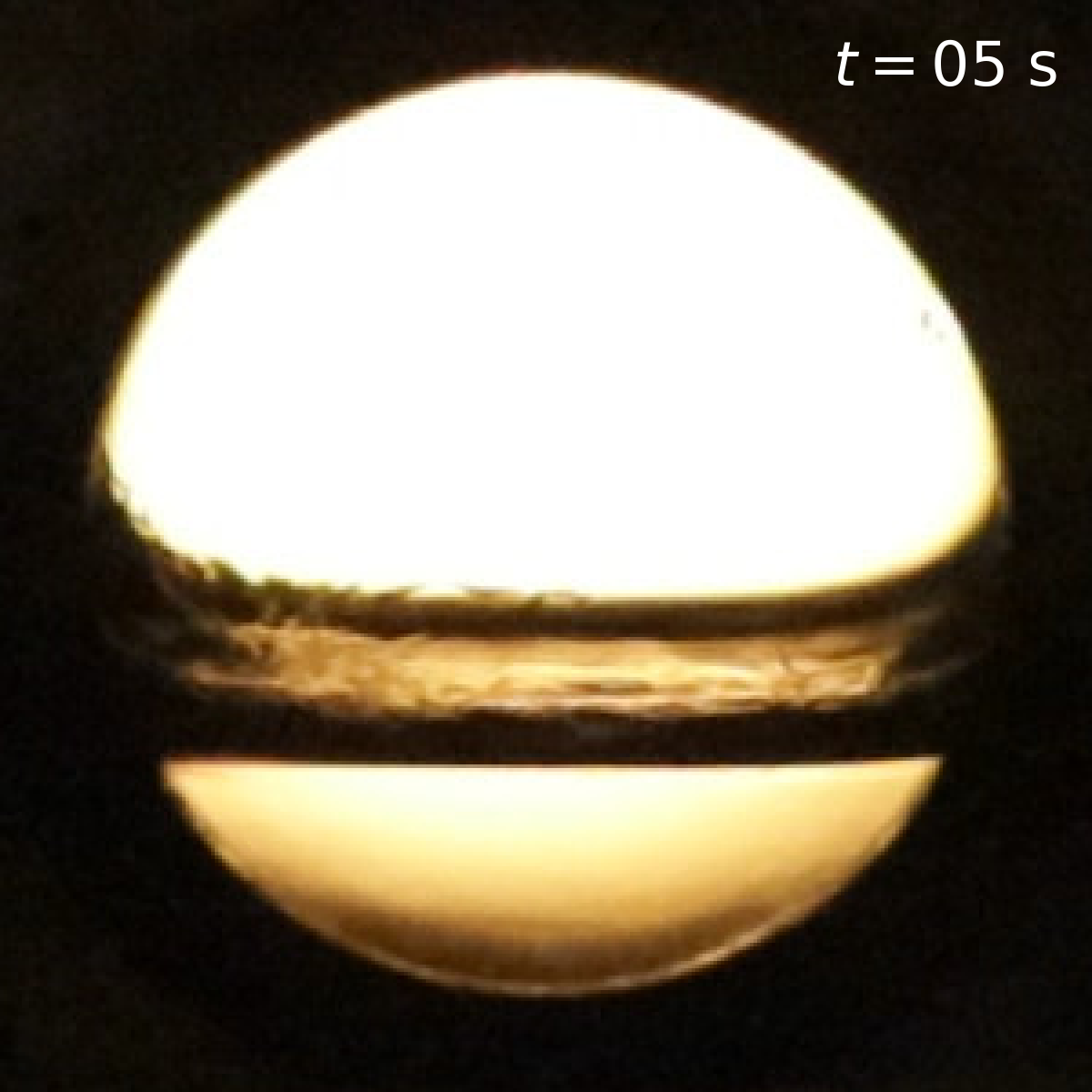}
\end{subfigure}%
\begin{subfigure}{.17\textwidth}
    \centering
    \includegraphics[width=\textwidth]{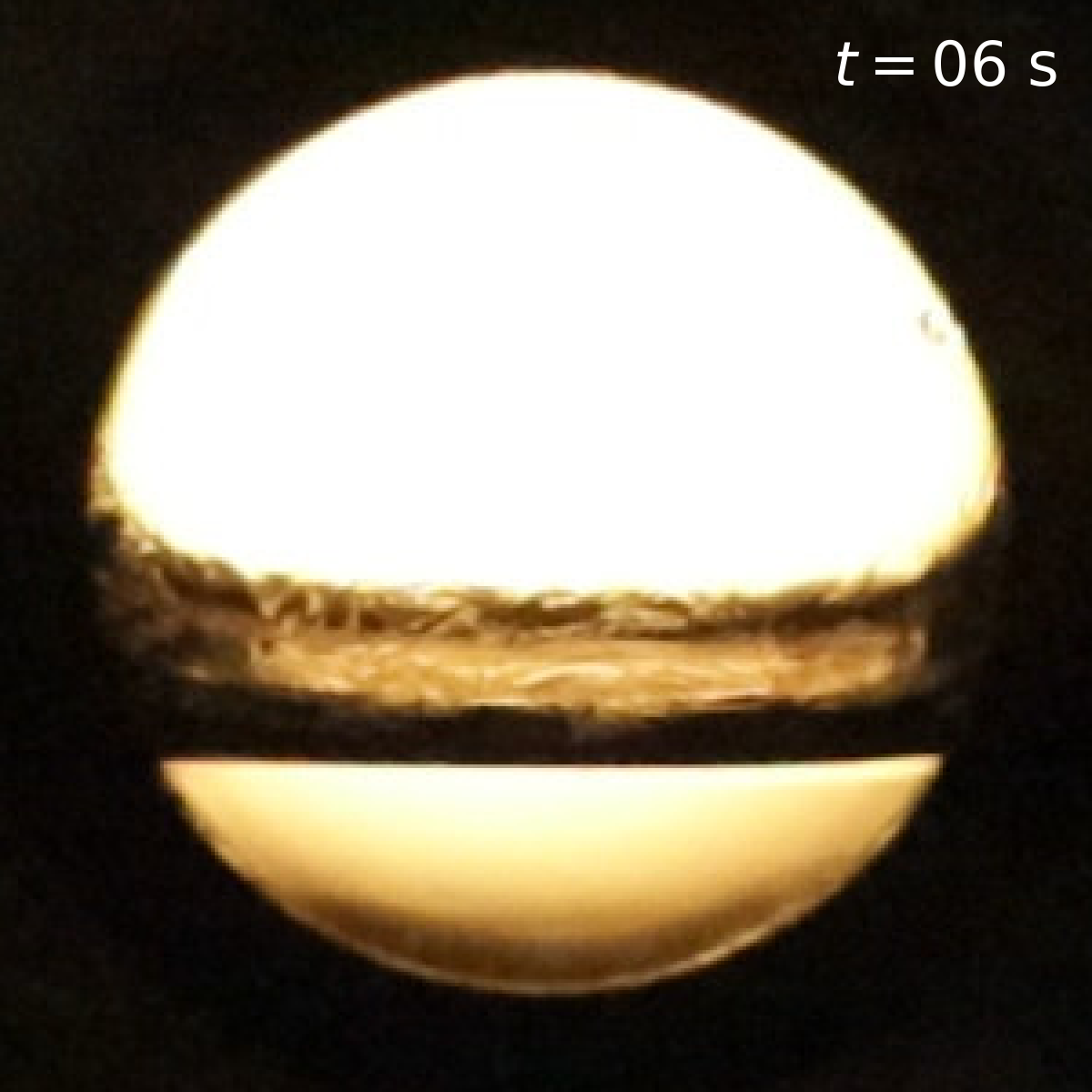}
\end{subfigure}%

\begin{subfigure}{.17\textwidth}
    \centering
    \includegraphics[width=\textwidth]{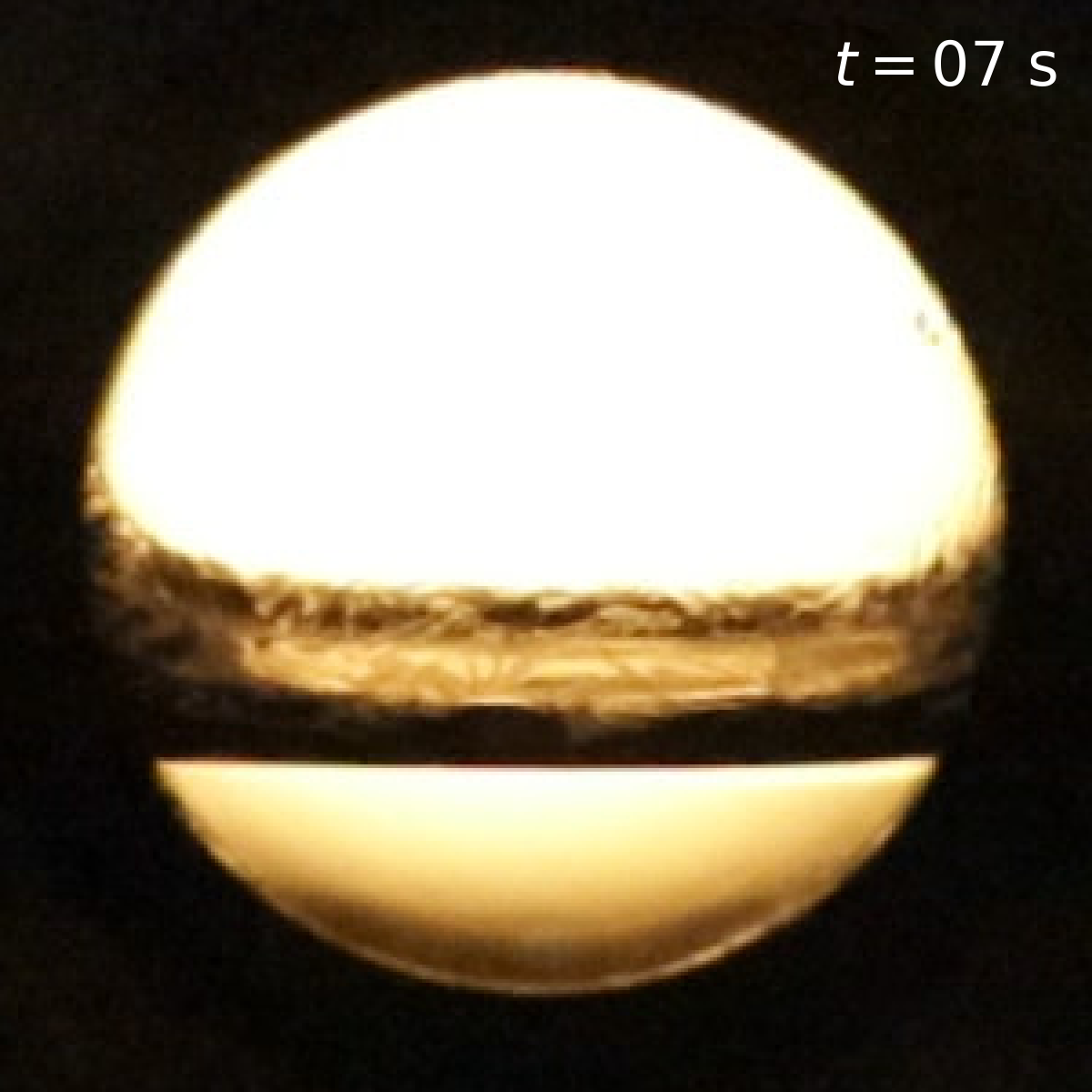}
\end{subfigure}%
\begin{subfigure}{.17\textwidth}
    \centering
    \includegraphics[width=\textwidth]{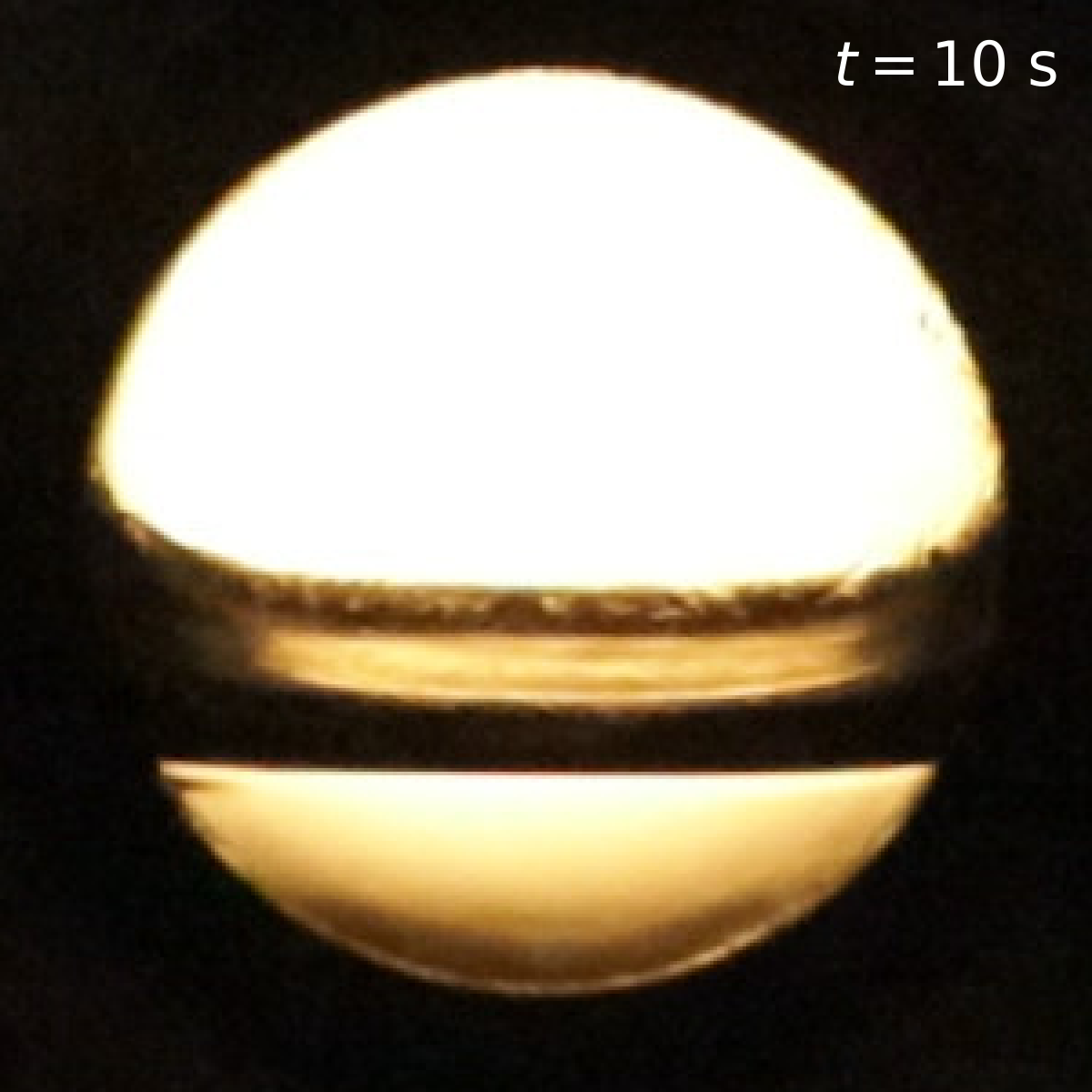}
\end{subfigure}%
\begin{subfigure}{.17\textwidth}
    \centering
    \includegraphics[width=\textwidth]{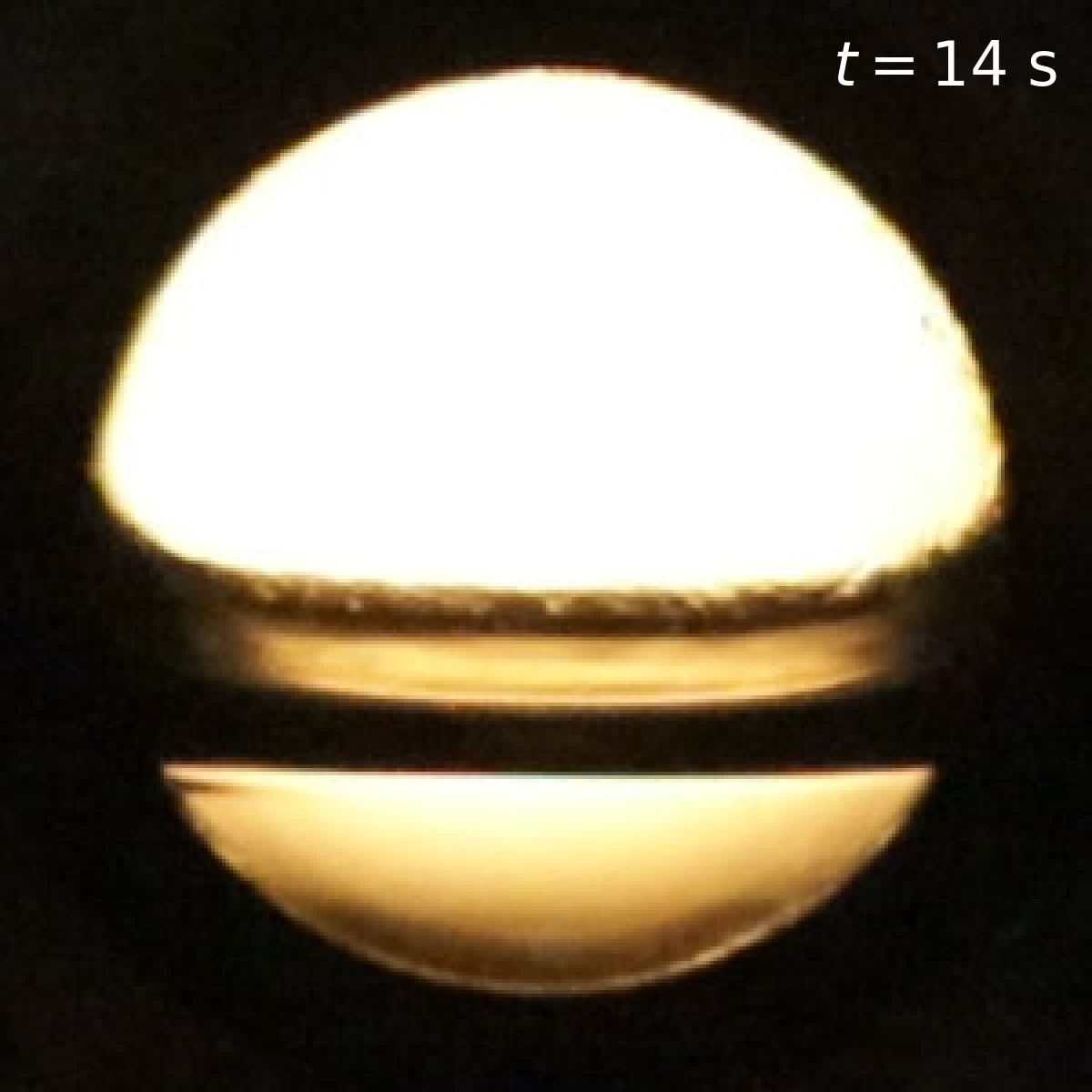}
\end{subfigure}%
\begin{subfigure}{.17\textwidth}
    \centering
    \includegraphics[width=\textwidth]{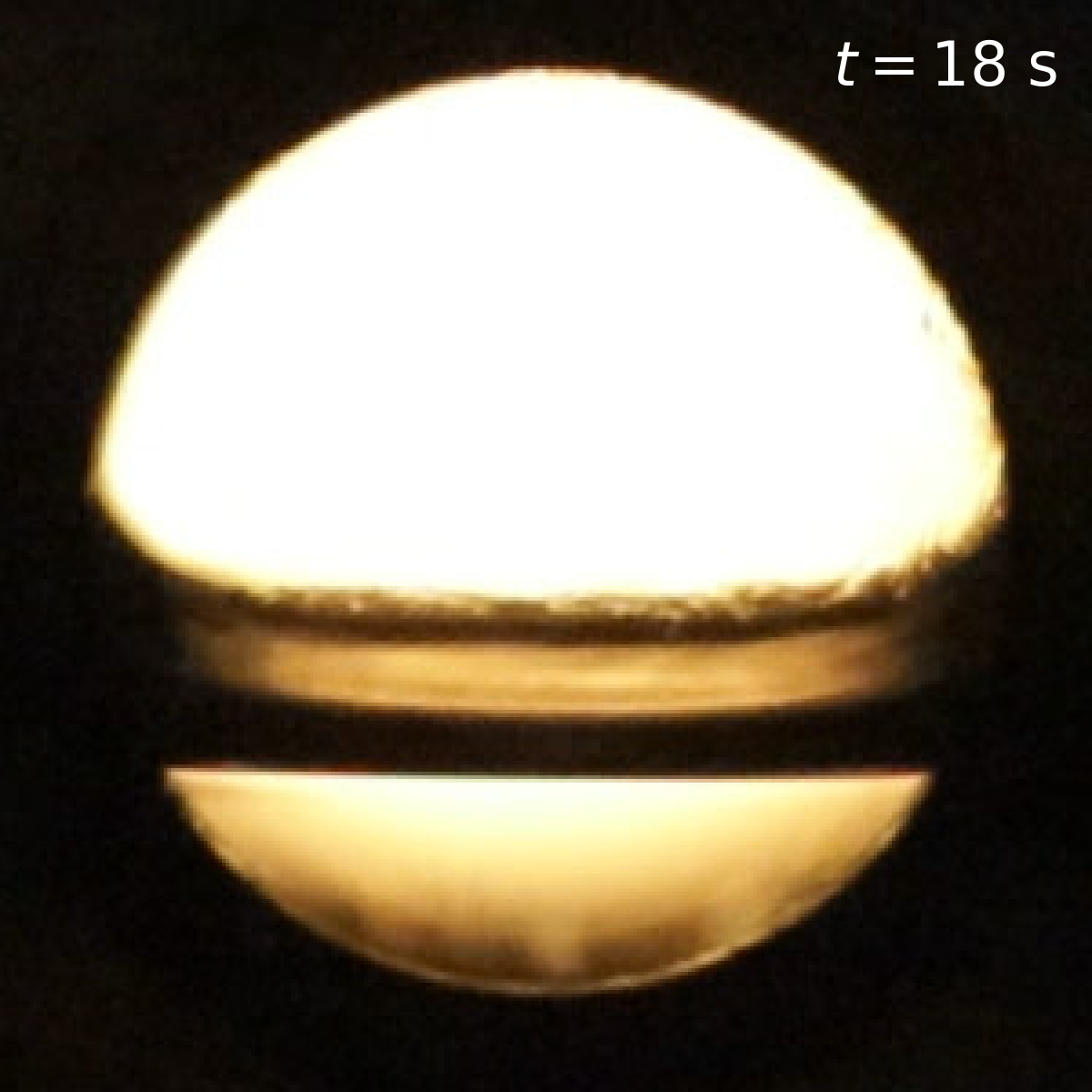}
\end{subfigure}%
\begin{subfigure}{.17\textwidth}
    \centering
    \includegraphics[width=\textwidth]{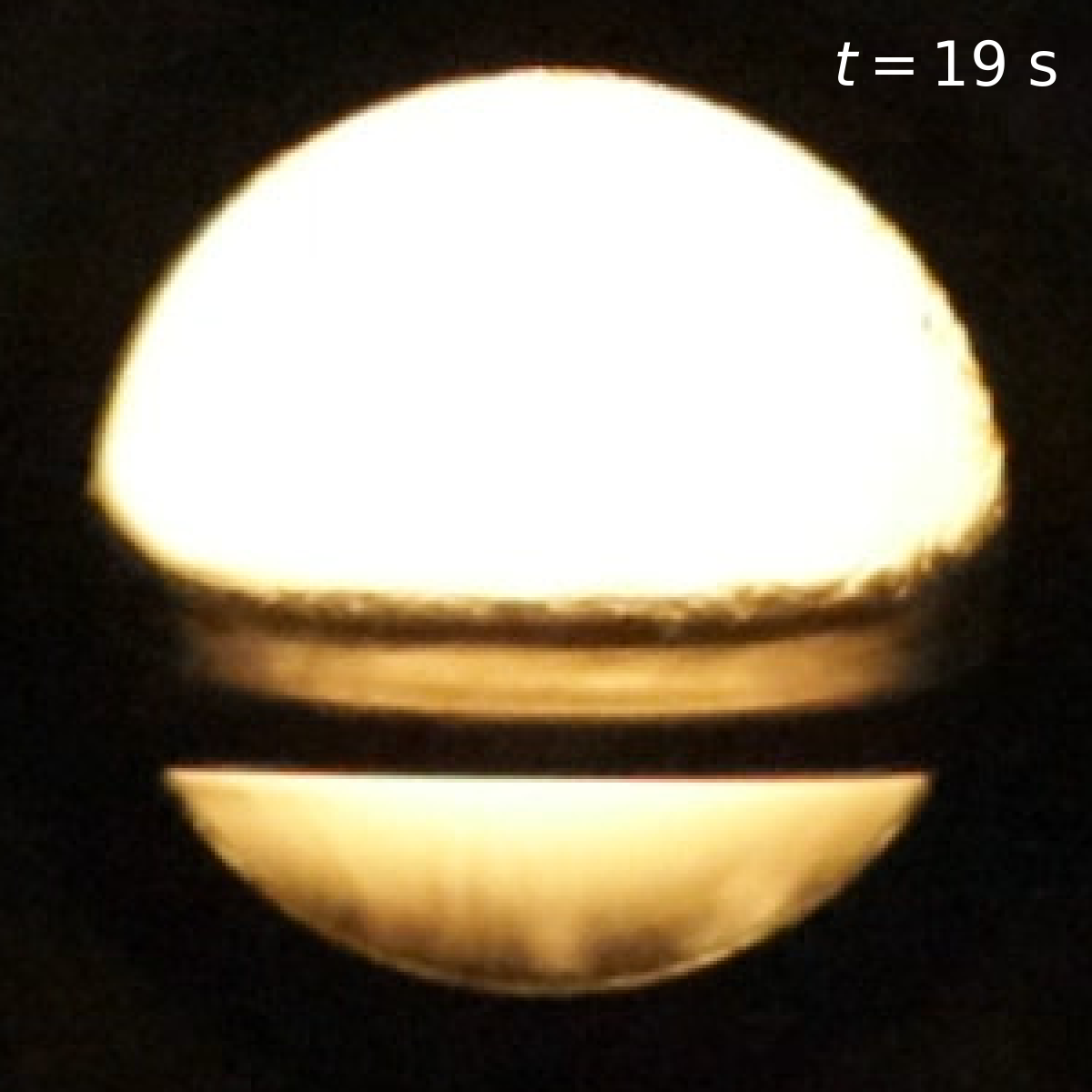}
\end{subfigure}%

\begin{subfigure}{.17\textwidth}
    \centering
    \includegraphics[width=\textwidth]{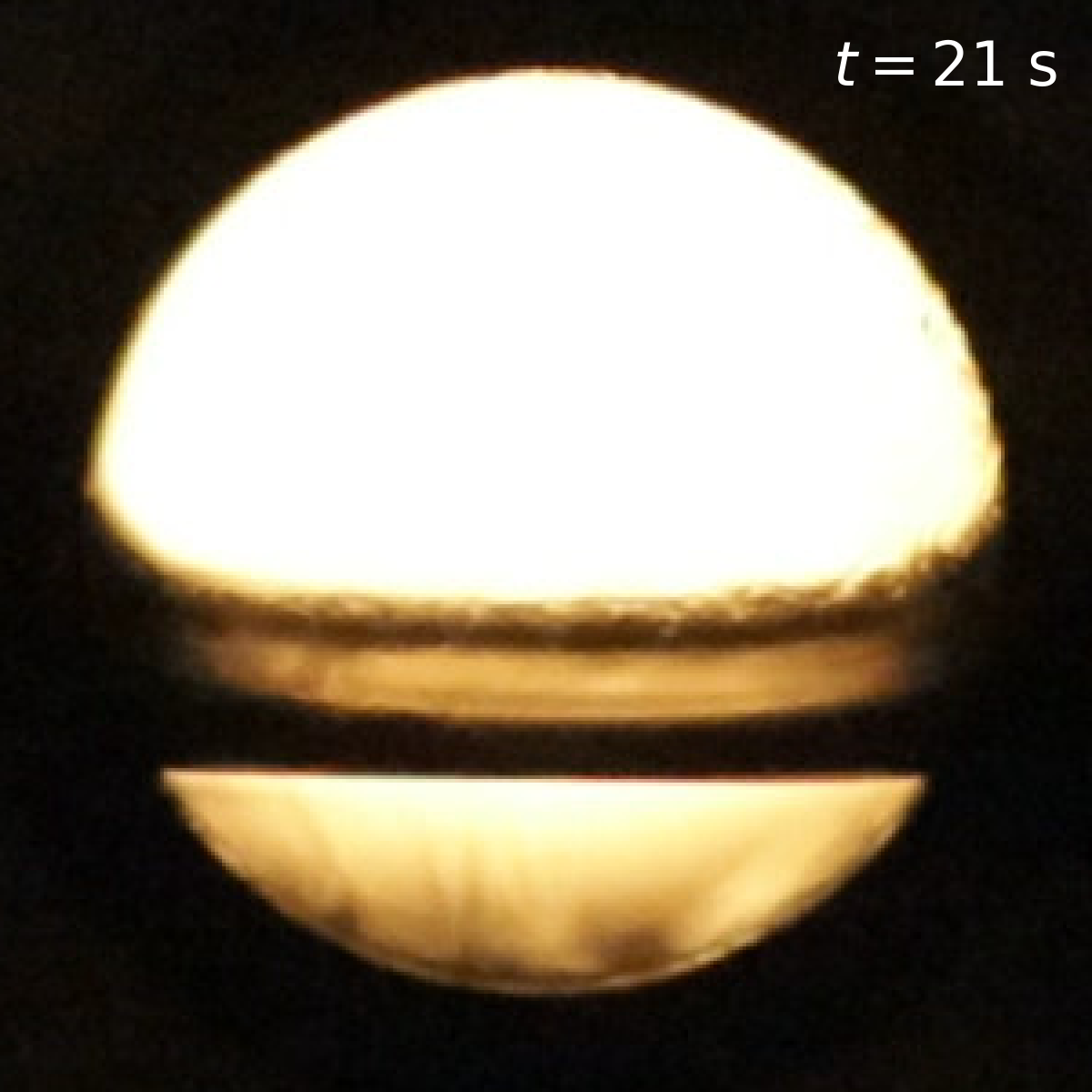}
\end{subfigure}%
\begin{subfigure}{.17\textwidth}
    \centering
    \includegraphics[width=\textwidth]{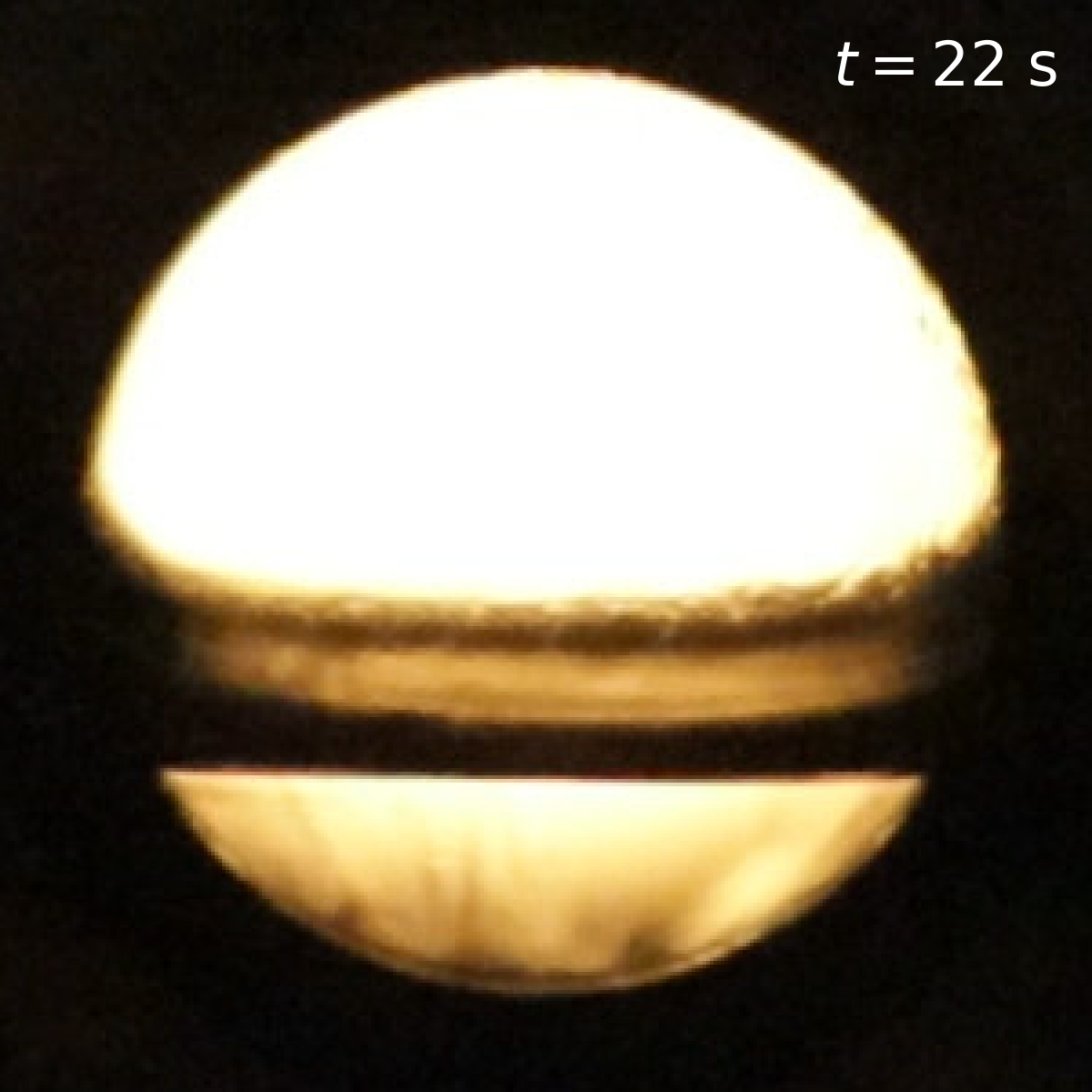}
\end{subfigure}%
\begin{subfigure}{.17\textwidth}
    \centering
    \includegraphics[width=\textwidth]{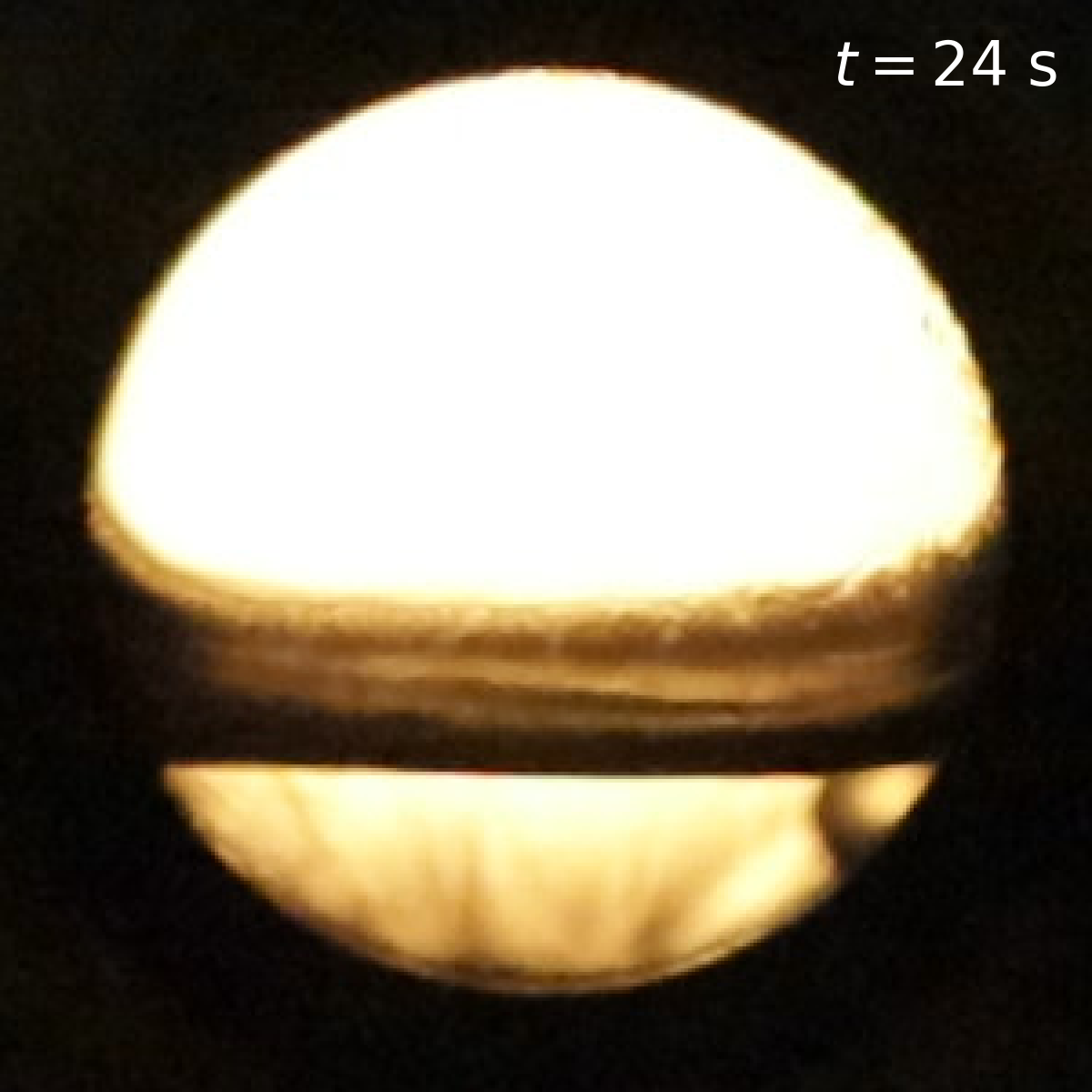}
\end{subfigure}%
\begin{subfigure}{.17\textwidth}
    \centering
    \includegraphics[width=\textwidth]{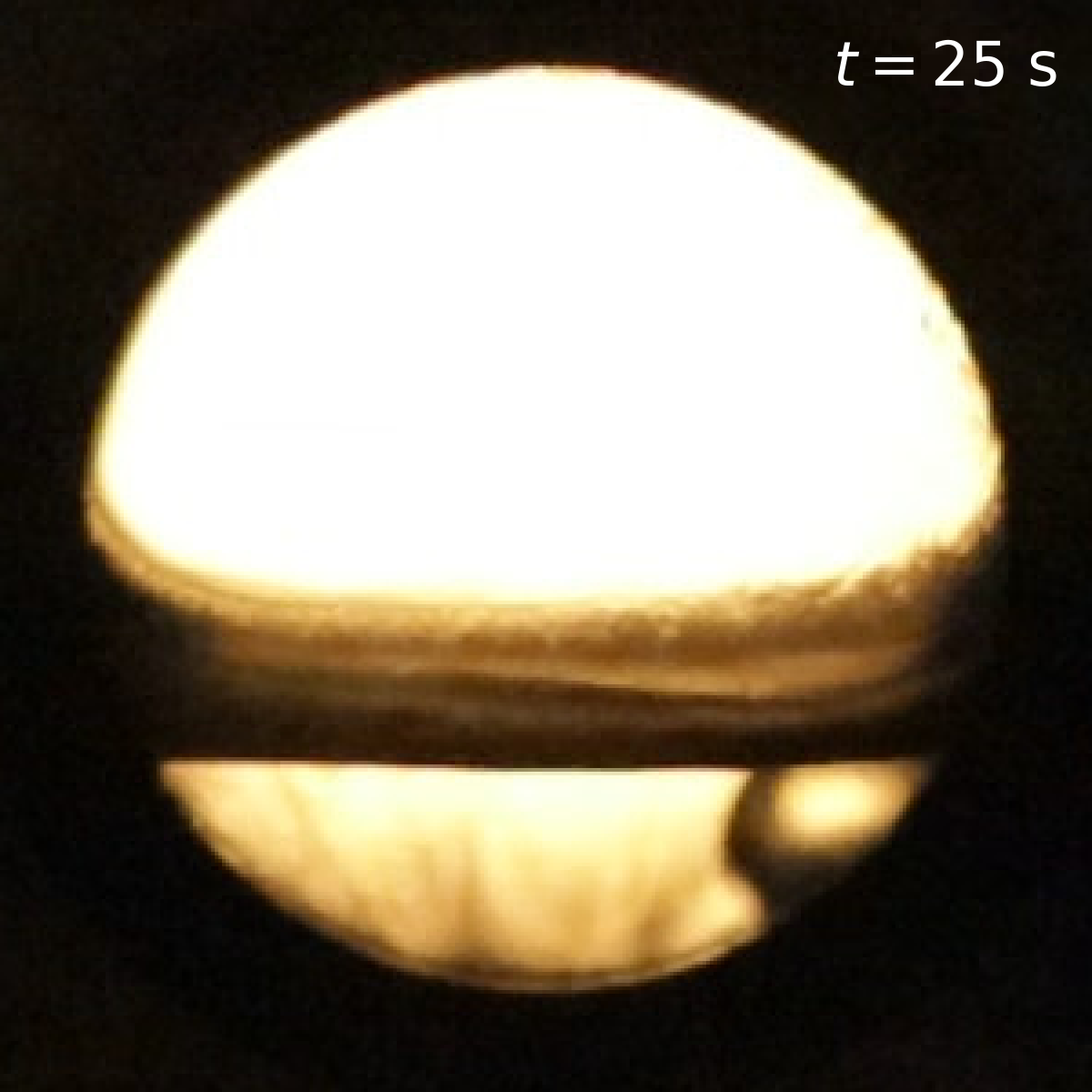}
\end{subfigure}%
\begin{subfigure}{.17\textwidth}
    \centering
    \includegraphics[width=\textwidth]{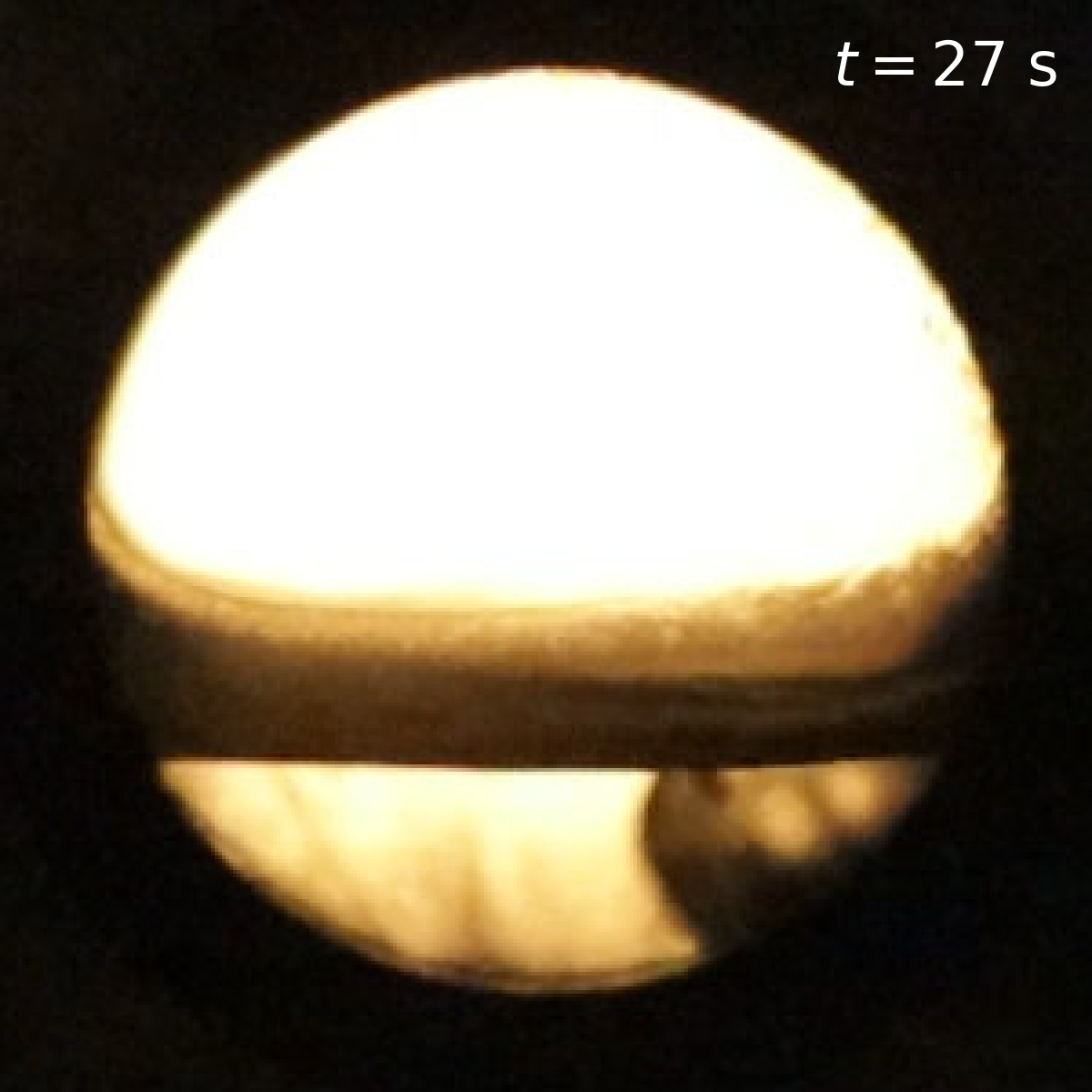}
\end{subfigure}%

\begin{subfigure}{.17\textwidth}
    \centering
    \includegraphics[width=\textwidth]{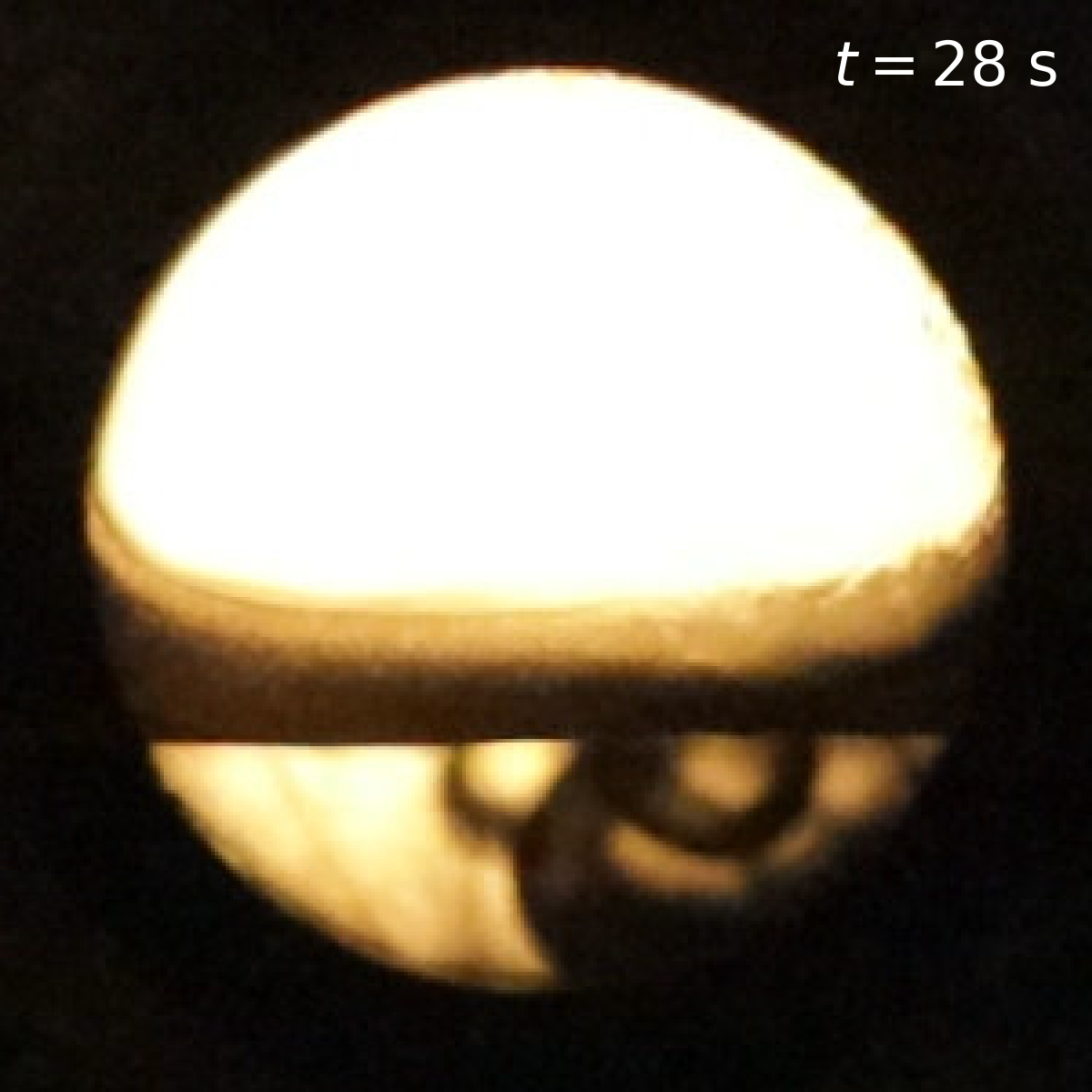}
\end{subfigure}%
\begin{subfigure}{.17\textwidth}
    \centering
    \includegraphics[width=\textwidth]{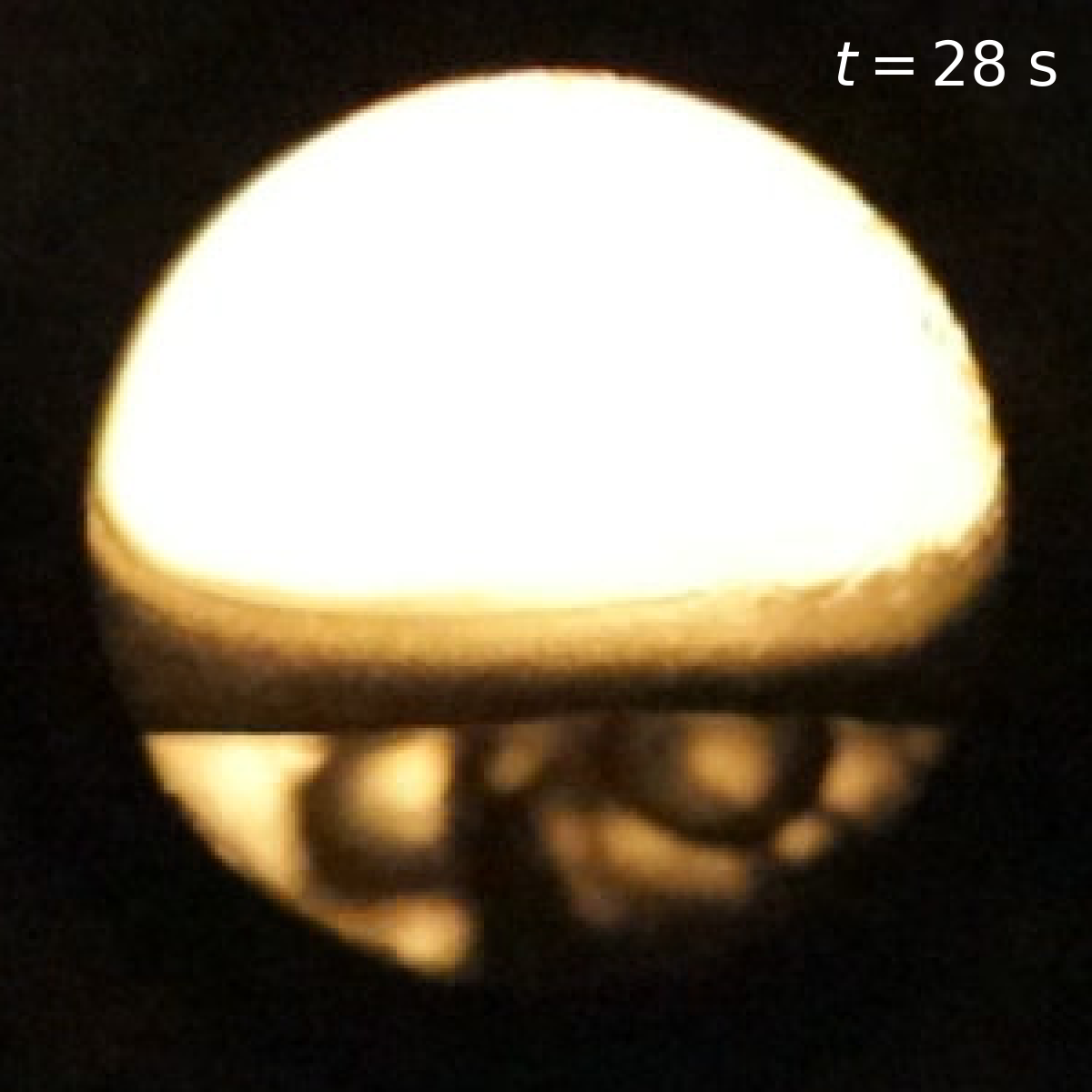}
\end{subfigure}%
\begin{subfigure}{.17\textwidth}
    \centering
    \includegraphics[width=\textwidth]{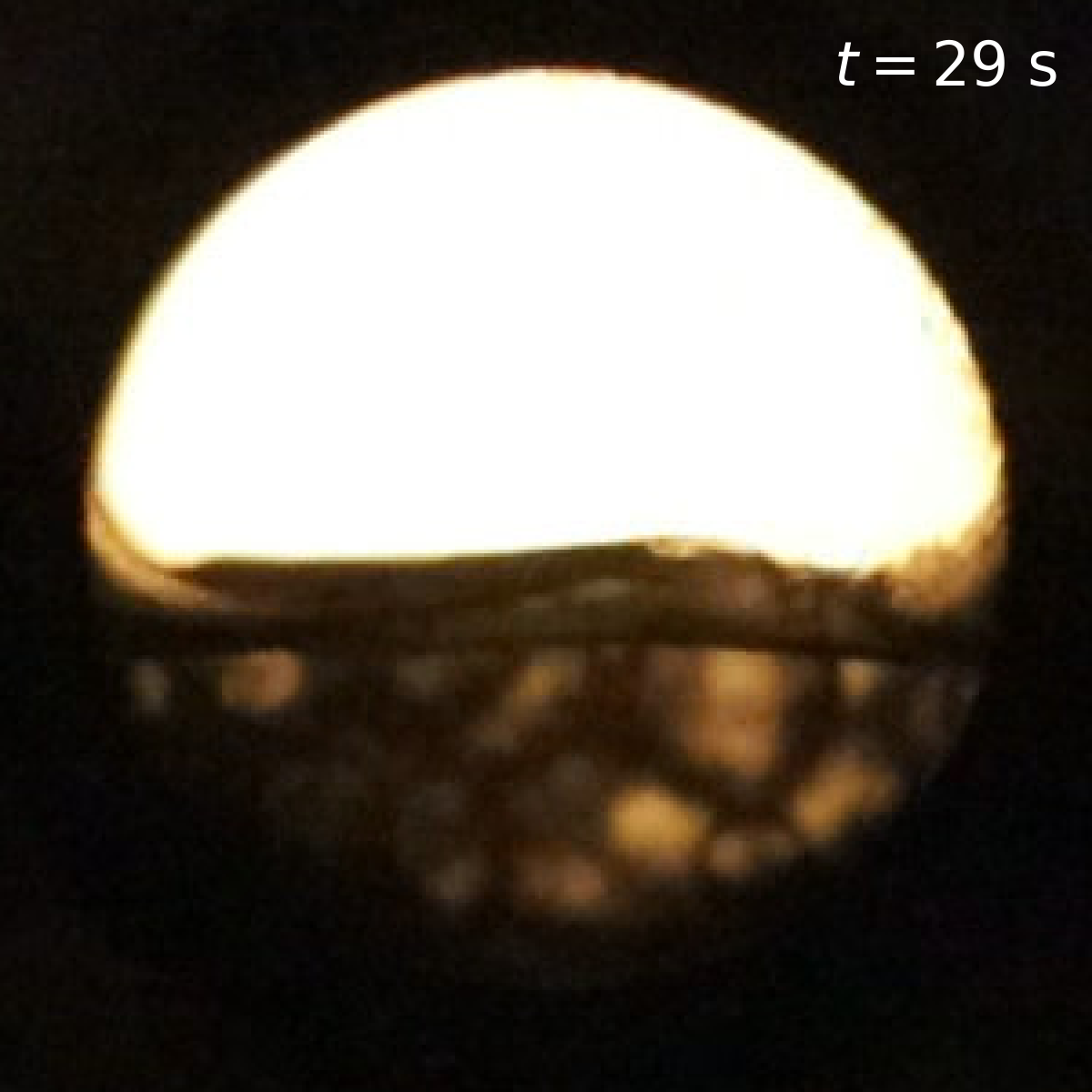}
\end{subfigure}%
\begin{subfigure}{.17\textwidth}
    \centering
    \includegraphics[width=\textwidth]{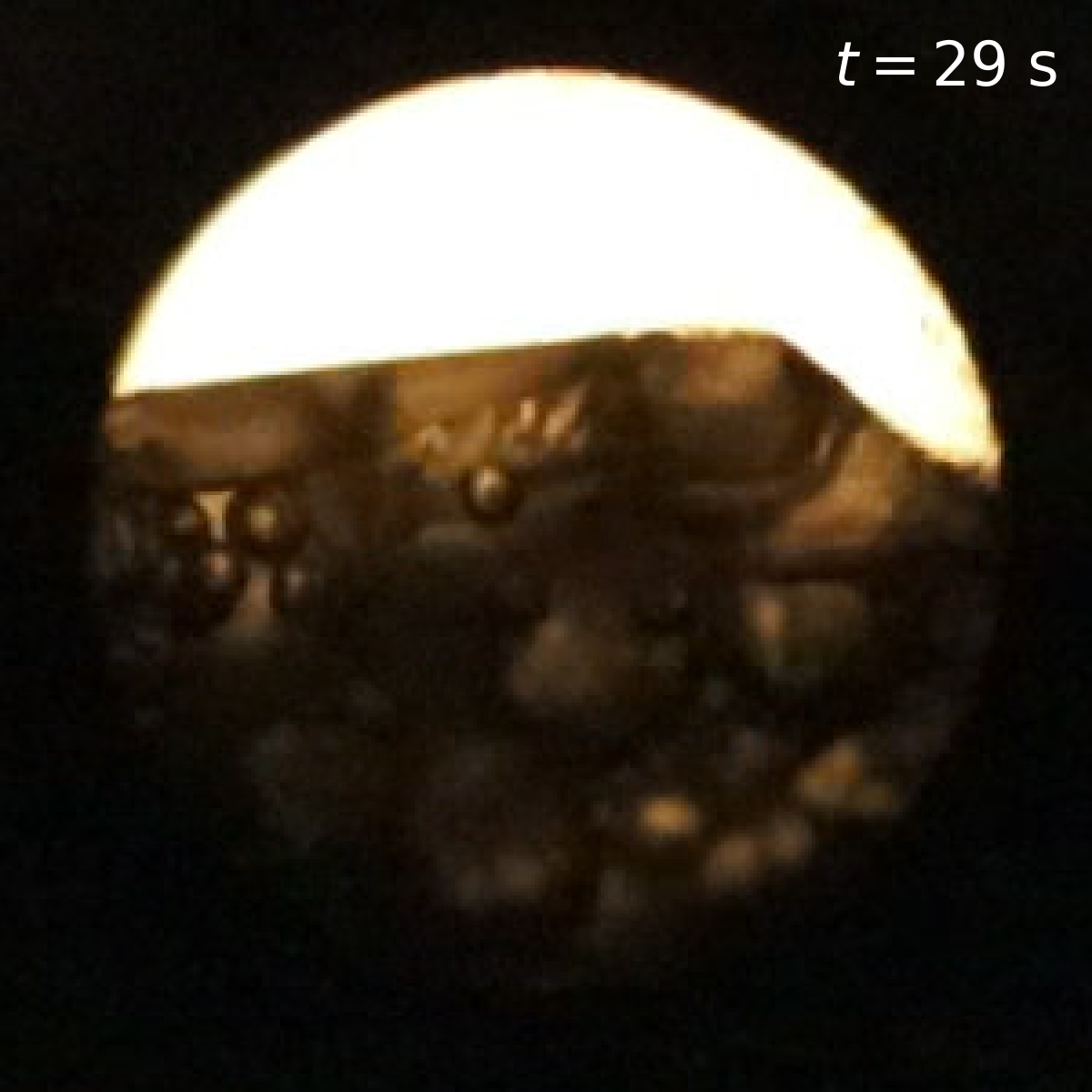}
\end{subfigure}%
\begin{subfigure}{.17\textwidth}
    \centering
    \includegraphics[width=\textwidth]{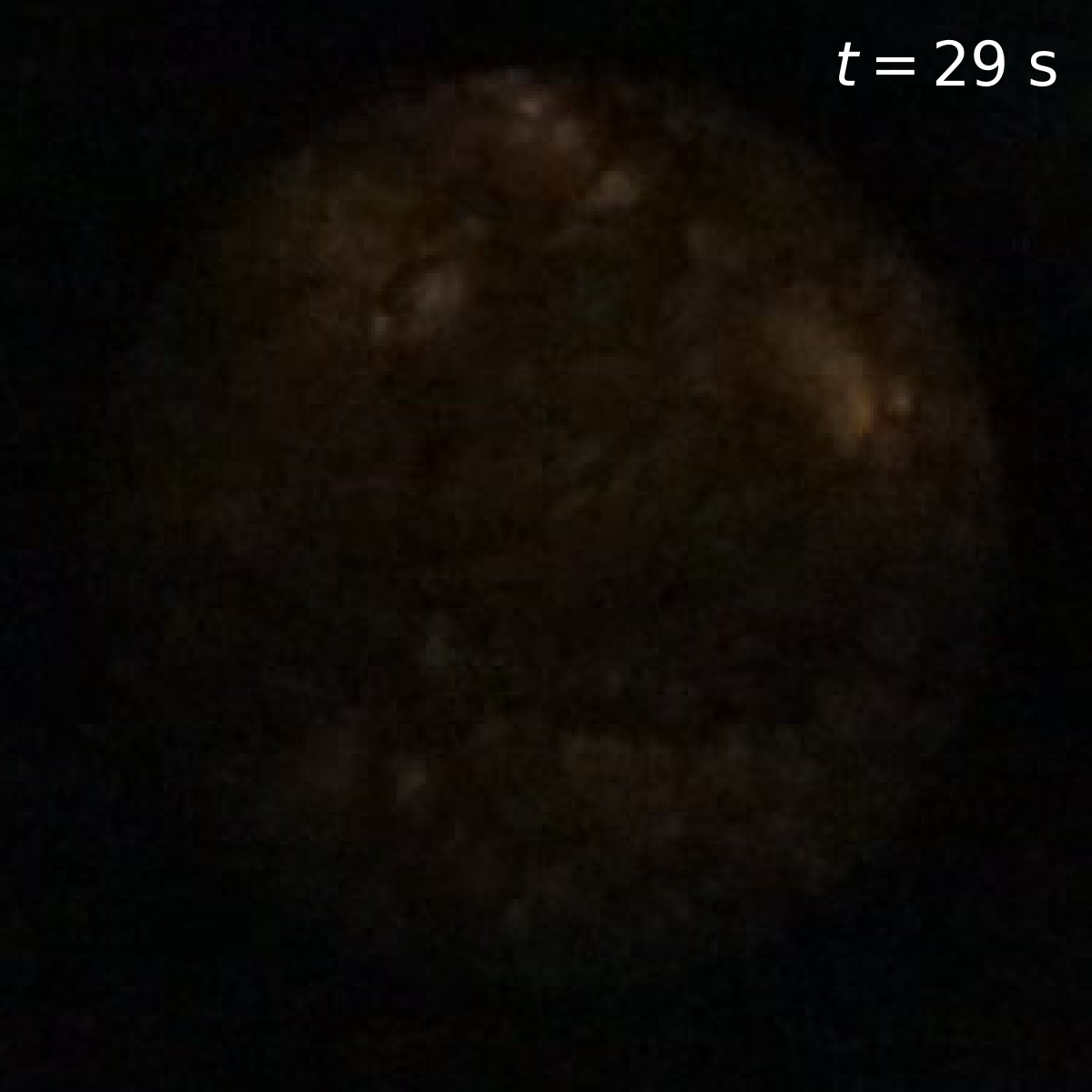}
\end{subfigure}%
\caption{
Rapid freezing of liquid \DTwo{}.
Images are sequentially from left to right, then top to bottom.
The entire sequence of images depicts a time span of \SI{\approx30}{\second}.
Some processes depicted above happened very quickly, requiring several images of the cell within the span of a single second to capture the change.
}
\label{fig:explosive_crystallization}
\end{figure}
When deuterium \DTwo{} is cooled down quickly below its freezing point, a different behavior can be seen, as is shown in \cref{fig:explosive_crystallization}.
At first a spontaneous formation of small nucleation sites along the interfaces of the liquid and the windows can be seen (see e.g. $t = \SI{06}{\second}$).
These nucleation sites appear to loose their sharply defined shape quickly  (see e.g. $t = \SI{14}{\second}$).
Shortly afterwards, the formation of a crystal structure in the bulk liquid can be seen (starting at  $t = \SI{18}{\second}$), which continues until the entire liquid phase is crystallized.
After remaining in this state for a short period of time, the formation of what appear to be fronts of phase change traveling across the solid can be observed (starting at  $t = \SI{24}{\second}$). 
These fronts rapidly fill the solid with interfaces, refracting light away from the camera.
This process happens on the time scale of several seconds and can leave the resulting solid behind as an opaque structure due to the multitude of internal surfaces. 
This phenomenon is likely to include a phase change to the gas phase, as it can lead to a large volume increase of the final solid, which is likely from gas-filled bubbles.
This behavior has been observed in rapidly freezing \HTwo{} as well as \DTwo{}.

\subsubsection{Crystal annealing}

\begin{figure}[t]
\centering
\begin{subfigure}{.16\textwidth}
    \centering
    \includegraphics[width=\textwidth]{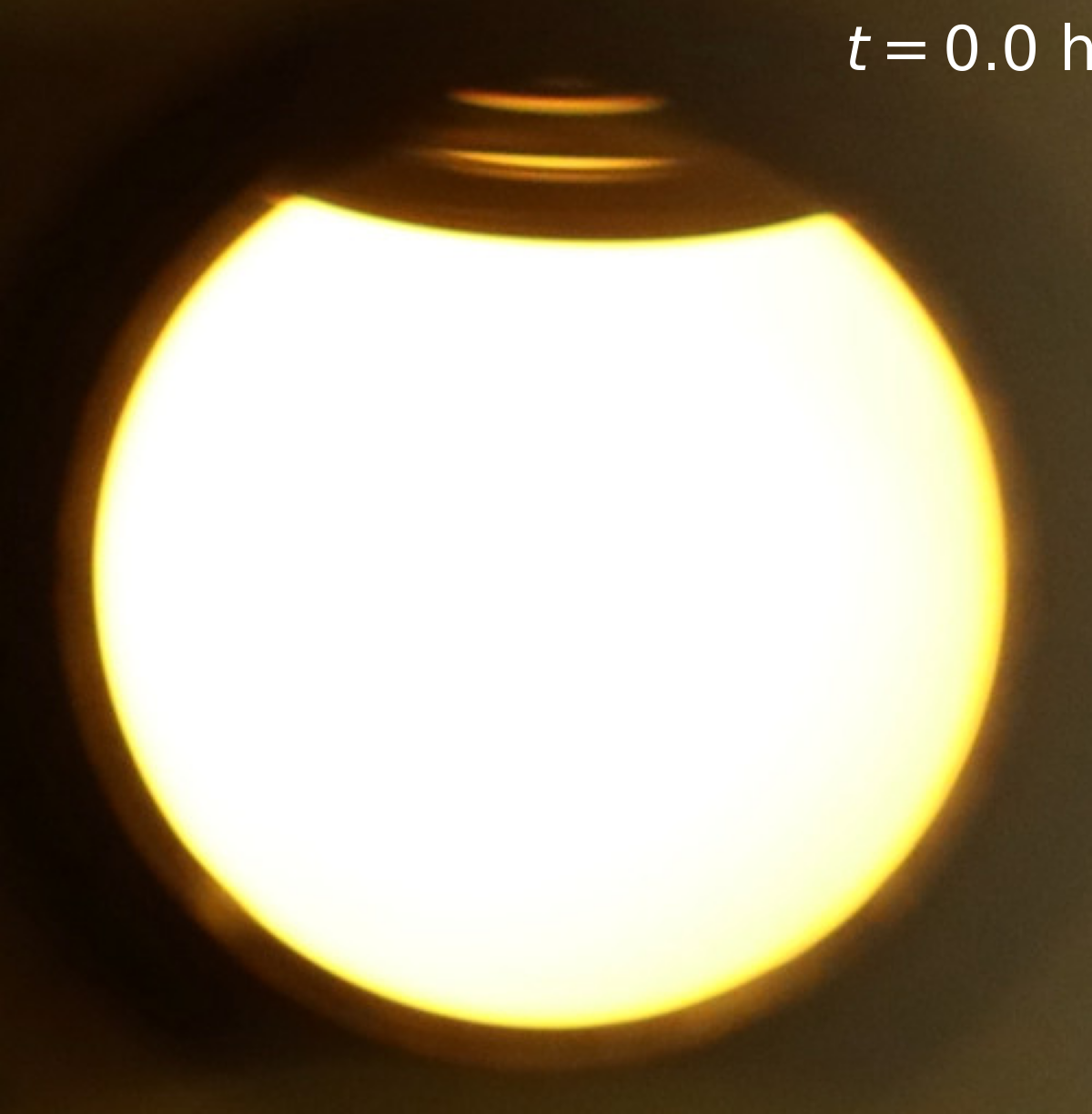}
\end{subfigure}%
\begin{subfigure}{.16\textwidth}
    \centering
    \includegraphics[width=\textwidth]{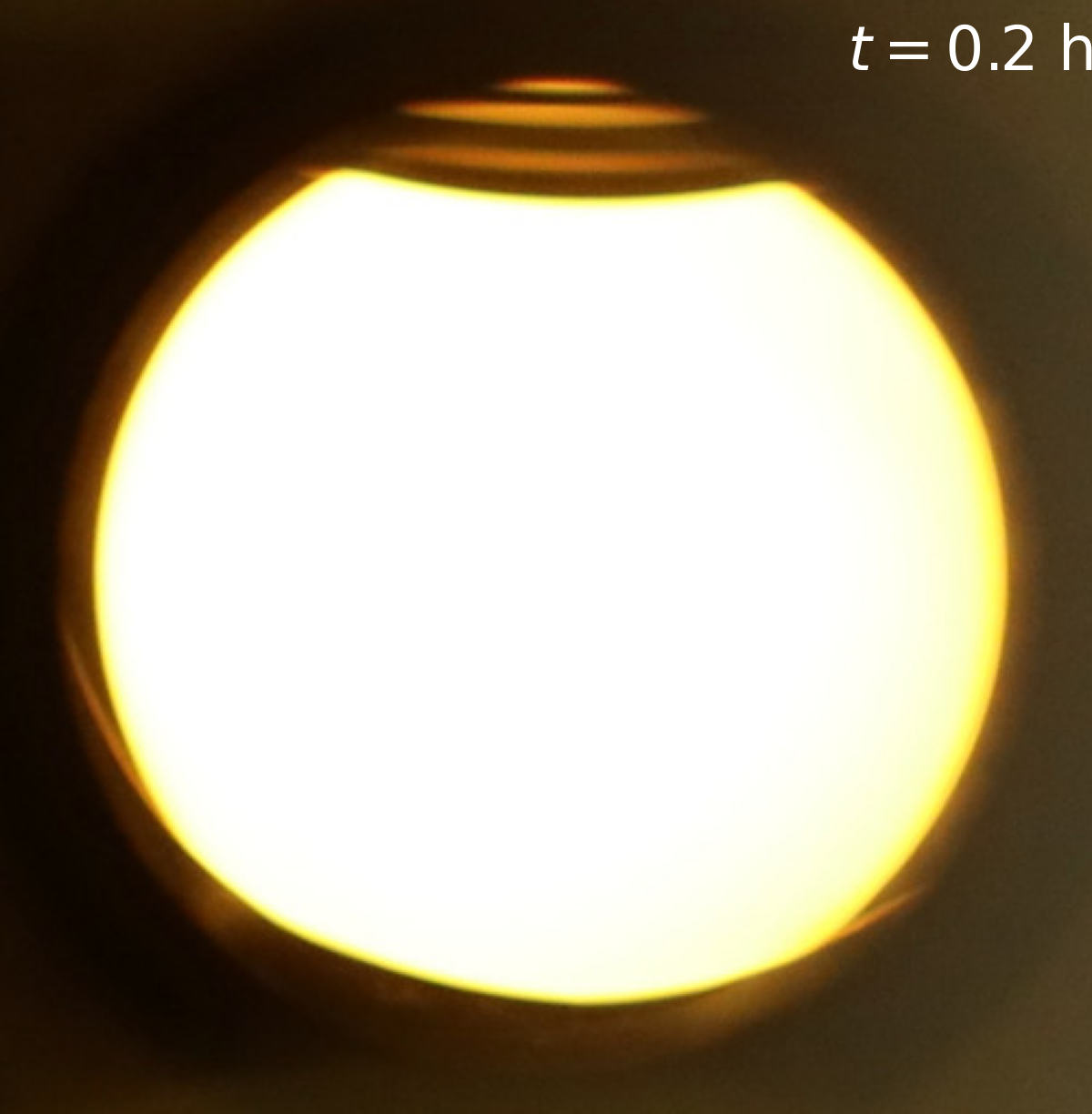}
\end{subfigure}%
\begin{subfigure}{.16\textwidth}
    \centering
    \includegraphics[width=\textwidth]{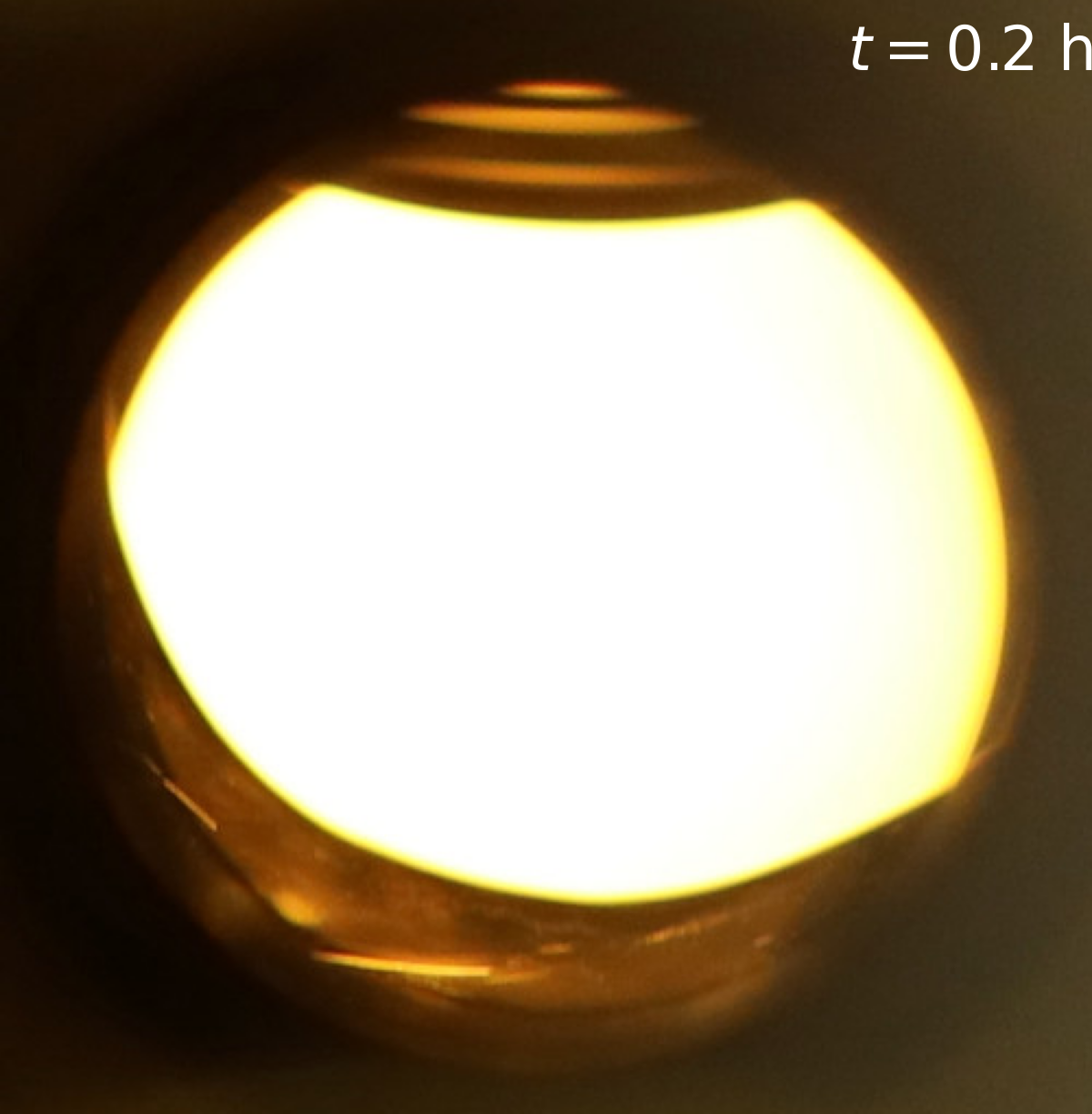}
\end{subfigure}%
\begin{subfigure}{.16\textwidth}
    \centering
    \includegraphics[width=\textwidth]{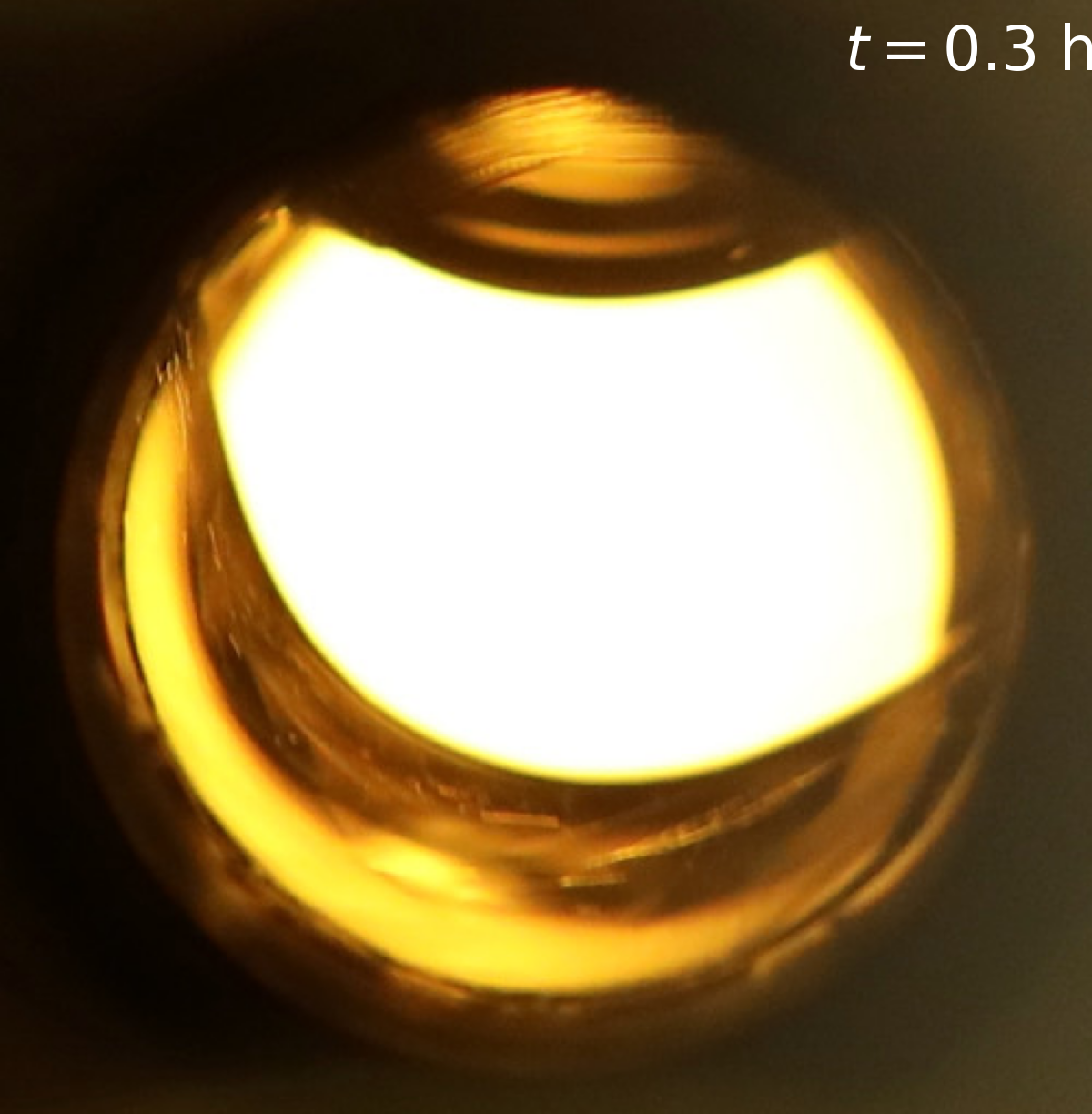}
\end{subfigure}%
\begin{subfigure}{.16\textwidth}
    \centering
    \includegraphics[width=\textwidth]{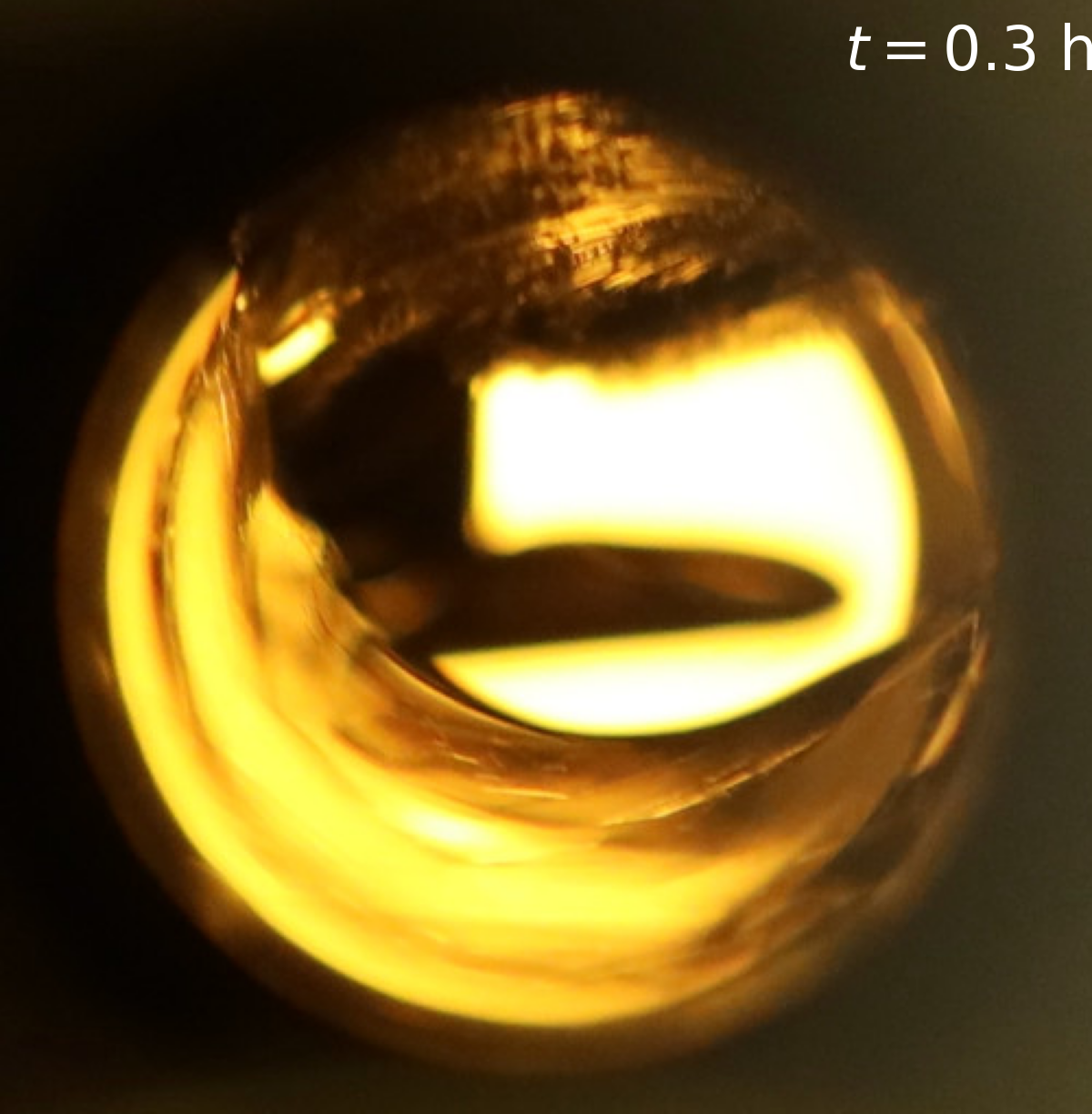}
\end{subfigure}%
\begin{subfigure}{.16\textwidth}
    \centering
    \includegraphics[width=\textwidth]{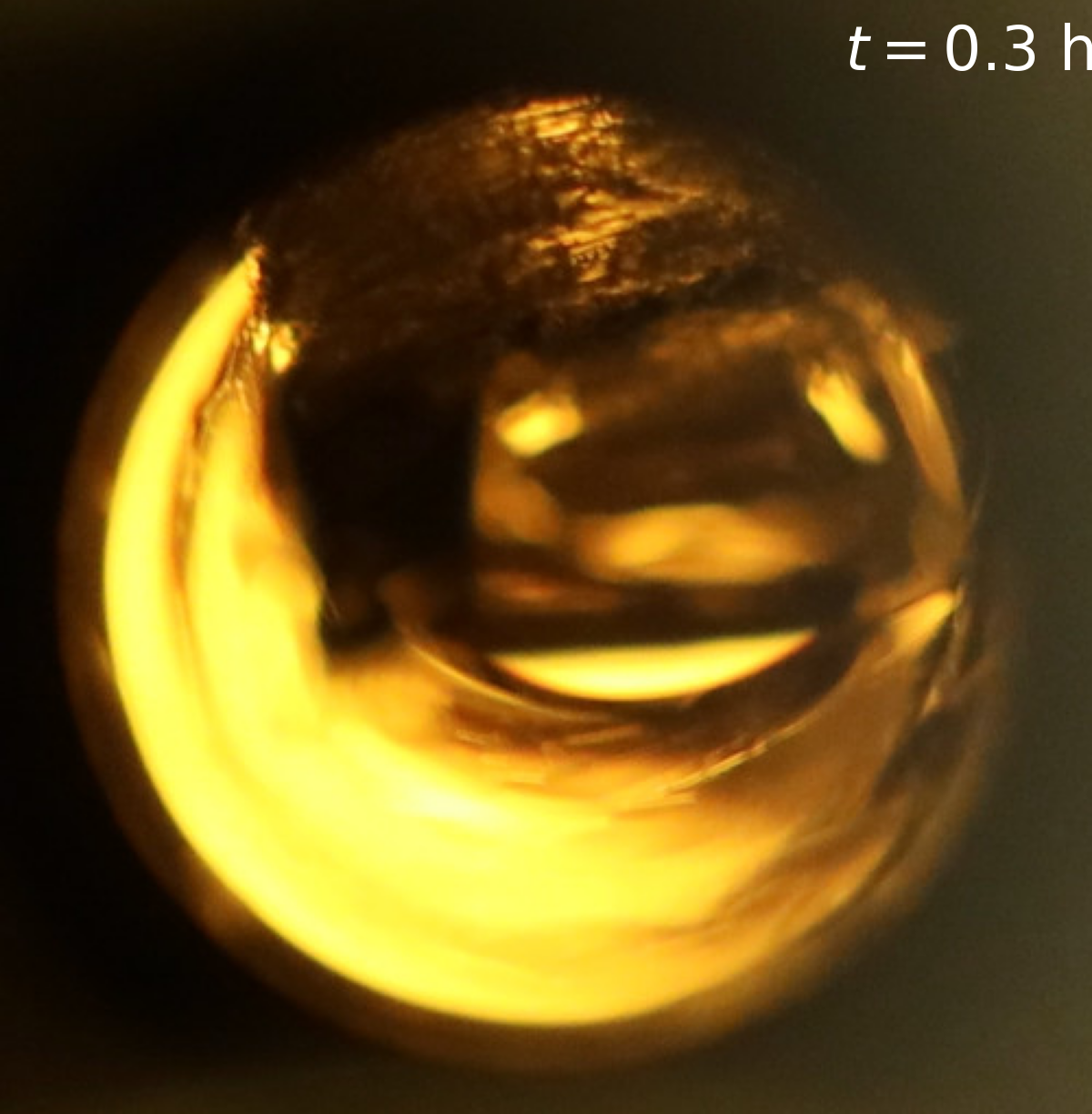}
\end{subfigure}%

\begin{subfigure}{.16\textwidth}
    \centering
    \includegraphics[width=\textwidth]{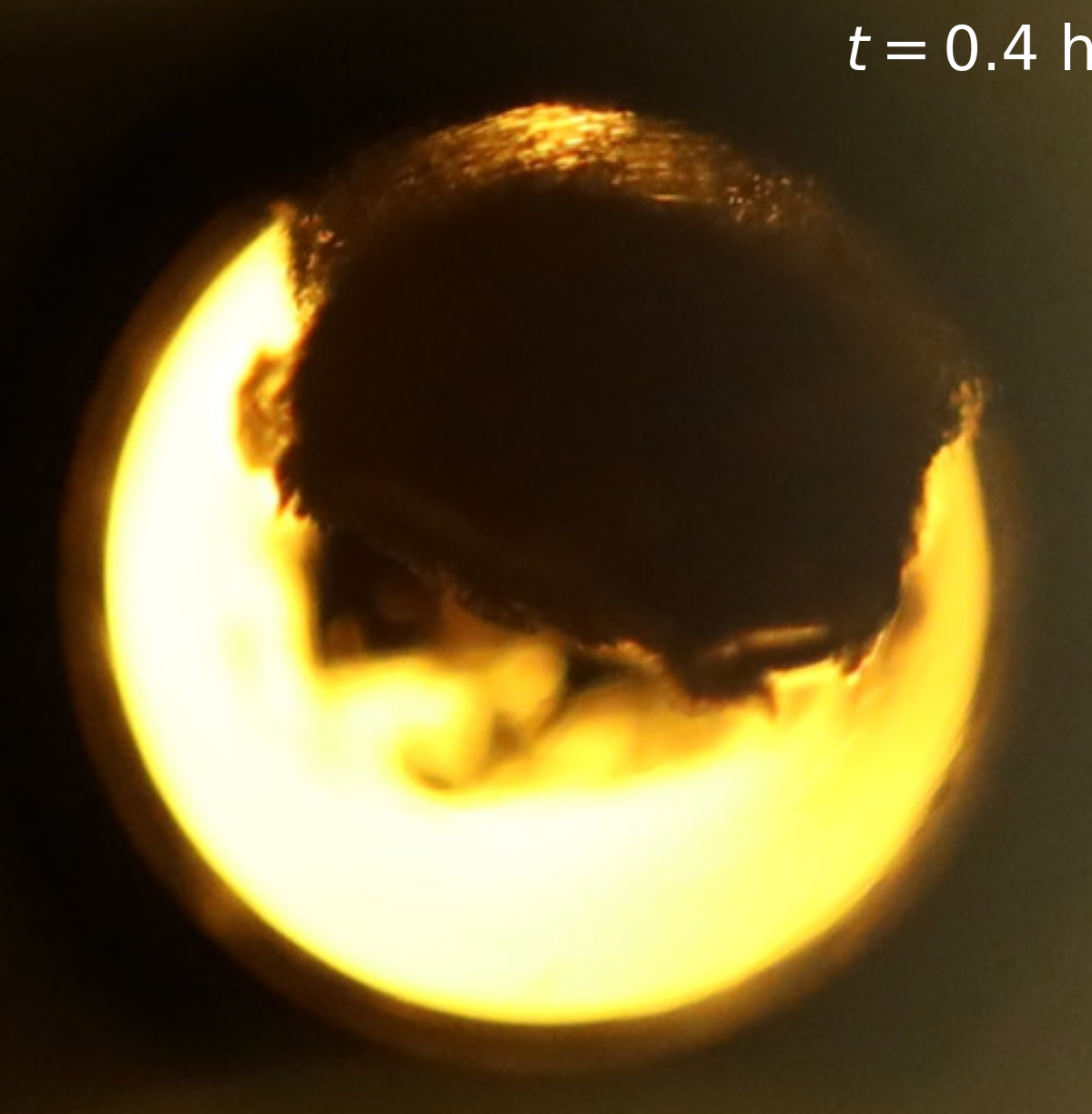}
\end{subfigure}%
\begin{subfigure}{.16\textwidth}
    \centering
    \includegraphics[width=\textwidth]{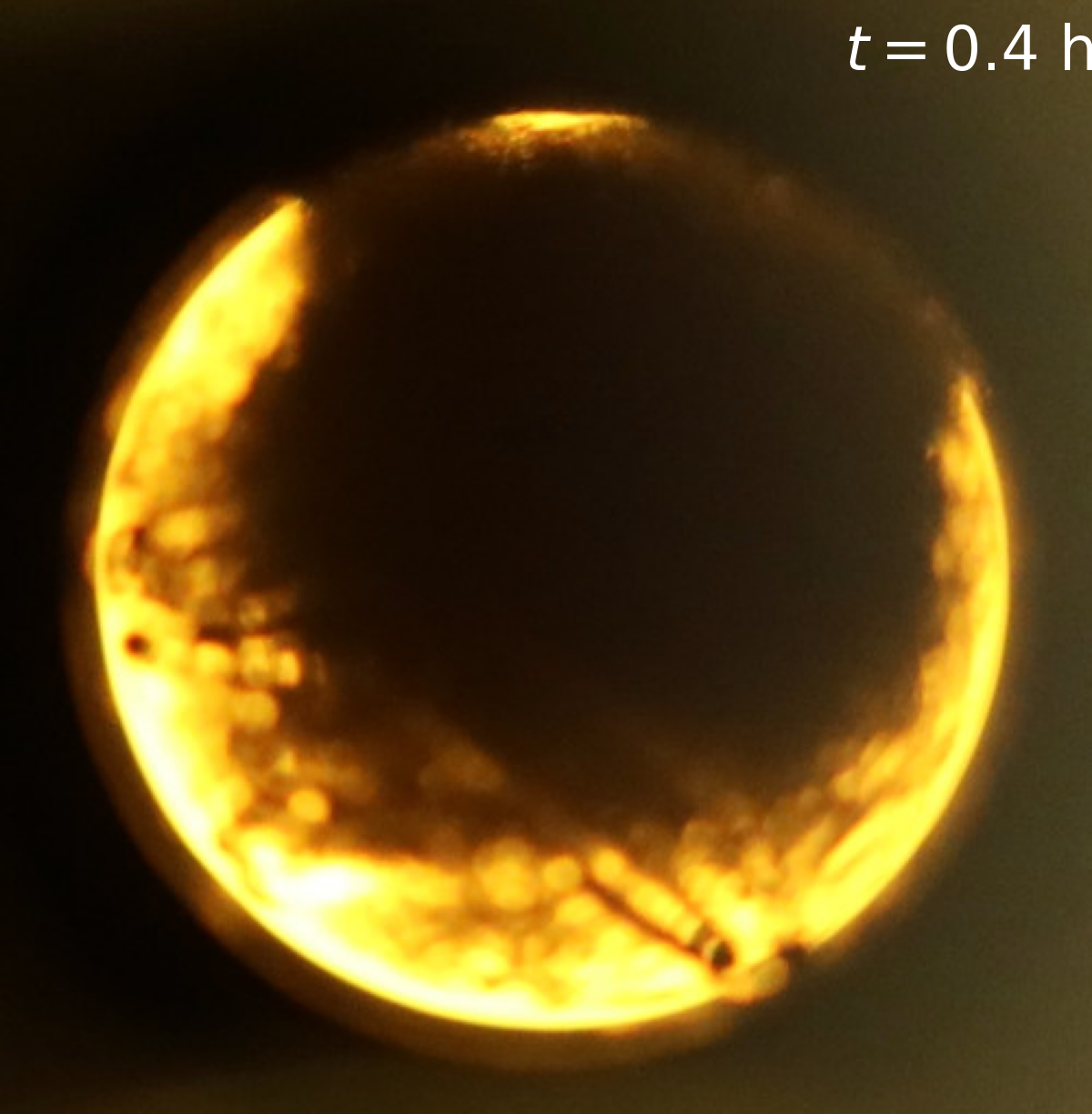}
\end{subfigure}%
\begin{subfigure}{.16\textwidth}
    \centering
    \includegraphics[width=\textwidth]{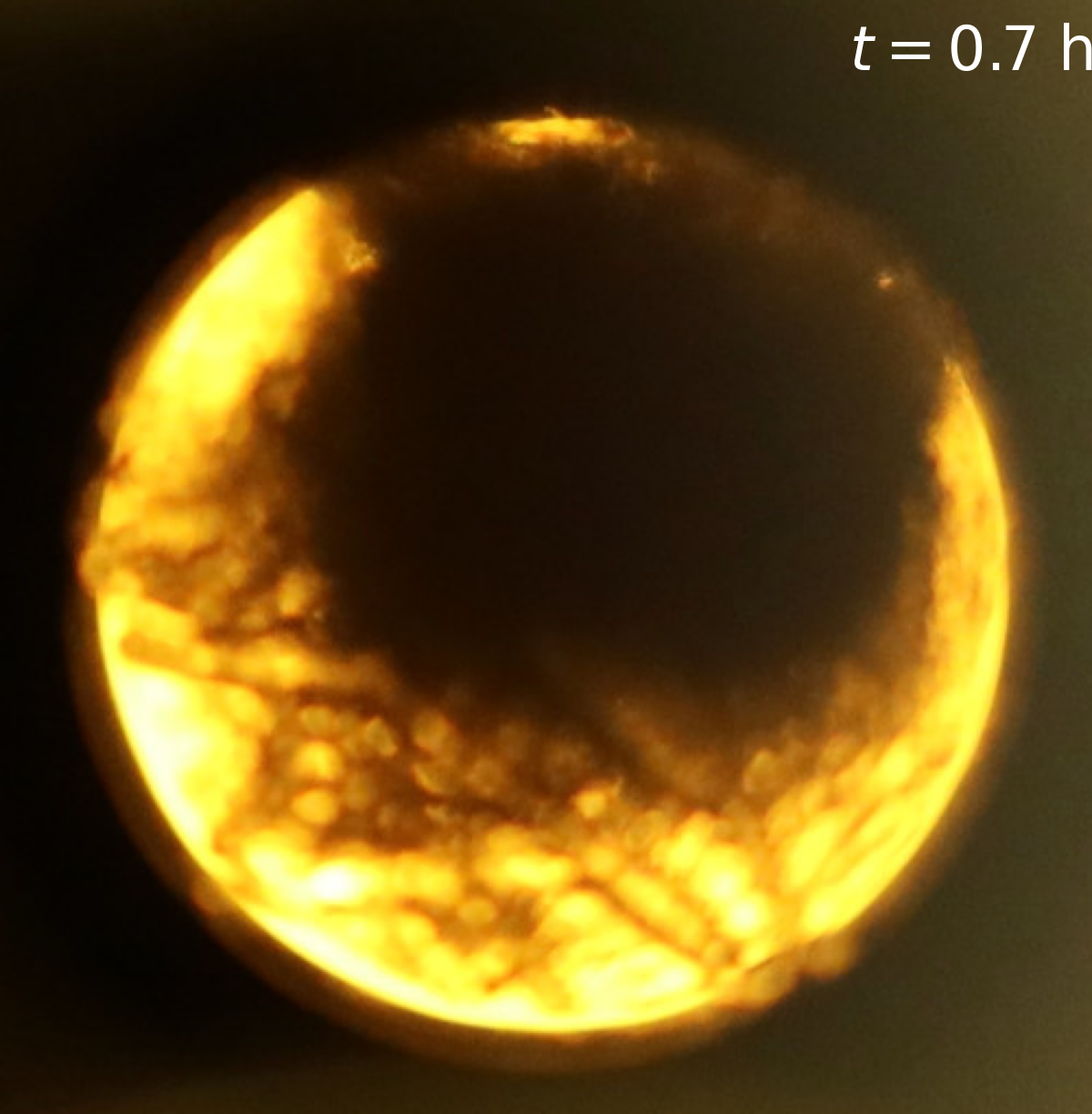}
\end{subfigure}%
\begin{subfigure}{.16\textwidth}
    \centering
    \includegraphics[width=\textwidth]{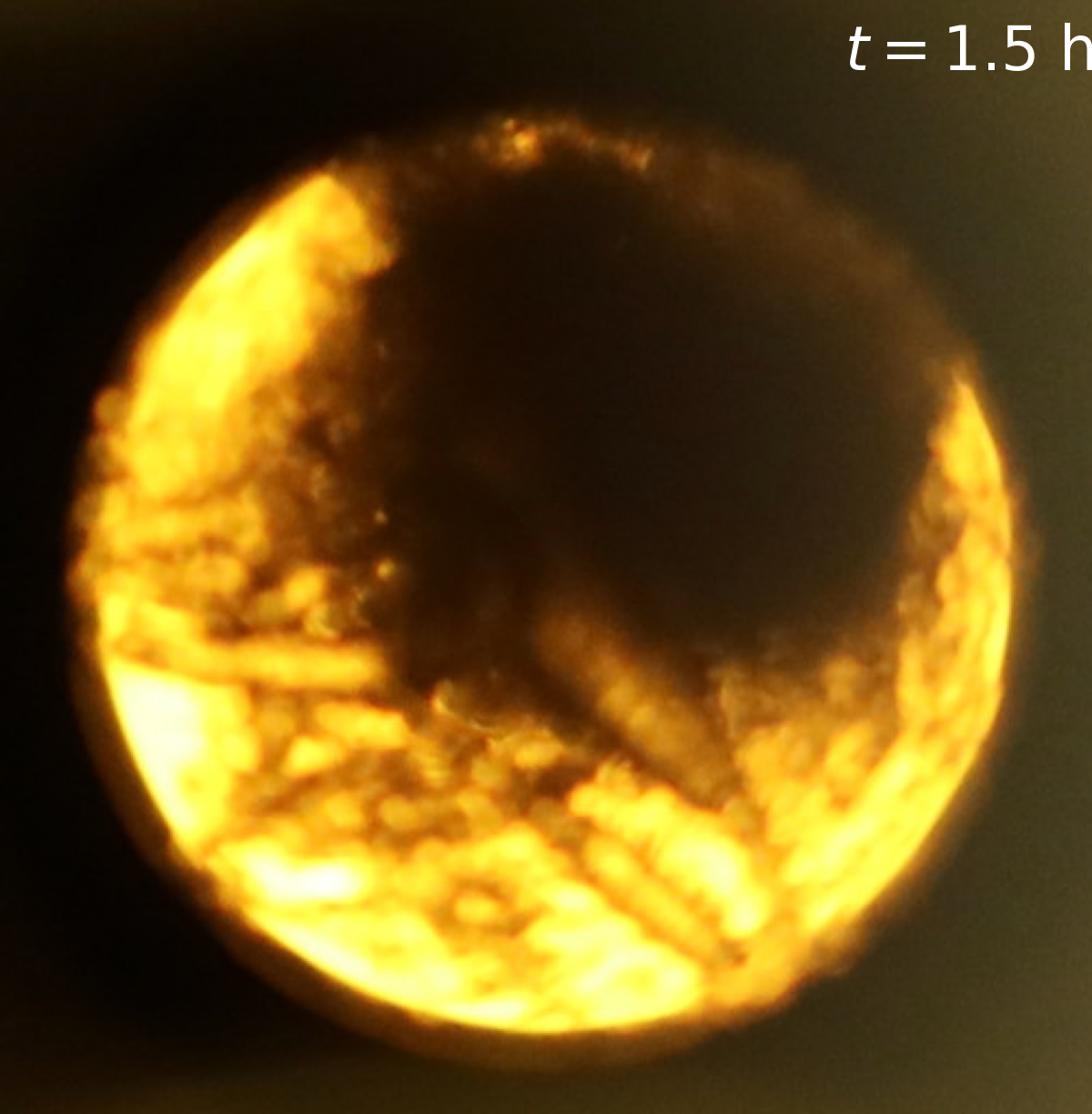}
\end{subfigure}%
\begin{subfigure}{.16\textwidth}
    \centering
    \includegraphics[width=\textwidth]{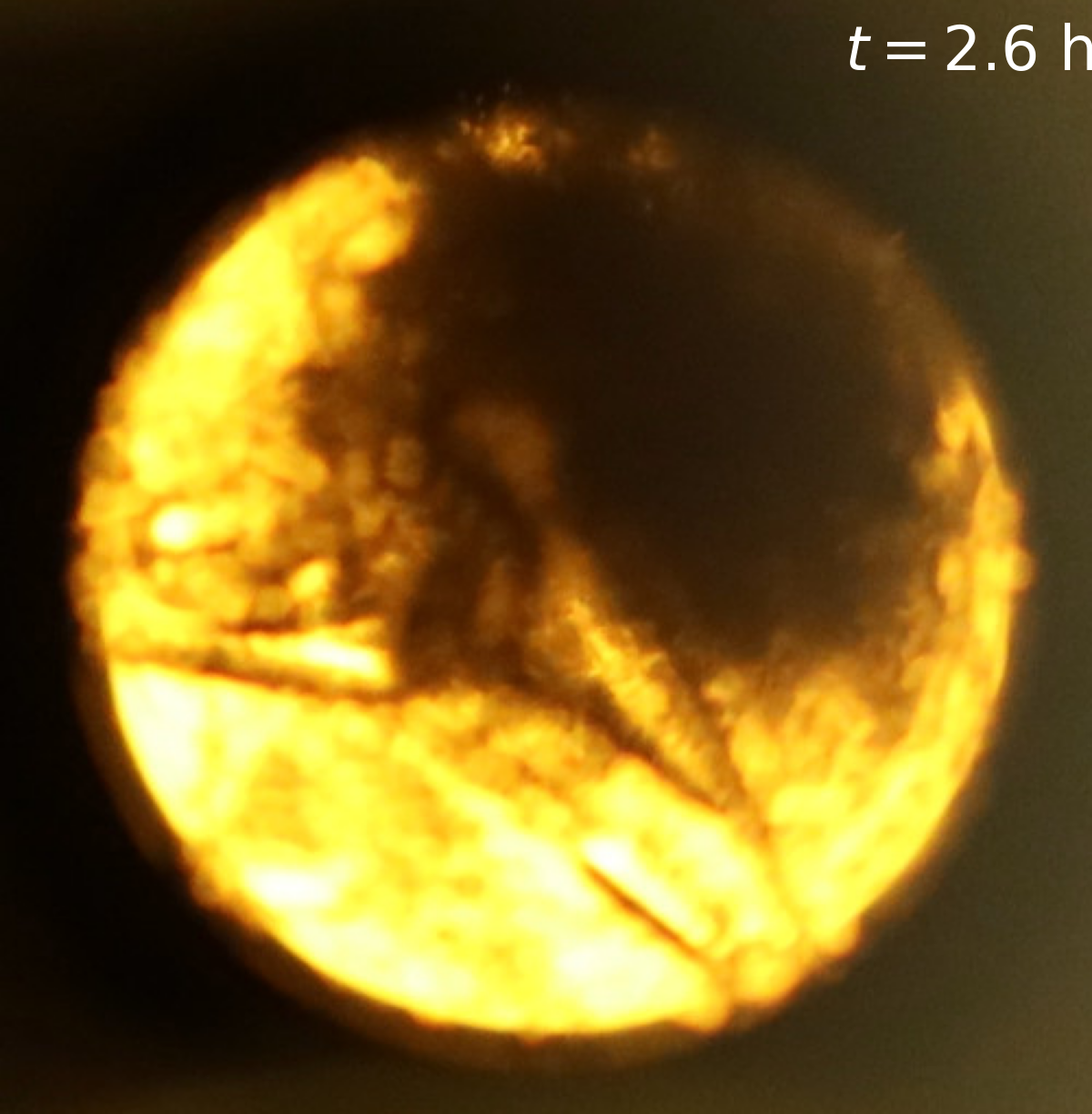}
\end{subfigure}%
\begin{subfigure}{.16\textwidth}
    \centering
    \includegraphics[width=\textwidth]{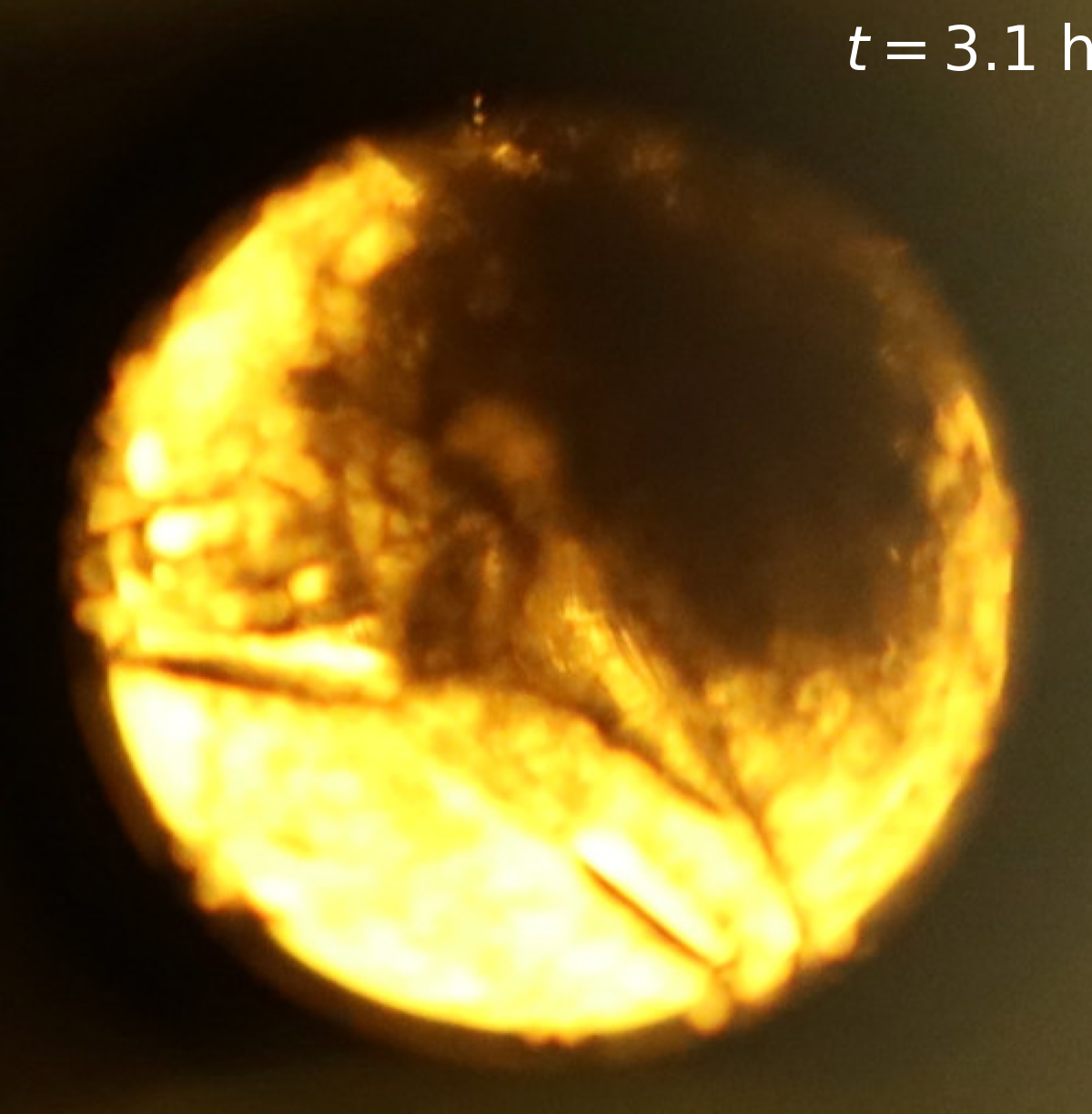}
\end{subfigure}%

\begin{subfigure}{.16\textwidth}
    \centering
    \includegraphics[width=\textwidth]{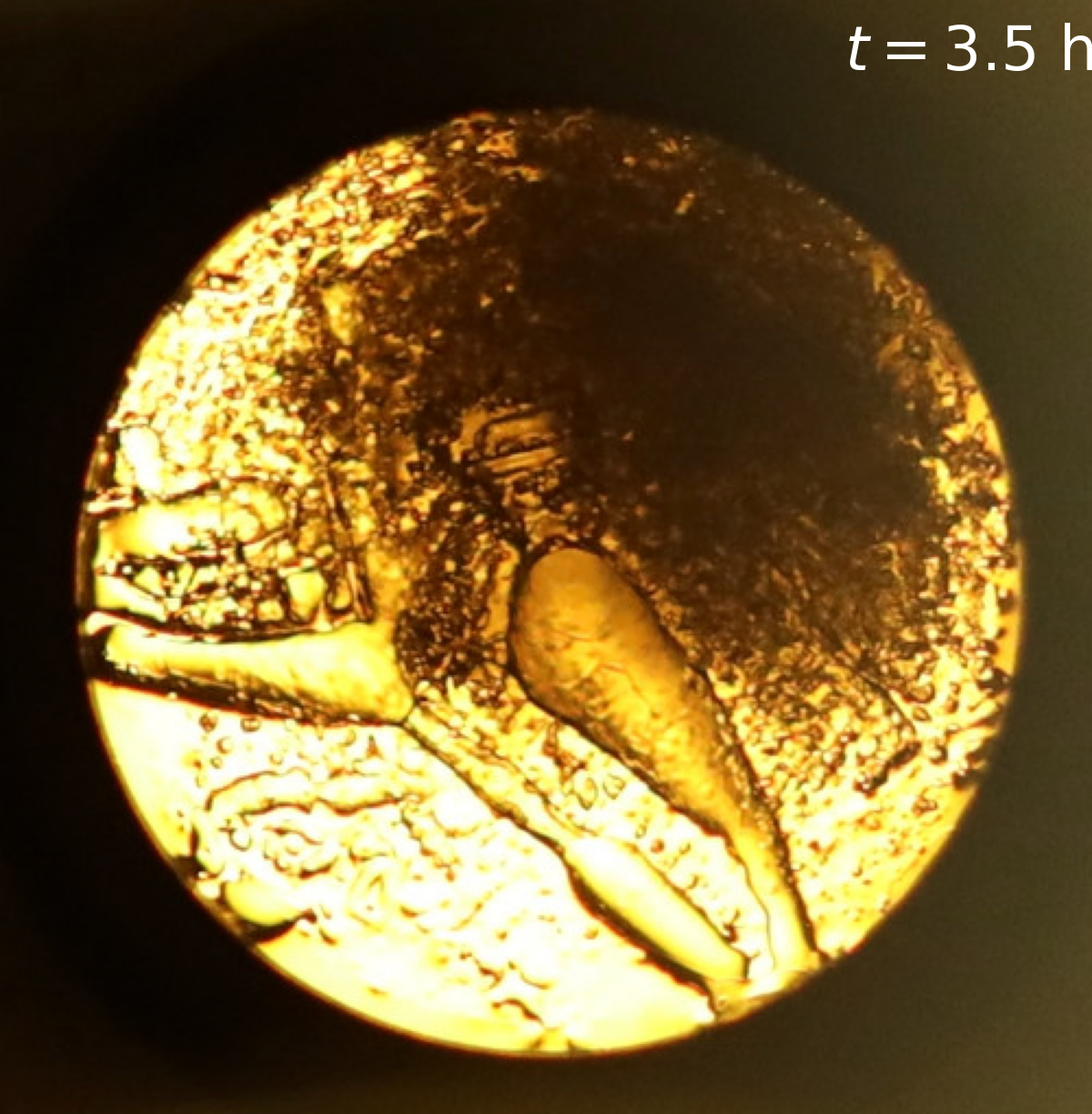}
\end{subfigure}%
\begin{subfigure}{.16\textwidth}
    \centering
    \includegraphics[width=\textwidth]{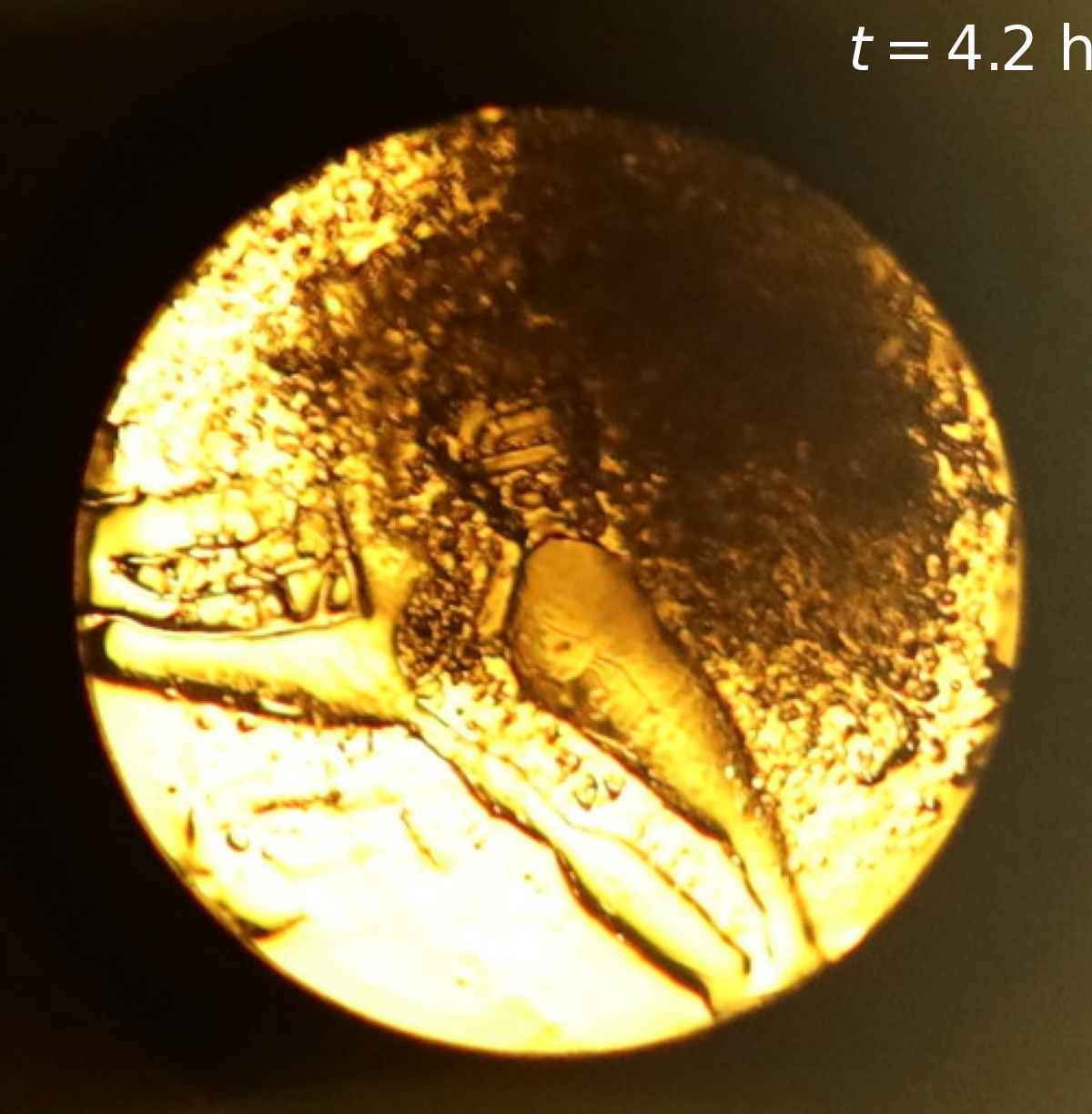}
\end{subfigure}%
\begin{subfigure}{.16\textwidth}
    \centering
    \includegraphics[width=\textwidth]{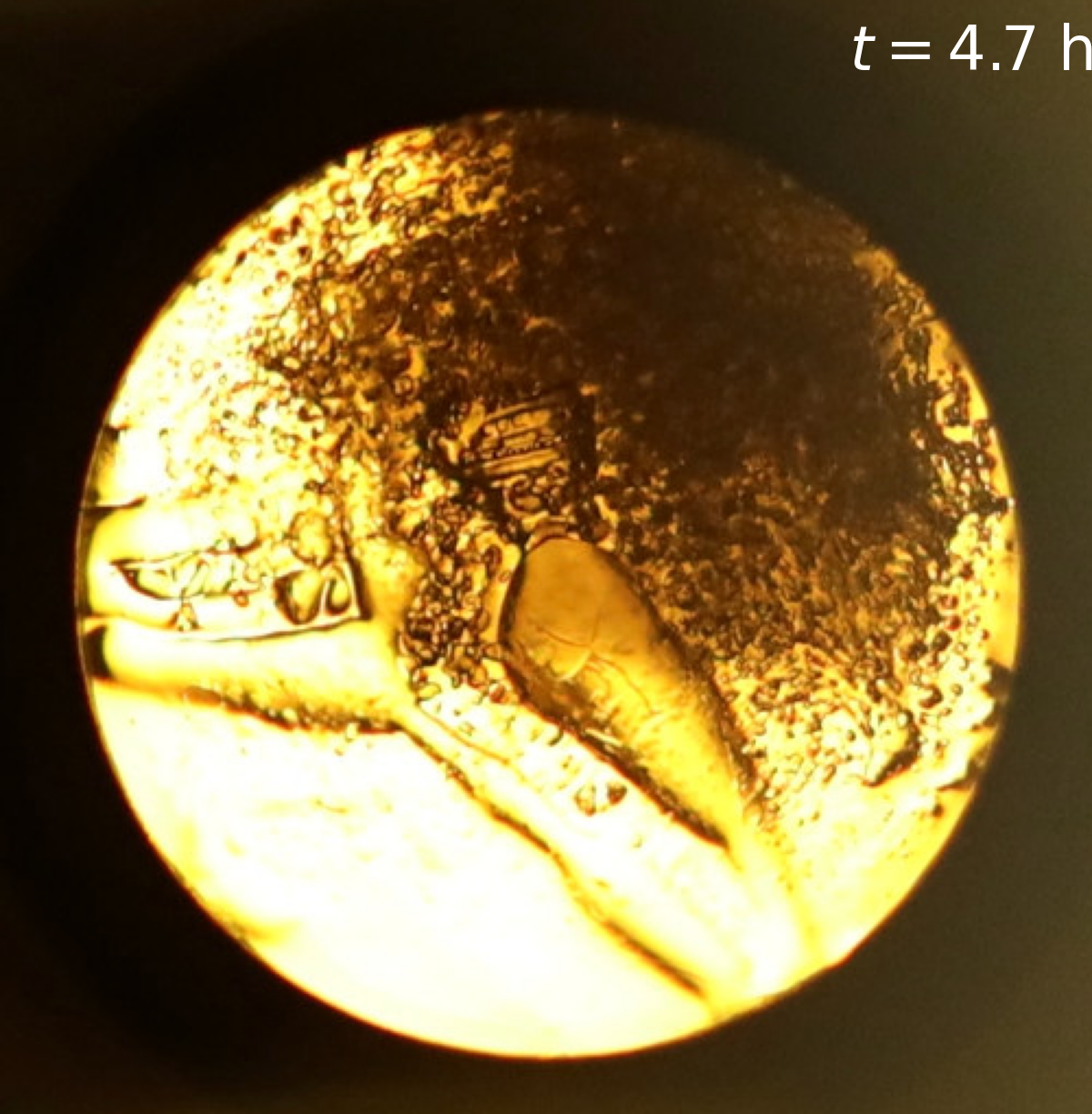}
\end{subfigure}%
\begin{subfigure}{.16\textwidth}
    \centering
    \includegraphics[width=\textwidth]{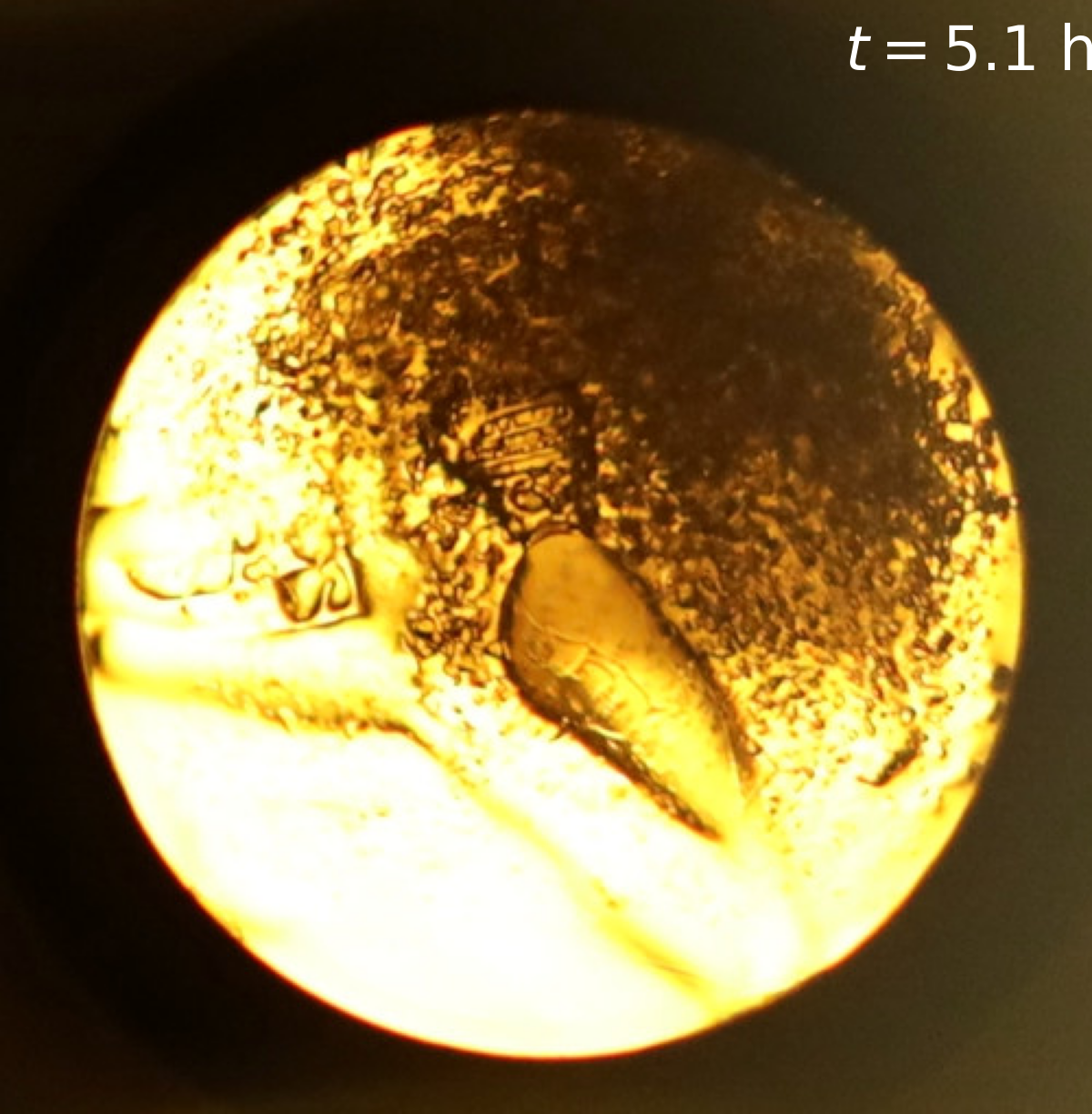}
\end{subfigure}%
\begin{subfigure}{.16\textwidth}
    \centering
    \includegraphics[width=\textwidth]{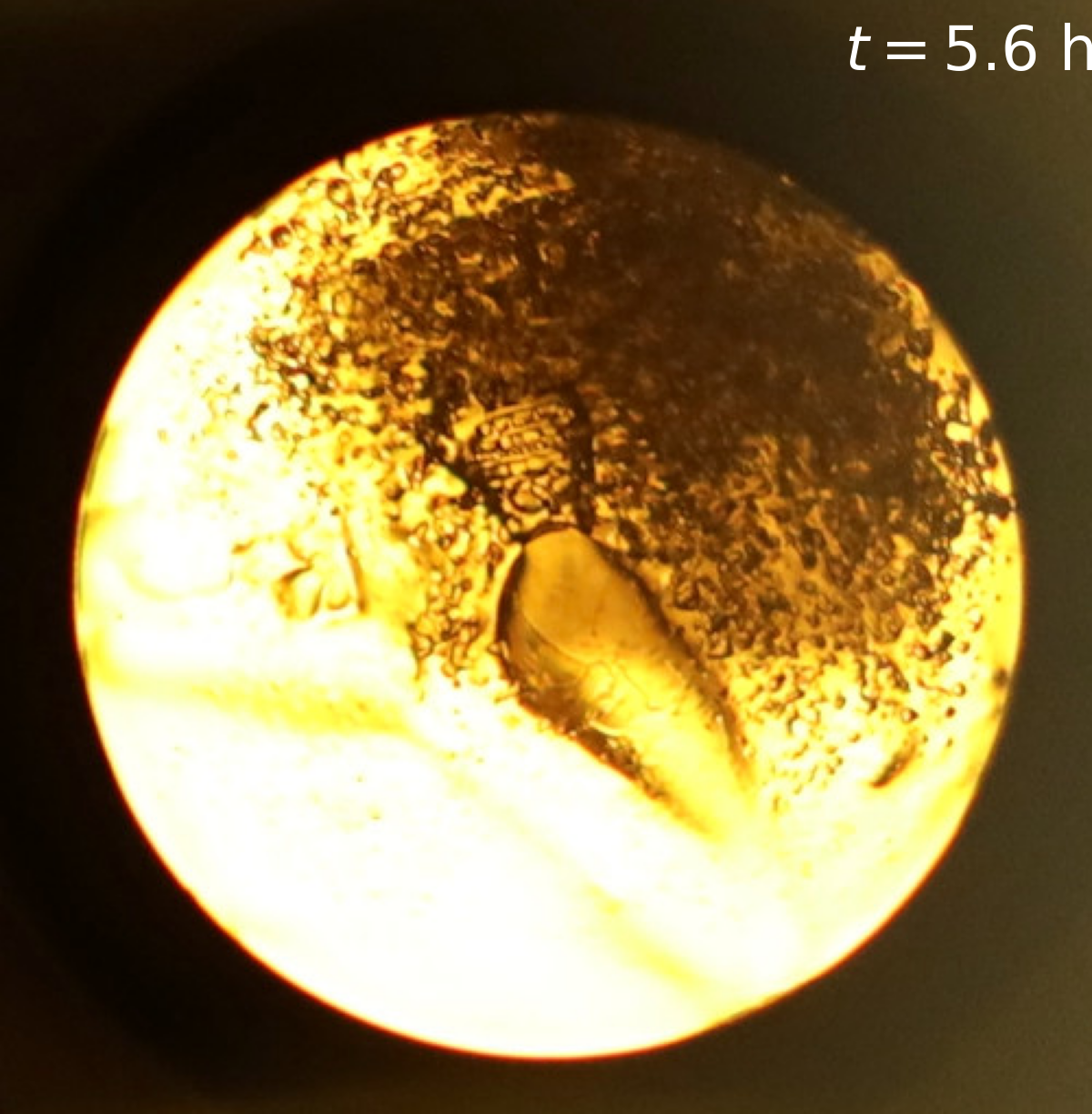}
\end{subfigure}%
\begin{subfigure}{.16\textwidth}
    \centering
    \includegraphics[width=\textwidth]{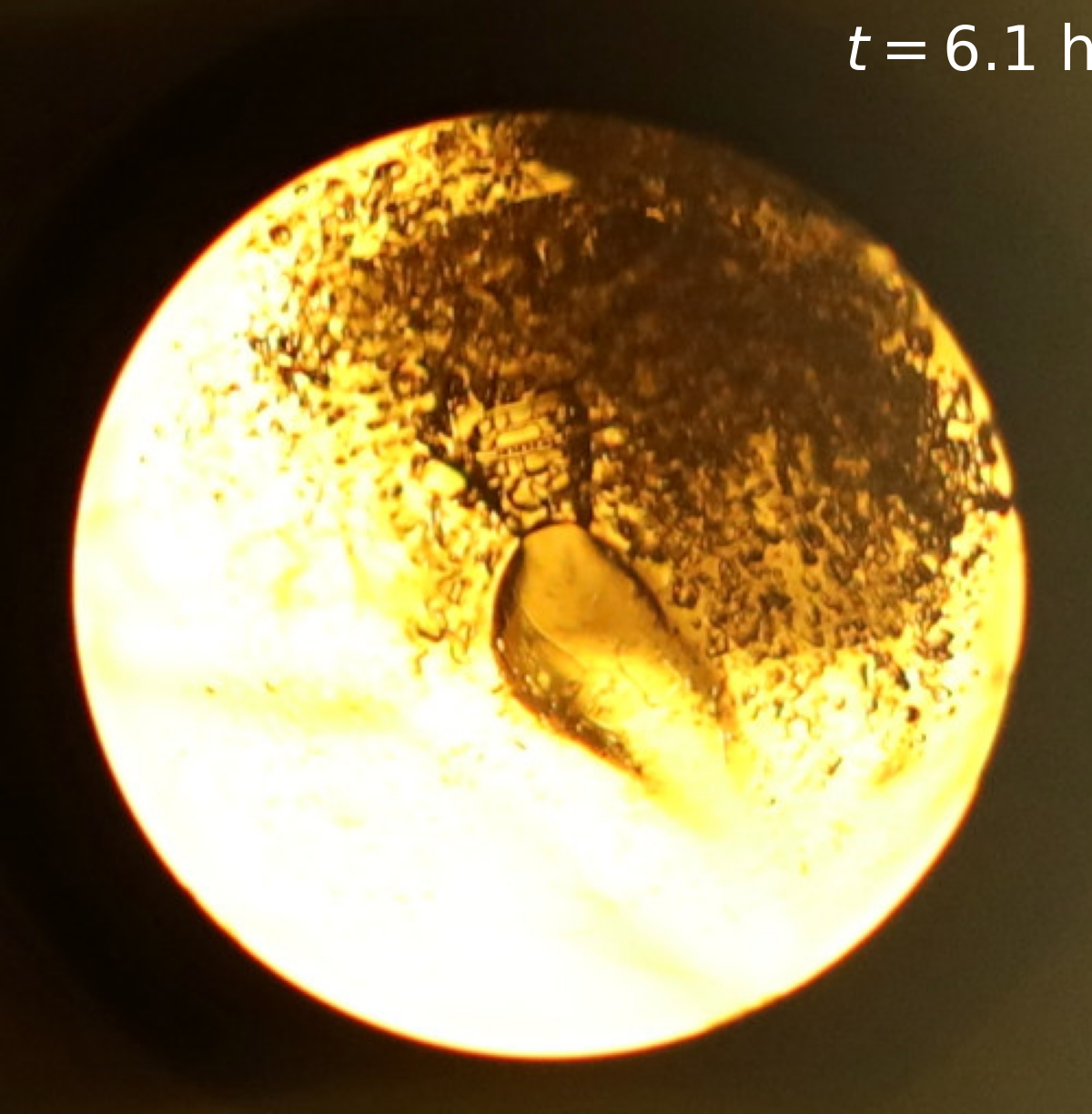}
\end{subfigure}%

\begin{subfigure}{.16\textwidth}
    \centering
    \includegraphics[width=\textwidth]{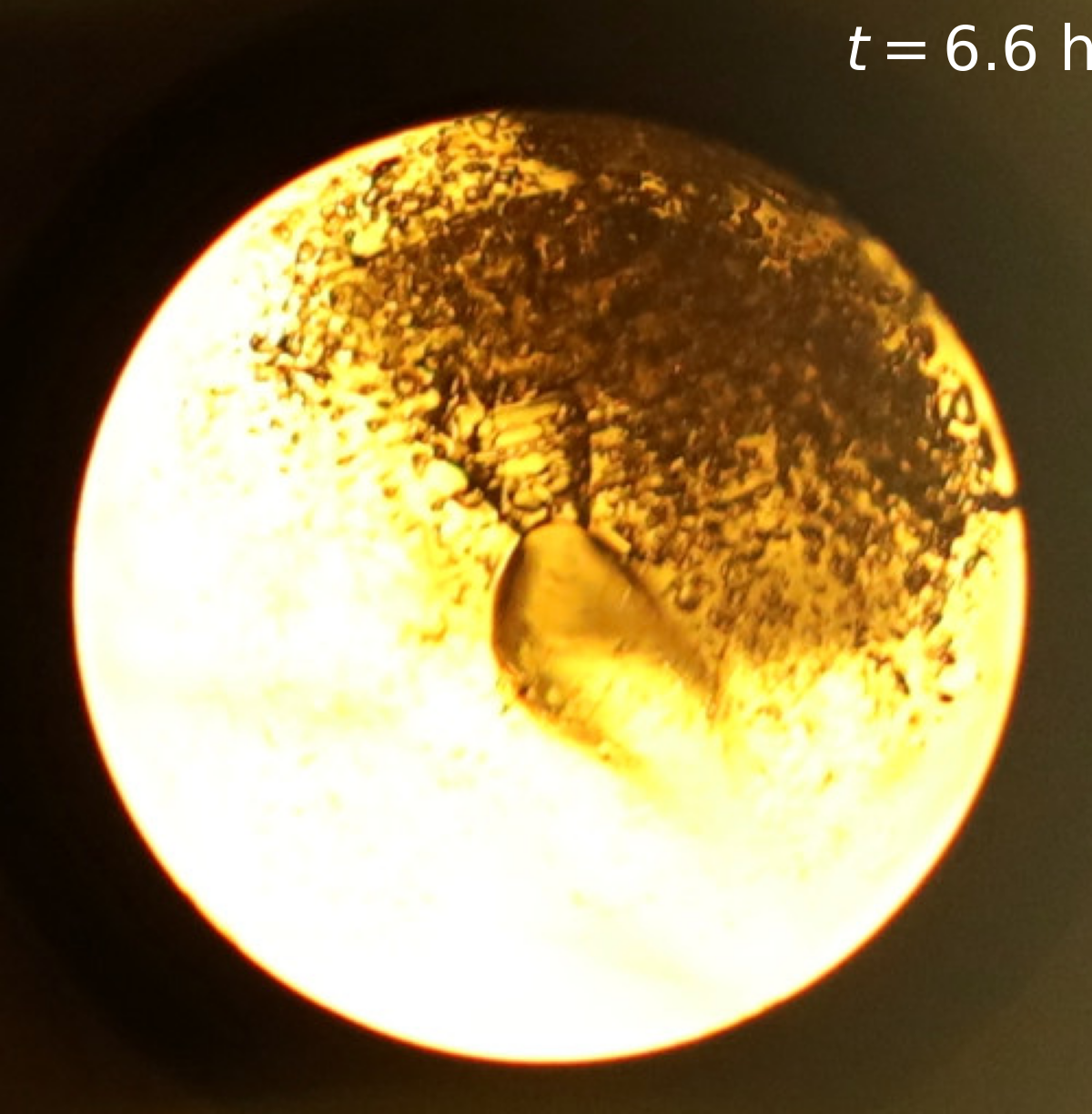}
\end{subfigure}%
\begin{subfigure}{.16\textwidth}
    \centering
    \includegraphics[width=\textwidth]{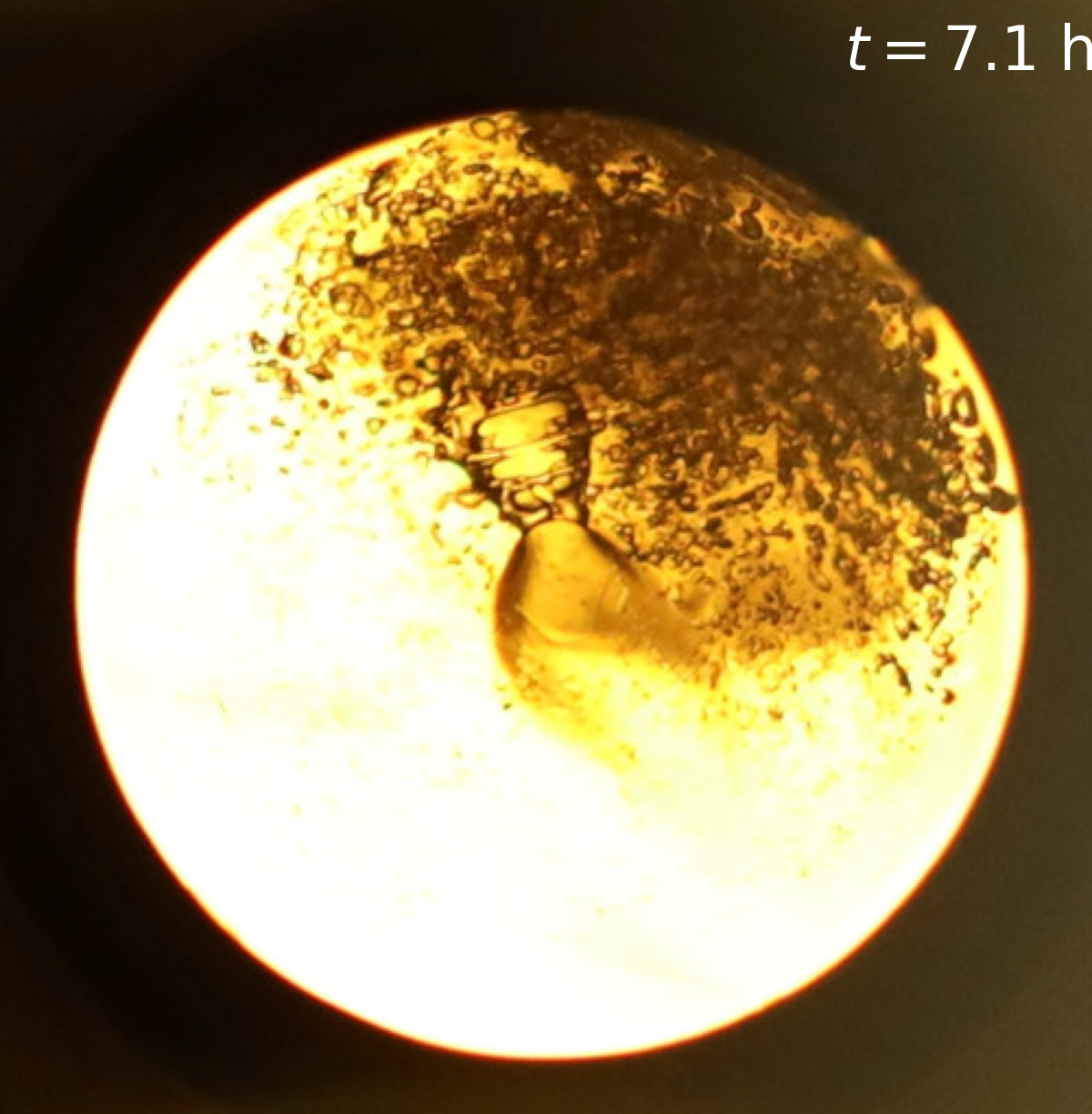}
\end{subfigure}%
\begin{subfigure}{.16\textwidth}
    \centering
    \includegraphics[width=\textwidth]{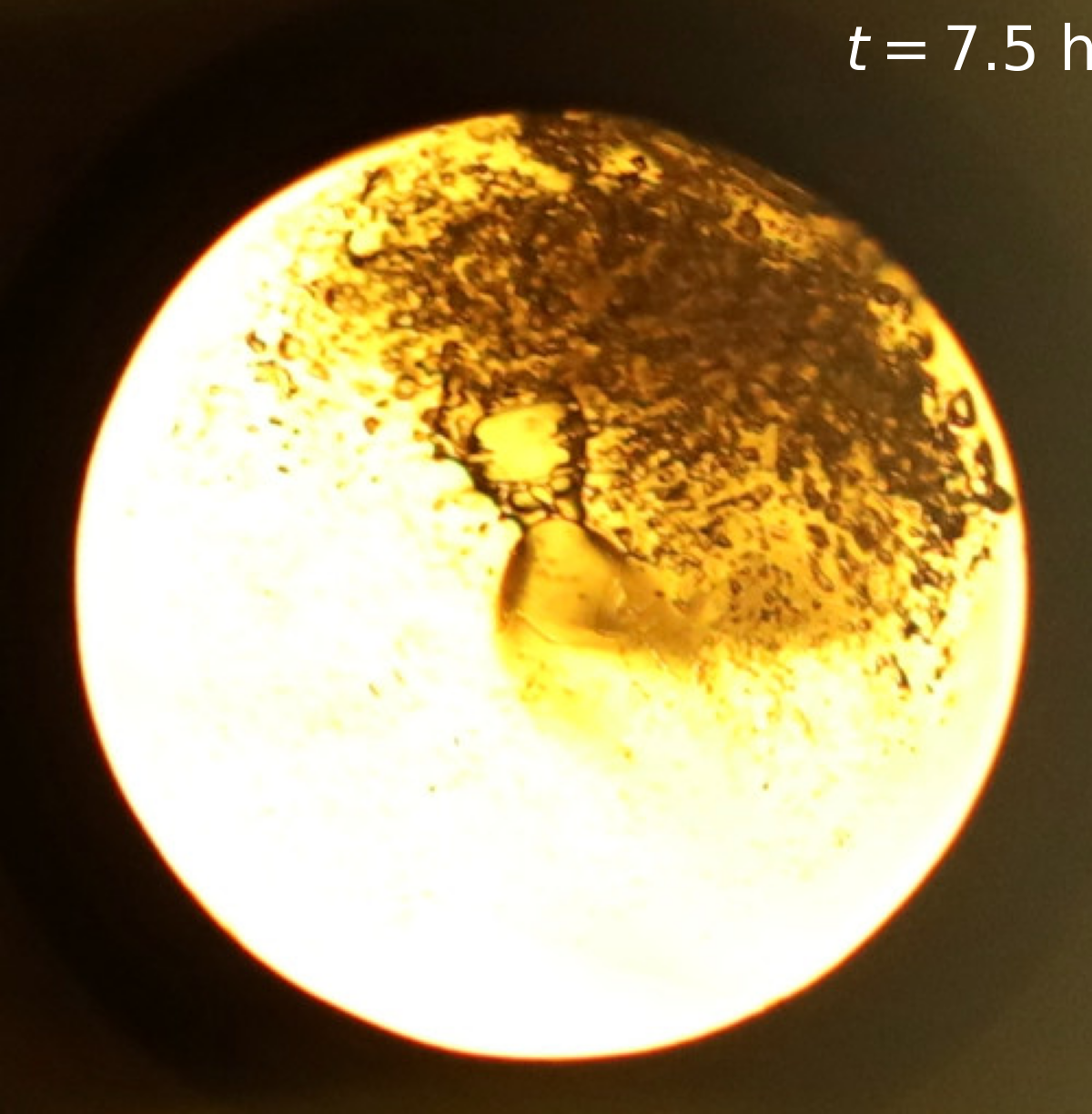}
\end{subfigure}%
\begin{subfigure}{.16\textwidth}
    \centering
    \includegraphics[width=\textwidth]{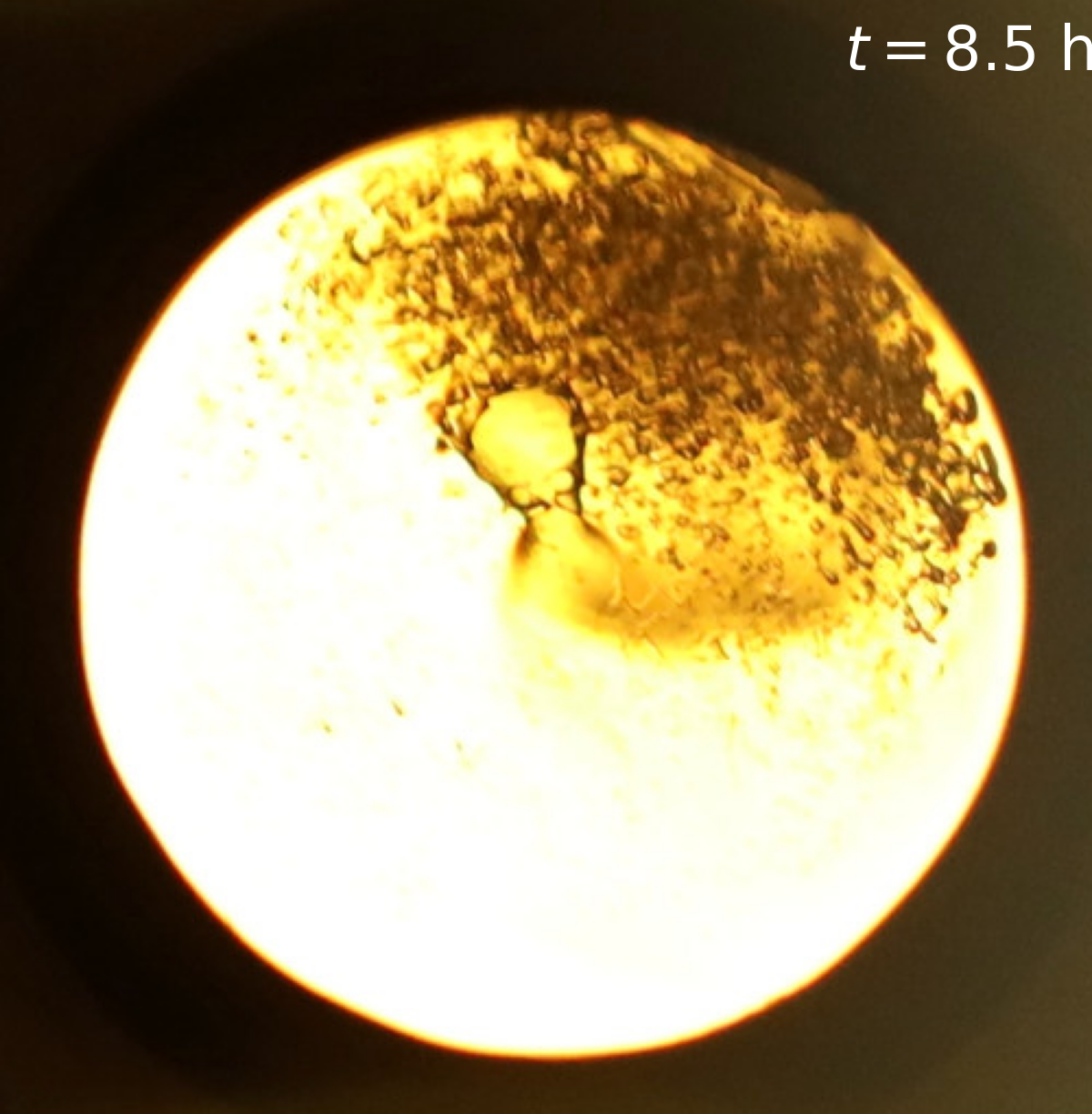}
\end{subfigure}%
\begin{subfigure}{.16\textwidth}
    \centering
    \includegraphics[width=\textwidth]{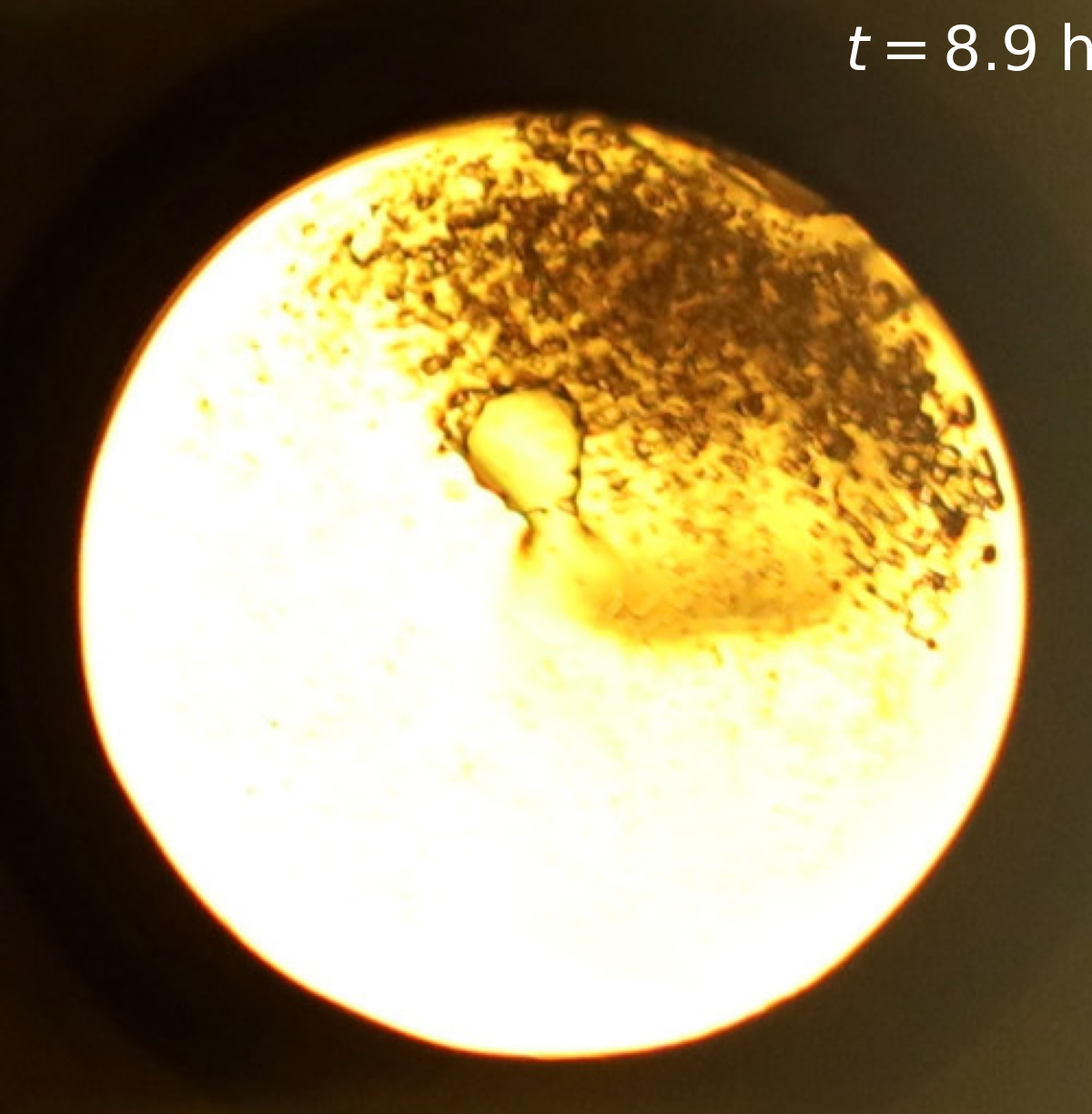}
\end{subfigure}%
\begin{subfigure}{.16\textwidth}
    \centering
    \includegraphics[width=\textwidth]{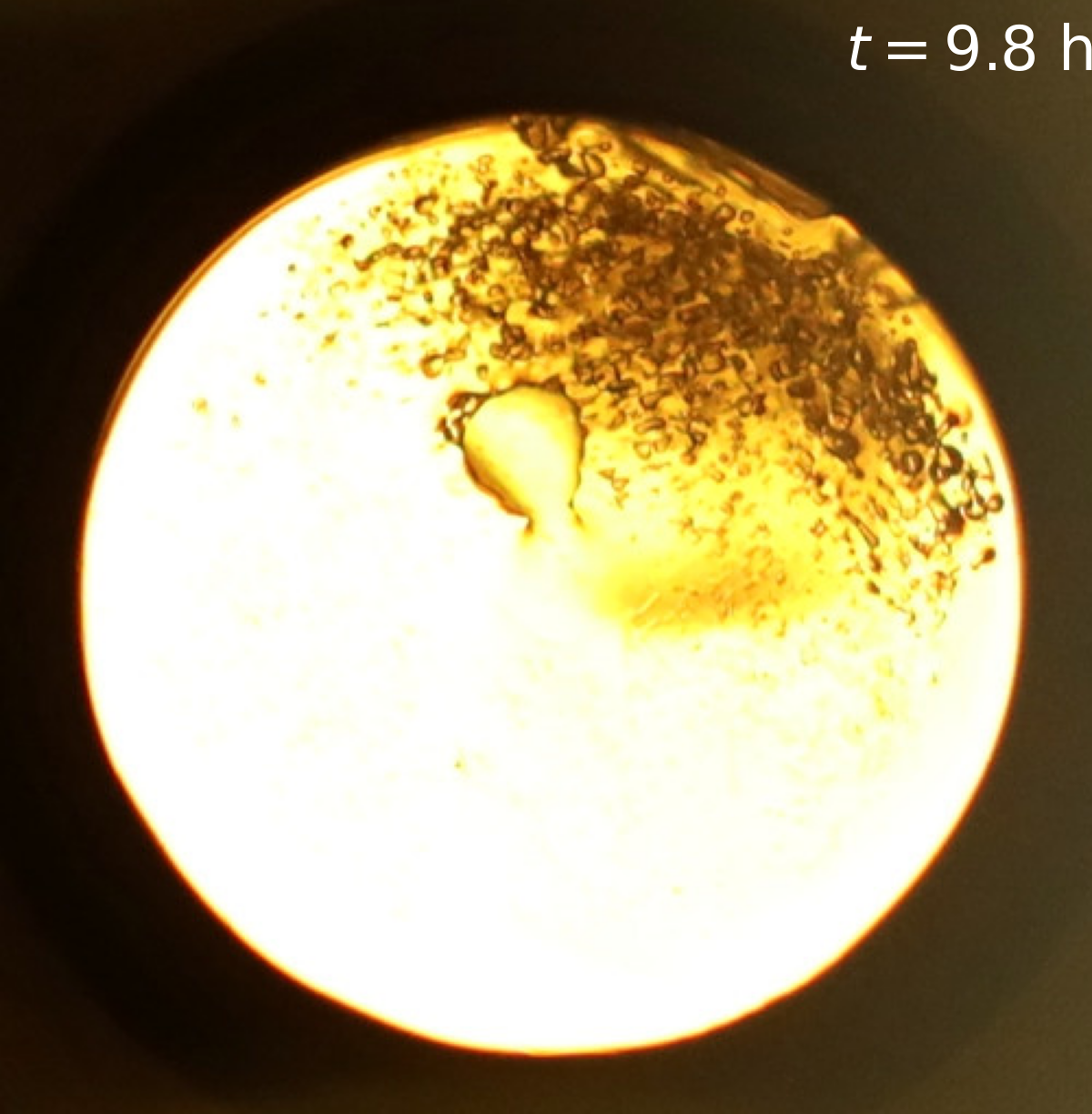}
\end{subfigure}%
\caption{
Freezing and annealing of solid \HTwo{}.
Images are sequentially from left to right, then top to bottom.
The focus was readjusted between the second and third row.
The entire sequence of images depicts a time span of \SI{\approx1}{\day}.
}
\label{fig:annealing}
\end{figure}

The processes of fast crystallization and crystal growth by desublimation leave behind an opaque solid.
In both cases, a subsequent annealing of the opaque crystal is observed resulting in a fully transparent solid.
Usually, the clearing starts at the bottom of the crystal and goes upwards as shown in \cref{fig:annealing}. 
The time-scale of this process for samples in the \gls{tapir2} measurement cell is on the order of \SI{1}{\day}. 
The annealing process thereby does not only influence the transparency of the crystal but also its shape. 
When the opaque crystal does not fill out the cell completely the annealing leaves behind a drop-shaped clear crystal. 
For a homogeneous but opaque cell filling crystal this shape transition is not possible. 
As the crystal gets transparent only during the annealing process the effects on the surface structure are not yet well observed. 

The parameters which influence the annealing process in the \gls{tapir2} measurement cell require more investigation. 
Presumably, the temperature in the cell, the thermal radiation through the cell windows and the inflowing gas if the cell is still connected to the buffer vessel of the setup have an effect. 
Also internal processes like the natural \orthopara{} conversion in \HTwo{} and \DTwo{} might contribute to the annealing process.

\subsection{Polariscopy to study stresses in \QTwo{} crystals}

\begin{figure}[t]
\centering
\begin{subfigure}{.5\textwidth}
    \centering
    \includegraphics[width=0.95\textwidth]{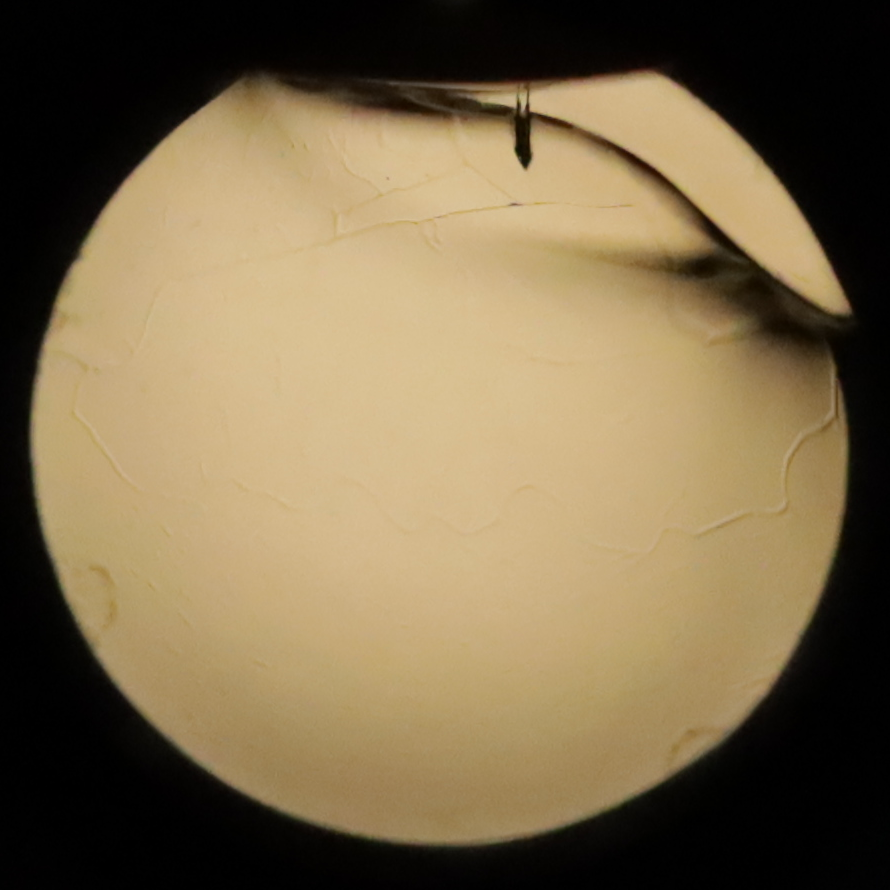}
    \caption{LED illumination}
    \label{fig:led_backlight}
\end{subfigure}%
\begin{subfigure}{.5\textwidth}
    \centering
    \includegraphics[width=0.95\textwidth]{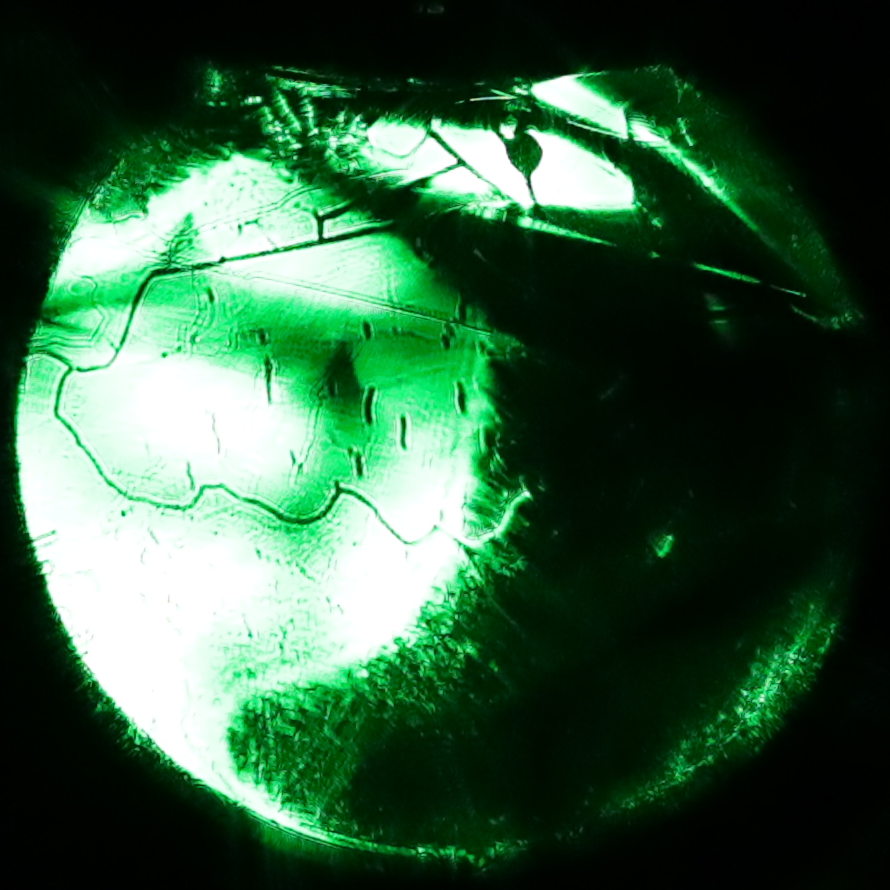}
    \caption{Collimated \SI{532}{\nano\meter} laser illumination}
    \label{fig:laser_backlight}
\end{subfigure}%
\caption{
Solid \HTwo{} with different back-illumination.
The phase boundary at the top right as well as cracks in the solid slightly refract light, appearing as dark lines.
}
\label{fig:polariscopy_reference}
\end{figure}

Using back-illumination, solid \QTwo{} can be imaged as shown in \cref{fig:polariscopy_reference}.
The liquid and solid phase can only be distinguished by the presence of a dark region across the phase boundary, as well as tiny imperfections in the solid.
With diffuse, white back-illumination from an LED, these imperfections consisting mainly of cracks, are only faintly visible.
They can be imaged much more clearly when using collimated laser light as illumination, as the stronger directionality of the light causes refractions of the light to cause a much stringer contrast.
This can be seen by comparing \cref{fig:led_backlight} and \cref{fig:laser_backlight}, which show the same solid \HTwo{} and identifying the pronounced black lines in the laser illuminated image with much fainter lines in the white light image.
However, both of these methods provide only insight into large scale defects of the crystal. 

\begin{figure}[t]
\centering
\begin{subfigure}{.33\textwidth}
    \centering
    \includegraphics[width=0.95\textwidth]{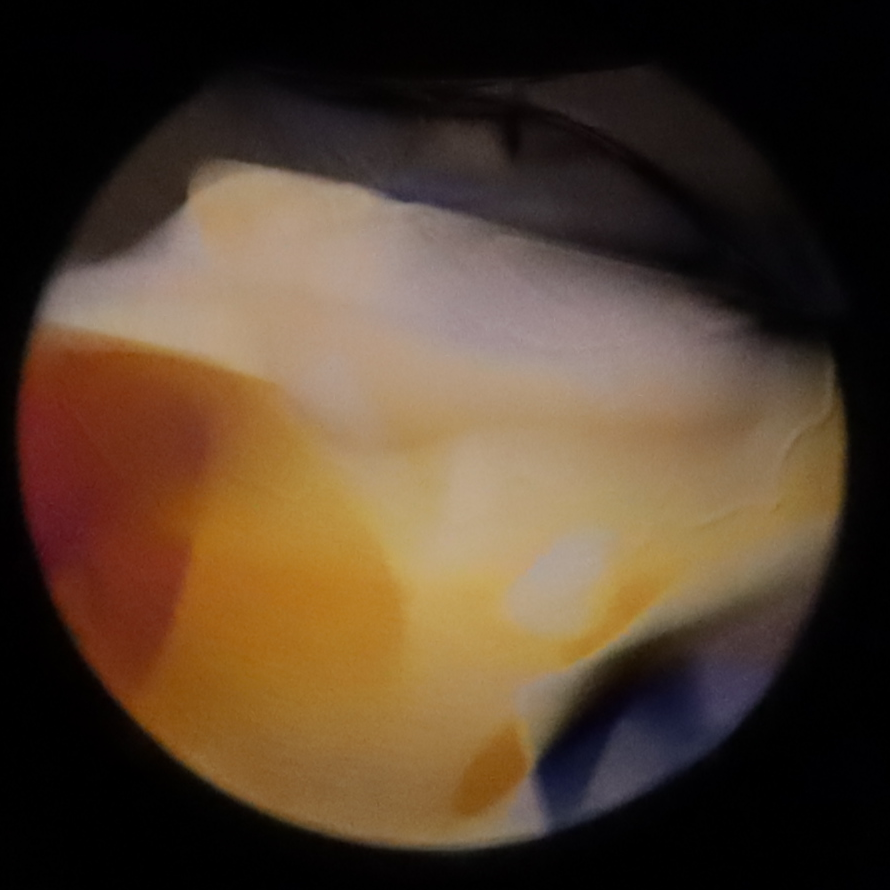}
    \caption{LED illumination\\ Polarization filter setting 1}
    \label{fig:polariscopy_with_colors1}
\end{subfigure}%
\begin{subfigure}{.33\textwidth}
    \centering
    \includegraphics[width=0.95\textwidth]{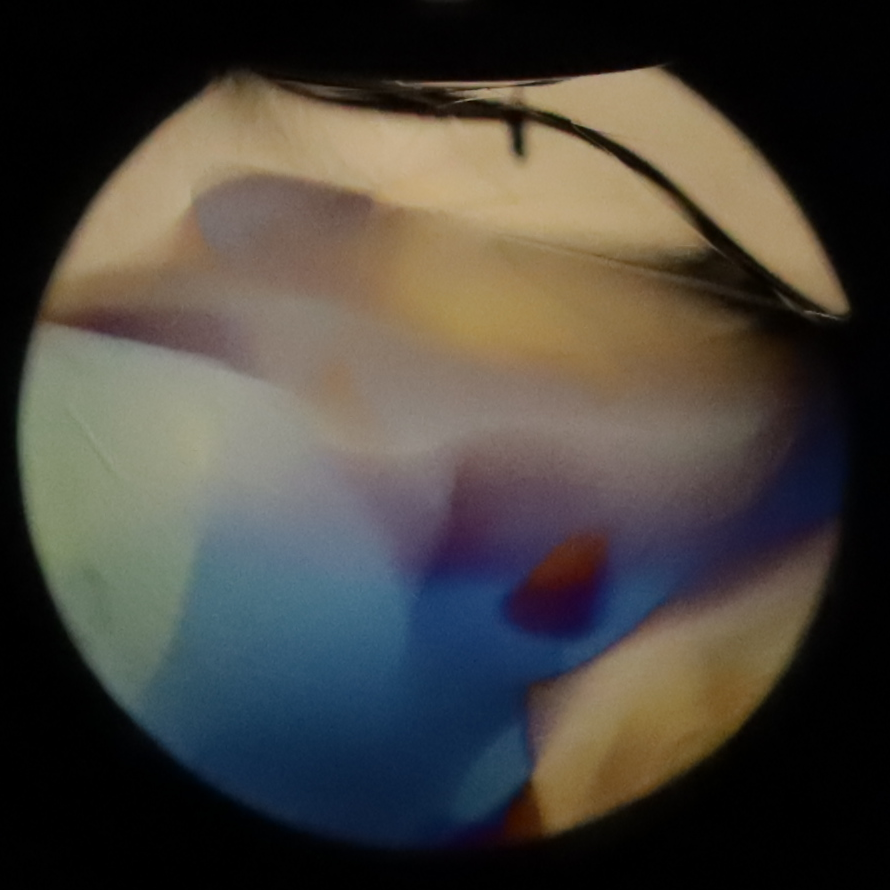}
    \caption{LED illumination\\ Polarization filter setting 2}
    \label{fig:polariscopy_with_colors2}
\end{subfigure}%
\begin{subfigure}{.33\textwidth}
    \centering
    \includegraphics[width=0.95\textwidth]{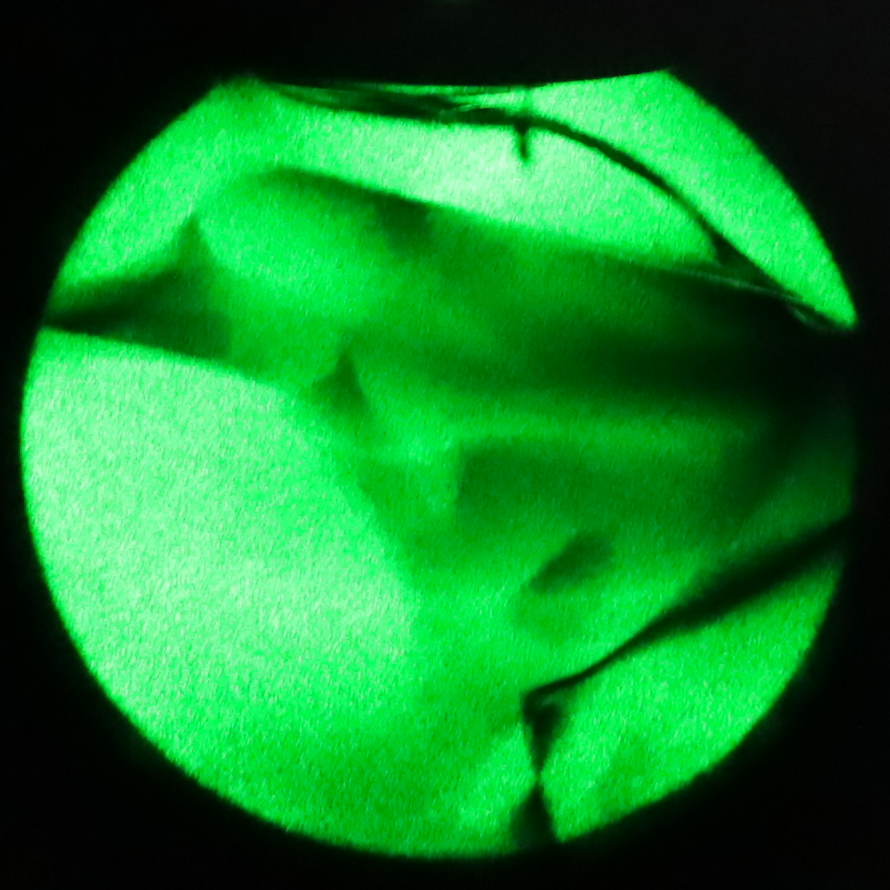}
    \caption{Diffuse laser illumination\\ Polarization filter setting 2}
    \label{fig:polariscopy_with_laser}
\end{subfigure}%
\caption{
The same solid \HTwo{} as shown in \cref{fig:polariscopy_reference} but polarization filters were inserted before and after the measurement cell.
}
\label{fig:polariscopy}
\end{figure}

The inclusion of linear polarization filters in front of and behind the cryogenic measurement cell allows a deeper insight into the crystal structure of solid \QTwo{} to be obtained.
Light transmitted through solid \QTwo{} which experiences anisotropic stress becomes optically active and causes a wavelength dependent rotation of the polarization plane via circular birefringence \cite{Zholonko2011}. 
For white light illumination, this rotation of the polarization plane causes areas of different stress to appear a different color, which can be seen in \cref{fig:polariscopy_with_colors1} and \cref{fig:polariscopy_with_colors2}, where the same \HTwo{} crystal as in \cref{fig:polariscopy_reference} is shown with two different relative orientations of the polarizers around the cell.
Distinct regions with sharp borders can be identified, strongly suggesting a polycrystalline structure of the solid.
For maximal contrast, monochromatic laser back-illumination can be used, in our case with a \SI{532}{\nano\meter} diode pumped solid state laser, which is shown in \cref{fig:polariscopy_with_laser}, allowing an easier assignment of different zones in the crystal compared to the color images.

While some features are visible in both the simple back-illuminated and polariscopy images, the latter show internal structures of the crystal which could not be seen previously. 
A further feature of measurements using polarized light is the possibility to exclude the liquid phase to a large degree by setting the polarizers to be perpendicular, blocking all light which hasn't experienced circular birefringence in the solid.
This can be seen in the upper right corner of \cref{fig:polariscopy_with_colors1}, where the light passing through the liquid \HTwo{} and adjacent solid \HTwo{} is strongly suppressed. 

\subsection{Infrared spectroscopy of isomer change and molecular interactions}

Due to the symmetry of the hydrogen molecule, its intrinsic dipole moment, as well as higher order moments, are negligible on the scale of the \gls{tapir2} setup. 
For  IR spectroscopy in the gaseous phase much longer absorption paths are required. 
However, with the high densities in the liquid and solid phase, intermolecular interactions can induce transition dipole moments, enabling IR spectroscopy even with short absorption path lengths \cite{McKellar1984,McKellar1988}. 
With regard to the interactions mentioned, three main contributions were identified. 
They are: molecule collisions, formation of Van-Der-Waals dimers (London dispersion interactions between polarizable molecules), and phonons due to the long range interactions in the liquid and solid phase. 
Based on these interactions a compact set of descriptors for the IR spectra is given in \cite{Groessle2020}. 
Since the IR spectra are driven by intermolecular interactions the changes of intensity with density and composition (including \orthopara{}-ratios) are highly non-linear.
Therefore, to use IR spectroscopy for quantitative analysis, an extensive calibration of the spectra is required \cite{Groessle2017,Mirz2020}. 

\begin{figure}[t]
\centering
\includegraphics[width=0.95\textwidth]{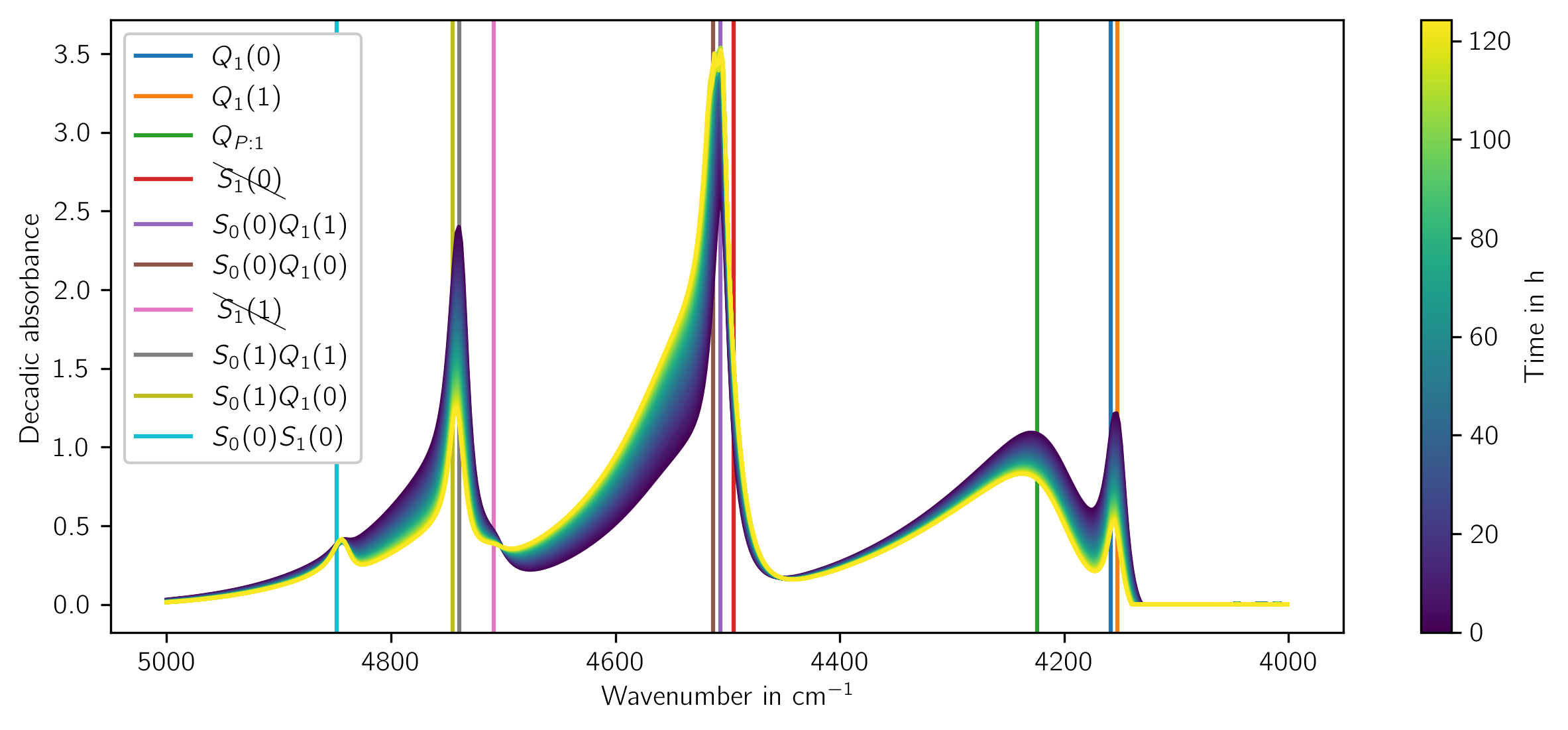}
\caption{
IR absorption spectra of liquid \HTwo{}. 
Shown is the time evolution of the first vibrational band over about \SI{120}{\hour} due to the natural \orthopara{}-conversion.
Line positions are taken from \cite{Groessle2020}.
}
\label{fig:ir_ortho_para_conversion}
\end{figure}

In \cref{fig:ir_ortho_para_conversion} IR absorption spectra taken over a time span of about \SI{120}{\hour} of the same liquid \HTwo{} sample are shown.
Displayed is the first vibrational band. 
Most peaks are a superposition of multiple ro-vibrational excitations of molecules or dimers, respectively.
During the spectra acquisition, the temperature in the measurement cell is kept constant and therefore the sample undergoes no density changes or phase transitions. 
The variations in the intensities in the spectra therefore result only from the natural \orthopara{} conversion of the \HTwo{}. 
Starting from the room temperature equilibrium of \SI{75}{\percent} ortho-\HTwo{} and \SI{25}{\percent} para-\HTwo{} (first spectrum) the temperature-related \orthopara{}-conversion shifts the corresponding ratio towards the para molecule. 
From the spectra it is clear that this conversion influences all the peaks with respect to intensity, line shape, and position. 
This applies even to the phonon peak \QPhonon{1}, which seems to be ortho dominated as its intensity decreases with the decreasing ortho-\HTwo{} proportion. 
The same holds for the collision peak (\QOne{0}, \QOne{1}) and the dimer peak of the transitions \SZero{1}\QOne{1} and \SZero{1}\QOne{0}. 
On the other hand, the \SZero{0}\QOne{1} + \SZero{0}\QOne{0} peak and the \SZero{0}\SOne{0} peak are para dominated and their intensity increases with increasing amount of para-\HTwo{} in the sample.

Most of the peaks are actually the superposition of multiple ro-vibrational transitions of the molecules and dimers \cite{Gustafsson2019}. 
One peak where the mentioned non-linearity becomes directly visible is the collision induced monomer peak \QOne{0} and \QOne{1} (around $\nu=4150$). 
Here, \QOne{0} is the contribution from para-\HTwo{} and \QOne{1} from ortho-\HTwo{}.
Due to the natural \orthopara{}-conversion the corresponding ratio is shifting towards the para molecule. 
While the sum of the occupational numbers of the rotational states $J=0$ and $J=1 $is constant during the ortho-para-conversion the total intensity decreases due to differences in the molecular interaction potential.
This peak is then followed by a phonon transition around $\nu=4250$ (labeled \QPhonon{1}).
All the spectral features from $\nu=5000 cm^{-1}$ to $\nu=4400 cm^{-1}$ are then based on dimer transitions.

\subsection{Raman spectroscopy of isotopic and isomer composition}

\begin{figure}[t]
\centering
\begin{subfigure}{.5\textwidth}
    \centering
    \includegraphics[width=0.95\textwidth]{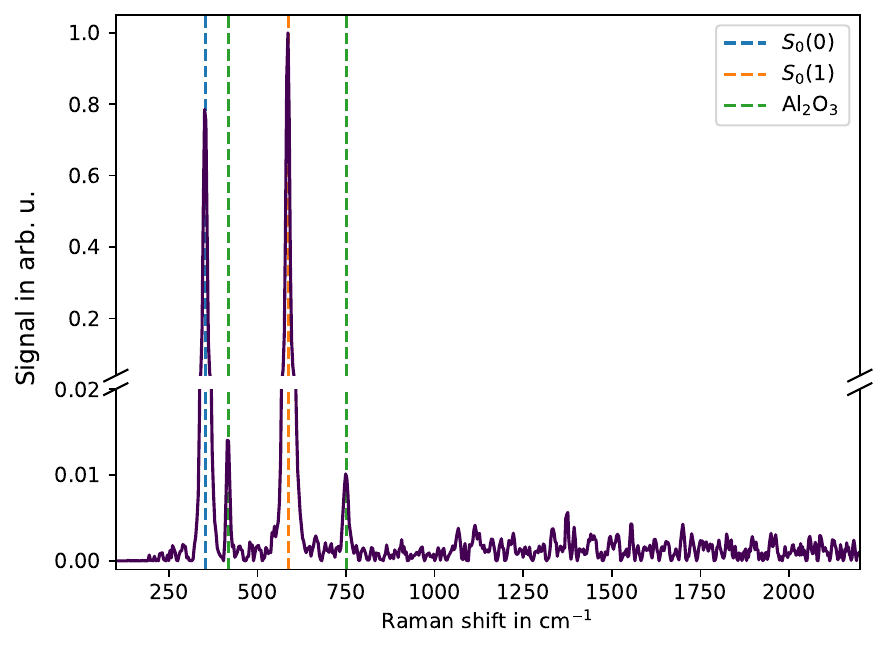}
    \caption{S$_0$-branch region}
    \label{fig:raman_s0}
\end{subfigure}%
\begin{subfigure}{.5\textwidth}
    \centering
    \includegraphics[width=0.95\textwidth]{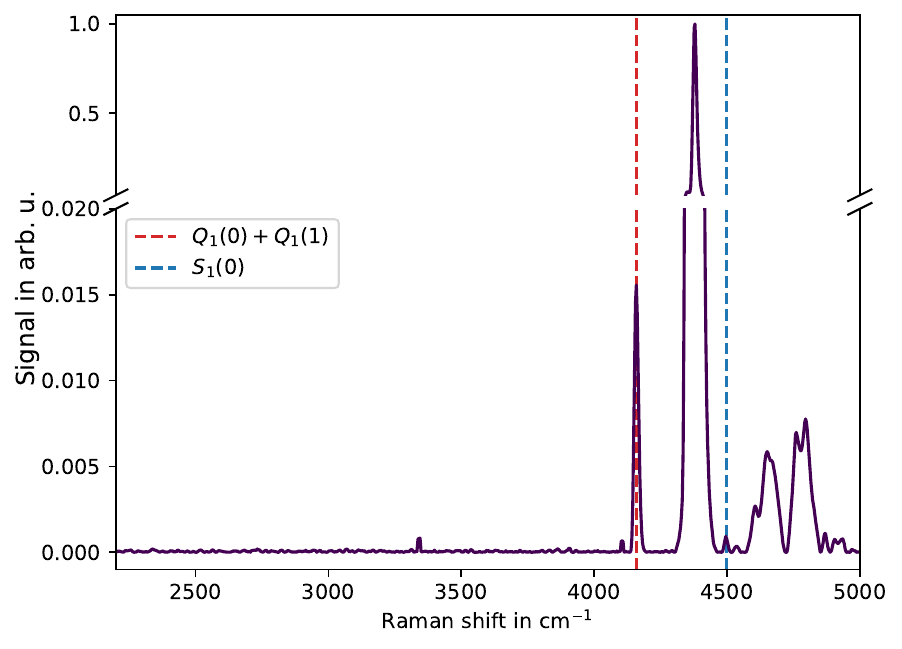}
    \caption{Q$_1$-branch region}
    \label{fig:raman_q1}
\end{subfigure}%
\caption{
Sample Raman spectra of liquid \HTwo{}.
Some background peaks are labeled for reference.
}
\label{fig:raman_reference}
\end{figure}

The Raman setup of the \gls{tapir2} experiment allows for spectroscopy of the cryogenic cell contents.
As mentioned in \cref{sec:setup}, two different spectrometers can be connected to the collection side, which allows the investigation of either the \SZeroBranch{} (rotational) or \QOneBranch{} (ro-vibrational) of the hydrogen isotopologues.
For the \SZeroBranch{}, the range of wavenumbers that needs to be covered, spans between the \SZero{0} peak of \TTwo{} at \TTwoSZeroZero{} on the lower end up to at least the \SZero{4} peak of \HTwo{} at \HTwoSZeroFour{}.
Higher rotational states $J$ are not significantly occupied at temperatures that can be reached with \gls{tapir2}.
For the \QOneBranch{}, the range of wavenumbers that needs to be covered spans between the \QOne{0} peak of \TTwo{} at \TTwoQOneZero{} on the lower end up to at least the \QOne{0} peak of \HTwo{} at \HTwoQOneZero{}.
Example spectra for both cases are shown in \cref{fig:raman_reference}.

In the \SZeroBranch{} shown in \cref{fig:raman_s0}, one can clearly see both the \SZero{0} line originating from para \HTwo{} at \HTwoSZeroZero{} and the \SZero{1} line from ortho \HTwo{} at \HTwoSZeroOne{}.
Besides these lines, some additional Raman lines at \SapphireRamanOne{} and \SapphireRamanThree{} resulting from the sapphire windows \cite{Bellafatto2025} are visible.

In the \QOneBranch{} shown in \cref{fig:raman_q1}, an overlapping line from the \HTwo{} \QOne{0} and \QOne{1} lines is visible at \SI{\approx 4160}{\per\centi\meter}.
Furthermore, the \SOne{0} line of the \SOneBranch{} at \HTwoSOneOne{} can be seen in-between a multitude of other lines.
These background lines are not Raman lines, but photoluminescence of chromium impurities in the sapphire windows, which gets excited by the incident laser light.

\begin{figure}[t]
\centering
\includegraphics[width=0.95\textwidth]{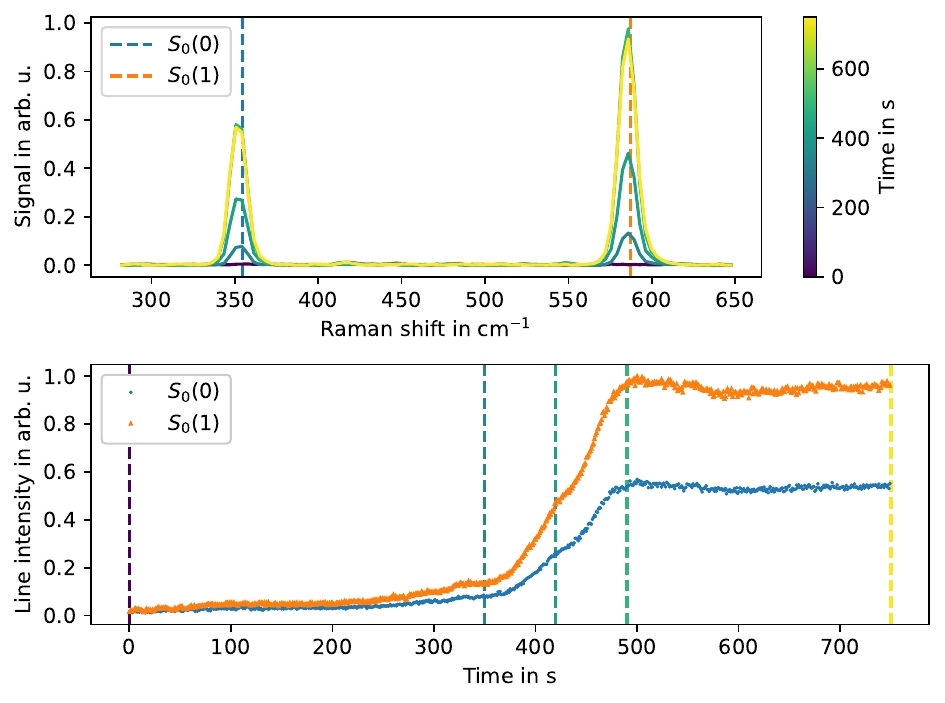}
\caption{ Raman spectra and line intensity as function of time for \HTwo{} condensing into the measurement cell.
}
\label{fig:raman_filling}
\end{figure}

Due to the much higher densities at low temperatures and especially in the liquid or solid phase, a much stronger Raman signal than in the gas phase at room temperature is observed.
As a result, measurement times can be drastically decreased, providing access to dynamic processes on the timescale of seconds.
An example of this is the condensation of liquid \HTwo{} into the measurement cell, which is shown in \cref{fig:raman_filling}.
At first, only the signal of \HTwo{} in the gas phase is visible, which increases in density as the gas cools down. 
Then condensation happens, and at around the \SI{350}{\second} mark the level of the condensed liquid rises into the laser beam path.
Up until around \SI{490}{\second} the filling level of the cell rises, until all available gas down to the saturation curve has been condensed.
In this example the filling level never covered the entire laser beam cross section.

\begin{figure}[t]
\centering
\includegraphics[width=0.95\textwidth]{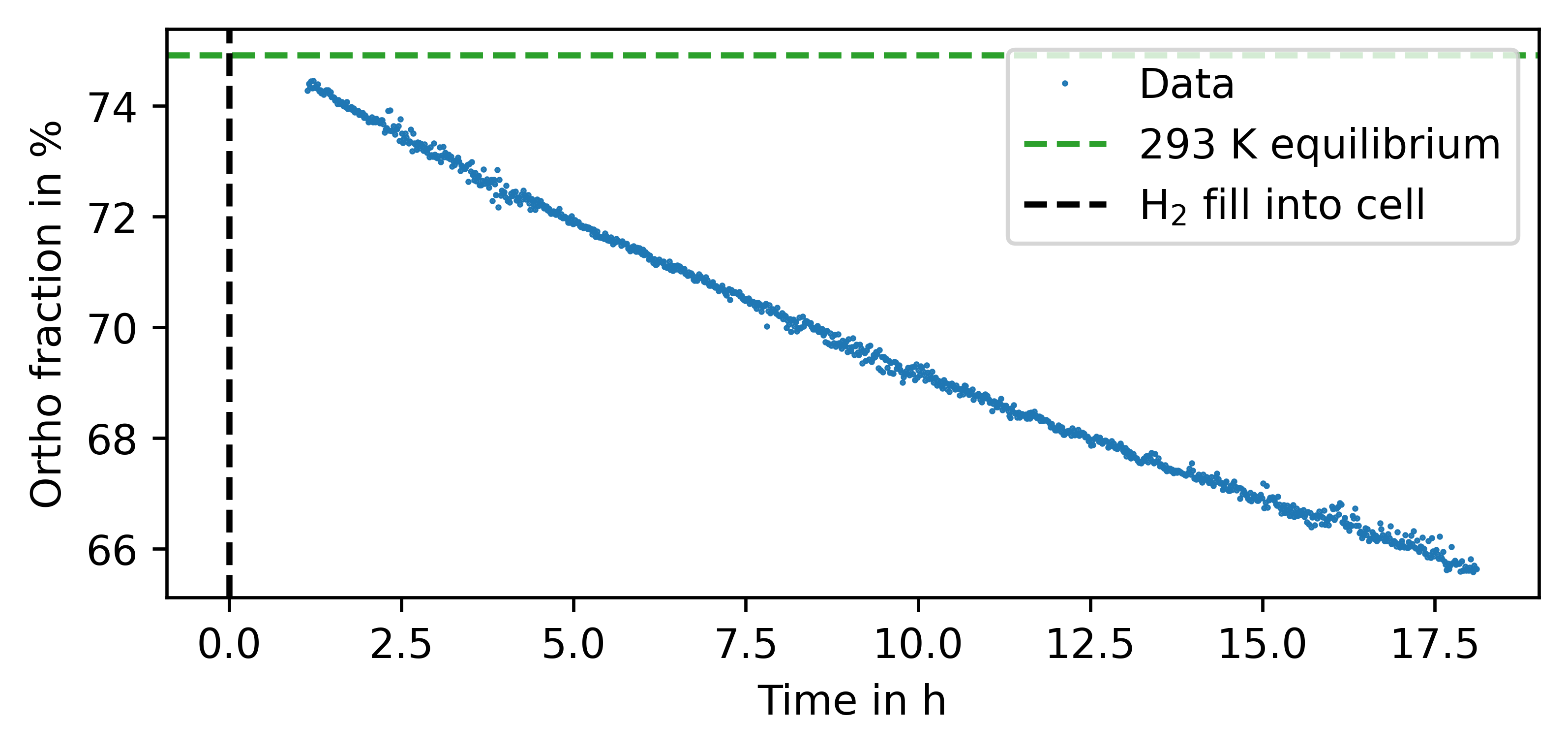}
\caption{
Evolution of ortho fraction of liquid \HTwo{}, condensed from room temperature equilibrium \orthopara{} ratio, due to natural conversion in the liquid phase.
Shown values are not calibrated, using only background subtraction and theoretical line strengths.
}
\label{fig:raman_ortho_para_conversion}
\end{figure}

Based on the different \SZeroBranch{} Raman lines of the hydrogen isotopologues, a monitoring of the \orthopara{} spin isomer state is possible.
This is done by integrating the peak areas for all the \SZero{J} peaks individually, weighting these areas with the transition matrix elements of the different transitions, and then summing up the even $J$ peaks for the para and the odd $J$ peaks for the ortho state. 
At the temperatures where hydrogen is in the liquid phase, states with $J \geq 2$ are not significantly occupied (\num{<e-6} at \SI{33}{\kelvin}), therefore only the \SZero{0} and \SZero{1} peak are present (as can be seen in \cref{fig:raman_s0}), and monitoring them is sufficient in order to determine the ortho para ratio.
An example of this is shown in \cref{fig:raman_ortho_para_conversion}, where \HTwo{} with a room temperature equilibrium \orthopara{} ratio of $\approx 75/25$ was condensed into the measurement cell.
The high densities in the liquid phase cause a gradual conversion from the ortho to the para state, which is the preferred state at low temperatures, at much higher rates than in the gas phase.

The \orthopara{} ratio measured in the cell will be calibrated against an external reference laser Raman system at room temperature to ensure accurate measurements of the chemical and spin isomer composition.
However, being able to monitor the \orthopara{} ratio in-situ will allow for a direct cross-calibration of IR spectra of the liquid or solid phase without the need to evaporate the sample into a room temperature measurement setup.
In particular for layered mixtures of liquids or solids where effects of diffusion can be of interest, this has the crucial advantage of preserving the layered structure as the Raman spectrum measurement is performed.

\section{Conclusion and outlook}

The \gls{tapir2} experiment aims to investigate the properties of all six hydrogen isotopologues, their spin isomers and their mixtures in the gaseous, liquid, and solid phase, as well as the dynamics of their phase changes.
These investigations will be primarily performed using optical methods including, polariscopy as well as infrared and Raman spectroscopy.
Commissioning measurements have been performed to verify the different methods. 

It has been shown that both IR and Raman spectroscopy can be used to obtain in-situ information about the isotopic and isomer concentration of hydrogen.
The higher density in the liquid compared to the gas phase allows for short acquisition times, enabling real-time monitoring of compositions, qualifying both spectroscopy methods as useful tools for process optimization and monitoring in applications using liquid hydrogen such as fundamental experimental physics, the hydrogen economy or the fusion fuel cycle.
Calibrating these in-situ measurement methods against each other and against an external reference will be a main focus of the upcoming activities of the \gls{tapir2} experiment.

In addition to spectroscopic methods, the option to image the cell contents and perform polariscopy measurements has already revealed interesting phenomena during the crystallization of \HTwo{} and \DTwo{}.
Based on the information obtained on the formed crystal, different procedures to grow crystals of hydrogen isotopologues and their mixtures will be investigated, aiming at finding a scalable, repeatable, and fast way of creating homogeneous crystals, as will be needed for the production of fuel pellets for magnetic confinement fusion or targets for inertial confinement fusion.

Combining both imaging and spectroscopic measurements will allow for the investigation of effects such as diffusion of hydrogen isotopologues into each other in layered systems.
In order to further improve these capabilities, an upgrade of the infrared detection system is planned, replacing the single pixel detector with a 2D focal plane array to achieve hyperspectral imaging in the infrared range.

\pagebreak

\section*{Acknowledgments}

 This work was partially supported by the Bundesministerium für Forschung, Technologie und Raumfahrt (BMFTR, German Federal Ministry of Research, Technology and Space) within the \textit{Verbundprojekt: Inertial Fusion Energy (IFE) Targetry HUB für die DT-Trägheitsfusion (IFE-Targetry-HUB)} - \textit{Teilvorhaben: Tritium-Phasenraum-Navigation (TriPaN)} provided under grant number 13F1013H.

\pagebreak
\bibliographystyle{style/ans_js}                                                                           %custom ANS journal submission template bibliography style
\bibliography{bibliography}

\end{document}